\pdfoutput=1

\documentclass[a4paper,11pt]{article}
\usepackage{jheppub}

\usepackage{bm}

\usepackage{color}
\usepackage[usenames,dvipsnames,svgnames,table]{xcolor}

\usepackage{amsmath}
\usepackage{amssymb}
\usepackage{graphicx}
\usepackage{grffile}
\usepackage{slashed}
\usepackage{soul}
\usepackage{multirow}
\usepackage{pdflscape}
\usepackage[section]{placeins}

\newcommand{\MET}{\ensuremath{E_{\rm T}^{\rm miss}}}

\def\lsim{\mathrel{\raise.3ex\hbox{$<$\kern-.75em\lower1ex\hbox{$\sim$}}}}
\def\gsim{\mathrel{\raise.3ex\hbox{$>$\kern-.75em\lower1ex\hbox{$\sim$}}}}

\newcommand{\bea}{\begin{eqnarray}}
\newcommand{\eea}{\end{eqnarray}}

\def\sw{s_W}
\def\cw{c_W}

\newcommand{\be}{\begin{equation}}
\newcommand{\ee}{\end{equation}}

\title{The NMSSM is within Reach of the LHC:\\
Mass Correlations \& Decay Signatures}

\author[a,b]{Sebastian~Baum,}
\author[c]{Nausheen~R.~Shah,}
\author[a,b,d]{Katherine~Freese}

\affiliation[a]{The Oskar Klein Centre for Cosmoparticle Physics, Department of Physics, Stockholm University, Alba Nova, 10691 Stockholm, Sweden}
\affiliation[b]{Nordita, KTH Royal Institute of Technology and Stockholm University, Roslagstullsbacken 23, 10691 Stockholm, Sweden}
\affiliation[c]{Department of Physics \& Astronomy, Wayne State University, Detroit, MI 48201, USA}
\affiliation[d]{Leinweber Center for Theoretical Physics, Department of Physics, University of Michigan, Ann Arbor, MI 48109, USA}

\emailAdd{sbaum@fysik.su.se}
\emailAdd{nausheen.shah@wayne.edu}
\emailAdd{ktfreese@umich.edu}

\preprint{NORDITA-2018-128
\\\phantom{0} \hfill LCTP-18-32
\\\phantom{0} \hfill WSU-HEP-1901}

\abstract{The Next-to-Minimal Supersymmetric Standard Model (NMSSM), the singlet extension of the MSSM which fixes many of the MSSM's shortcomings, is shown to be within reach of the upcoming runs of the Large Hadron Collider (LHC). A systematic treatment of the various Higgs decay channels and their interplay has been lacking due to the seemingly large number of free parameters in the NMSSM's Higgs sector. We demonstrate that due to the SM-like nature of the observed Higgs boson, the NMSSM's Higgs and neutralino sectors have highly correlated masses and couplings and can effectively be described by four physically intuitive parameters: the physical masses of the two CP-odd states and their mixing angle, and $\tan\beta$, which plays a minor role. The heavy Higgs bosons in the NMSSM have large branching ratios into pairs of lighter Higgs bosons or a light Higgs and a $Z$ boson. Search channels arising via these Higgs cascades are unique to models like the NMSSM with a Higgs sector larger than that of the MSSM. In order to cover as much of the NMSSM parameter space as possible, one must combine conventional search strategies employing decays of the additional Higgs bosons into pairs of SM particles with Higgs cascade channels. We demonstrate that such a combination would allow a significant fraction of the viable NMSSM parameter space containing additional Higgs bosons with masses below 1\,TeV to be probed at future runs of the LHC.}

\begin{document}

\maketitle
\flushbottom

%*********************************************************
\section{Introduction}
%*********************************************************
The discovery of the 125 GeV Standard Model (SM)-like Higgs boson~\cite{Aad:2012tfa,Chatrchyan:2012xdj} has prompted the search for additional Higgs bosons at the Large Hadron Collider (LHC). The most straightforward context for such searches is provided by two Higgs Doublet Models (2HDMs) \cite{Branco:2011iw}, which extend the SM's particle content by a second Higgs $SU(2)_L$ doublet. The simplest supersymmetric realization of a 2HDM is the Minimal Supersymmetric Standard Model (MSSM). The collider signatures of such heavy Higgs boson at the LHC have been extensively studied in the literature, see e.g. Refs.~\cite{Carena:2013ytb,Djouadi:2015jea,Craig:2015jba,Bechtle:2016kui,Tanabashi:2018oca}.

The discovery of the SM-like 125\,GeV Higgs boson also sparked renewed attention in the Next-to-Minimal Supersymmetric Standard Model (NMSSM)~\cite{Ellwanger:2009dp,Maniatis:2009re} since it not only solves the $\mu$-problem~\cite{Kim:1983dt} of the MSSM but also alleviates the fine-tuning associated with the 125\,GeV Higgs boson and the tension implied by the current lack of evidence for superpartners below the weak scale (see e.g.~\cite{Barbieri:2006bg,Hall:2011aa,Perelstein:2012qg,Farina:2013fsa}). The NMSSM augments the field content of the MSSM by a SM-singlet chiral superfield $\widehat{S}$; this extends the particle content by singlet scalar and pseudo-scalar bosons $H^{\rm S}$ and $A^{\rm S}$, which mix with their corresponding Higgs-doublet counterparts, and a singlet fermion, the singlino $\widetilde{S}$, which mixes with the neutralinos. One of the three CP-even states in the NMSSM must be identified with the 125 GeV SM-like state observed at the LHC. In the following, we reserve the notation $h_{125}$ for this SM-like Higgs boson.

The presence of these singlet states introduces new interactions and decay channels, enriching the collider phenomenology of the NMSSM Higgs sector compared to the MSSM. In particular, so-called {\it Higgs cascade decays} appear prominently, where a heavy Higgs decays into two lighter Higgs bosons or a light Higgs and a $Z$ boson, see Fig.~\ref{fig:Hdiagrams}. The presence of Higgs cascade decays warrants the extension of search strategies for additional Higgs bosons beyond the conventional search channels for heavy Higgs bosons developed mostly for models with a Higgs sector consisting of only two Higgs doublets, such as the MSSM~\cite{Kang:2013rj,King:2014xwa,Carena:2015moc,Ellwanger:2015uaz,Costa:2015llh,Baum:2017gbj,Ellwanger:2017skc,Heng:2018kyd}\footnote{Note that Higgs cascade decays have also been discussed in the context of general 2HDMs~\cite{Enberg:2018pye, Kling:2018xud}. However, the required mass splittings between the non SM-like Higgs bosons are difficult to achieve in consistent 2HDMs~\cite{Krauss:2018thf}.}. Many authors have studied the Higgs and neutralino LHC phenomenology in the NMSSM (see e.g. Refs.~\cite{Kang:2013rj,King:2014xwa,Carena:2015moc,Ellwanger:2015uaz,Costa:2015llh,Baum:2017gbj,Ellwanger:2017skc,Gherghetta:2012gb,Christensen:2013dra,Cheung:2014lqa,Dutta:2014hma,Ellwanger:2016sur} and references therein). However, these studies only cover specific regions of the NMSSM parameter space and typically consider one search channel at a time. A systematic study of the possible signals of the NMSSM Higgs sector and their correlations in parameter space has been perceived to be a challenging task due to the seemingly large number of free parameters controlling the theory. 

In this paper, we provide the first systematic approach towards categorizing the NMSSM Higgs sector. We simplify the parameter space of the theory by making use of the SM-like nature of the observed 125 GeV Higgs boson. In the region of NMSSM parameter space where the non SM-like Higgs bosons are light enough to be potentially accessible at the LHC, approximate {\it alignment without decoupling}~(see e.g. Refs.~\cite{Gunion:2002zf, gunion2008higgs, Carena:2013ooa,Carena:2014nza, Carena:2015moc}) must be realized. Such alignment implies correlations between the masses, mixing angles, and couplings in the Higgs and neutralino sector of the NMSSM. We stress that while our analytical understanding of the physically viable parameter space in the NMSSM is guided by assuming perfect alignment in the Higgs sector, we have verified our claims by extensive numerical scans over the NMSSM parameter space using \texttt{NMSSMTools}~\cite{NMSSMTools,Ellwanger:2004xm,Ellwanger:2005dv,Das:2011dg, Muhlleitner:2003vg} where alignment was {\it not} assumed a priori, see Figs.~\ref{fig:masscorrelations_Higgs} and~\ref{fig:masscorrelations_neu}.

The correlations between masses, mixing angles, and couplings become rather clouded when parameterizing the NMSSM in terms of the 7 parameters appearing in the Higgs scalar potential. We show that the region of NMSSM parameter space containing a SM-like 125\,GeV Higgs boson and additional Higgs bosons with masses below $\sim 1\,$TeV, i.e. the region most relevant for Higgs searches at the LHC, can be effectively described by only {\it four} physically intuitive parameters: the two physical masses of the CP-odd states, one mixing angle in the CP-odd sector, and $\tan\beta$. Note that the low $\tan\beta$ regime is of particular relevance for the NMSSM; there, modifying $\tan\beta$ has only minor effects on the NMSSM's phenomenology. Hence, we find that the phenomenology of the entire Higgs and neutralino sectors is governed largely by only three physical parameters in the CP-odd sector, see Eq.~\eqref{eq:Relparams}.

Our NMSSM re-parameterization in terms of the masses and mixing angles allows for transparent identification of the most relevant search strategies for different regions of parameter space. These insights allow us to analytically and numerically study, for the first time, the potential of a combination of different search channels arising via Higgs cascade decays. This categorization and coordination of possible Higgs decay channels is important both for extending the coverage of the NMSSM parameter space, as well as identifying the underlying model giving rise to a potential discovery of additional Higgs bosons in the future, e.g. distinguishing the MSSM from the NMSSM. We show that combining Higgs cascade searches with more conventional search modes via decays of non SM-like Higgs bosons into pairs of SM particles, the LHC collaborations will be able to probe $\approx 50\,\%$ of the currently remaining viable NMSSM parameter space containing additional Higgs bosons below 1\,TeV in future runs of the LHC. We also entertain the scenario that the LHC collaborations can improve the sensitivity of the Higgs cascade decay based searches by an order of magnitude with respect to our projections and the sensitivity in conventional search channels by two orders of magnitude with respect to current limits based on $\mathcal{O}(30)\,$fb of data. Then, $\approx 90\,\%$ of the remaining parameter space containing Higgs bosons below 1\,TeV could be probed in the upcoming runs of the LHC, see Figs.~\ref{fig:standardvcascades}, \ref{fig:combinedreach} and~\ref{fig:combinedreach2}. While this latter scenario is optimistic, such sensitivities should be understood as a target for the experimental collaborations which would allow them to probe much of the remaining phenomenologically interesting NMSSM parameter space.

Note that a similar approach can be used to tackle a generic 2HDM+(complex)~singlet model~\cite{Baum:2018zhf}. However, the lack of relations in the Higgs sector's parameters prevents making concrete predictions for LHC phenomenology and the interplay of the search modes.

The remainder of this paper is organized as follows. In section~\ref{sec:model}, we describe the NMSSM parameter space, the correlations in the Higgs and neutralino sector, and our re-parameterization. We validate our analytic claims with extensive numerical parameter scans. In section~\ref{sec:signals} we discuss various decay channels, their correlations, and their sensitivity at the high luminosity LHC. The coordination of search strategies to cover the parameter space of the NMSSM is presented in section~\ref{sec:reach}. We reserve section~\ref{sec:conclusion} for our conclusions. Details regarding the implemented LHC constraints, benchmark points, collider simulations, and analytic expression for the Higgs trilinear couplings are presented in Appendices~\ref{app:LHCsearch}-\ref{app:ratios}.

%*********************************************************
\section{NMSSM Parameter Space}\label{sec:model}
%*********************************************************

The Next-to-Minimal Supersymmetric Standard Model augments the MSSM particle content with a chiral superfield $\widehat{S}$ uncharged under any of the SM gauge groups. In this paper, we study the scale-invariant NMSSM, where all dimensionful parameters in the superpotential are set to zero. This model enjoys an accidental $\mathbb{Z}_3$ symmetry under which all chiral superfields transform by $e^{2\pi i/3}$. The additional terms in the superpotential with respect to the MSSM are
\begin{equation}
	W \supset \lambda \widehat{S} \widehat{H}_u \cdot \widehat{H}_d + \frac{\kappa}{3} \widehat{S}^3,
\end{equation}
where $\widehat{H}_u$, $\widehat{H}_d$ are the up- and down-type Higgs doublets and $\lambda$ and $\kappa$ are dimensionless coefficients. The $\mu \widehat{H}_u \cdot \widehat{H}_d$ term of the MSSM is forbidden in the superpotential of the scale-invariant NMSSM; however, an effective $\mu$-term is generated in the scalar potential when the scalar component of the superfield $\widehat{S}$ gets a vacuum expectation value (vev), $\mu=\lambda\langle S \rangle$. If the vev of the singlet is induced by the breaking of supersymmetry, $\langle S \rangle$ is of the order of the supersymmetry breaking scale, thereby alleviating the $\mu$-problem for low-scale supersymmetry.\footnote{We denote superfields with a hat, e.g. $\widehat{S}$, the bosonic component with the bare letter e.g. $S$, and the fermionic component with a tilde, e.g. $\widetilde{S}$.}

The terms in the scalar potential involving only the Higgs doublets and the singlet are given by~\cite{Carena:2015moc}
\begin{equation} \begin{split}
	V^{H_u, H_d, S} &= m_S^2 S^\dagger S + m_{H_u}^2 H_u^\dagger H_u + m_{H_d}^2 H_d^\dagger H_d + \left( \lambda A_\lambda S H_u \cdot H_d + \frac{\kappa}{3} A_\kappa S^3 + {\rm h.c.} \right) \\
	& \qquad + \frac{g_1^2 + g_2^2}{8} \left( H_u^\dagger H_u - H_d^\dagger H_d \right)^2 + \frac{g_2^2}{2} \left| H_d^\dagger H_u \right|^2 + \lambda^2 \left| H_u \cdot H_d \right|^2 \\
	& \qquad + \lambda^2 S^\dagger S \left( H_u^\dagger H_u + H_d^\dagger H_d \right) + \kappa^2 \left( S^\dagger S \right)^2 + \kappa \lambda \left( S^2 H_u^* \cdot H_d^* + {\rm h.c.} \right) \;, 
\end{split} \end{equation}
where the $m_i^2$ and $A_i$ are soft SUSY-breaking parameters of dimension mass squared and mass, respectively, and $g_1$ and $g_2$ are the $U(1)_Y$ and $SU(2)_L$ gauge couplings.

Trading the parameters $\{ m_{H_d}^2, m_{H_u}^2, m_S^2 \}$ for the corresponding vevs via the minimization equations, fixing\footnote{Note that we use $v = 174\,$GeV while \cite{Carena:2015moc} uses the $v=246\,$GeV convention.} $v = \sqrt{v_u^2 + v_d^2} = 174\,$GeV, and defining $\tan\beta \equiv v_u/v_d$, the scalar potential is controlled by the following parameters
\begin{equation} \label{eq:NMSSMparams}
	\{ \lambda, \kappa, \tan\beta, \mu, A_\lambda, A_\kappa \} .
\end{equation}
Note that all parameters are real in the CP-conserving NMSSM. Of the dimensionless parameters, $\lambda$ and $\tan\beta$ can be chosen positive without loss of generality, while $\kappa$ and the dimensionful parameters can have both signs.

It it useful to rotate the doublet-like states to the (extended) Higgs basis~\cite{Georgi:1978ri, Donoghue:1978cj, gunion2008higgs, Lavoura:1994fv, Botella:1994cs, Branco99, Gunion:2002zf, Carena:2015moc}\footnote{Note that there are different conventions in the literature for the Higgs basis differing by an overall sign of $H^{\rm NSM}$ and $A^{\rm NSM}$.} defined in terms of the basis $\{\rm SM, NSM, S\} = \{$SM doublet, Non-SM doublet, Singlet$\}$,
\begin{align}
	H^{\rm SM} &= \sqrt{2} {\rm Re} \left( \sin\beta H_u^0 + \cos\beta H_d^0 \right), \label{eq:Hbasis1} \\
	H^{\rm NSM} &= \sqrt{2} {\rm Re} \left( \cos\beta H_u^0 - \sin\beta H_d^0 \right), \\
	A^{\rm NSM} &= \sqrt{2} {\rm Im} \left( \cos\beta H_u^0 + \sin\beta H_d^0 \right),
	\label{eq:Hbasis-1}
\end{align} 
where the $H_i^0$ are the neutral components of the corresponding doublet fields $H_i$. The couplings to pairs of SM particles take the particularly simple form
\begin{align}
	H^{\rm SM}({\rm down}, {\rm up}, {\rm VV}) &= \left( g_{\rm SM}, g_{\rm SM}, g_{\rm SM} \right), \\
	H^{\rm NSM}({\rm down}, {\rm up}, {\rm VV}) &= \left( - g_{\rm SM} \tan\beta, g_{\rm SM}/\tan\beta, g_{\rm SM} \right), \\
	A^{\rm NSM}({\rm down}, {\rm up}, {\rm VV}) &= \left( g_{\rm SM} \tan\beta, g_{\rm SM}/\tan\beta, g_{\rm SM} \right),
\end{align}
where ``down'' (``up'') stands for pairs of down-type (up-type) SM fermions, ``VV'' for pairs of vector bosons, and $g_{\rm SM}$ is the coupling of an SM Higgs boson of the same mass to such particles. The CP-even and CP-odd interaction states from the singlet $S$ do not couple to SM particles and are defined via
\begin{equation}
	S = \frac{1}{\sqrt{2}}\left(H^{\rm S} + i A^{\rm S} \right).
\end{equation}
The charged Higgs is defined by
\begin{equation}
	H^\pm = \cos\beta H_u^\pm + \sin\beta H_d^\pm \;.
\end{equation}
The remaining degrees of freedom make up the longitudinal polarization of the $W^\pm$ and $Z$ bosons after electroweak symmetry breaking.

The elements of the symmetric squared mass matrix for the CP-even Higgs bosons in the extended Higgs basis $\{ H^{\rm SM}, H^{\rm NSM}, H^S \}$ at tree-level are
\begin{align} \label{eq:NMSSM_MS11}
	\mathcal{M}_{S,11}^2 &= m_Z^2 c_{2\beta}^2 + \lambda^2 v^2 s_{2\beta}^2 \;,
	\\ \mathcal{M}_{S,12}^2 &= - \left(m_Z^2 - \lambda^2 v^2 \right) s_{2\beta}c_{2\beta} \;,
	\\ \mathcal{M}_{S,13}^2 &= 2 \lambda v \mu \left( 1 - \frac{M_A^2}{4\mu^2} s_{2\beta}^2 - \frac{\kappa}{2 \lambda} s_{2\beta} \right) ,
	\\ \mathcal{M}_{S,22}^2 &= M_A^2 + \left( m_Z^2 - \lambda^2 v^2 \right) s_{2\beta}^2 \;,
	\\ \mathcal{M}_{S,23}^2 &= - \lambda v \mu c_{2\beta} \left( \frac{M_A^2}{2\mu^2} s_{2\beta} + \frac{\kappa}{\lambda} \right), 
	\\ \mathcal{M}_{S,33}^2 &= \frac{\lambda^2v^2}{2} s_{2\beta} \left( \frac{M_A^2}{2\mu^2} s_{2\beta} - \frac{\kappa}{\lambda} \right) + \frac{\kappa \mu}{\lambda} \left( A_\kappa + \frac{4 \kappa \mu}{\lambda} \right) , \label{eq:NMSSM_MS33}
\end{align}
where we traded $A_\lambda$ for $M_A^2$, defined as
\begin{equation} \label{eq:MA}
	M_A^2 \equiv \frac{2 \mu}{s_{2\beta}} \left( A_\lambda + \frac{\kappa \mu}{\lambda} \right) ,
\end{equation}
and used the short-hand notation
\begin{equation}
	s_\beta \equiv \sin\beta\,, \quad c_\beta \equiv \cos\beta \;.
\end{equation}

The tree-level elements of the symmetric squared mass matrix for the CP-odd Higgs boson in the basis $\{ A^{\rm NSM}, A^S \}$ are given by
\begin{align}
	\mathcal{M}_{P,11}^2 &= M_A^2 \;, \label{eq:AMat}
	\\ \mathcal{M}_{P,12}^2 &= \lambda v \left( \frac{M_A^2}{2\mu} s_{2\beta} - \frac{3 \kappa \mu}{\lambda} \right) , \label{eq:AMat12}
	\\ \mathcal{M}_{P,22}^2 &= \lambda^2 v^2 s_{2\beta} \left( \frac{M_A^2}{4\mu^2} s_{2\beta} + \frac{3\kappa}{2\lambda} \right) - \frac{3 \kappa \mu}{\lambda} A_\kappa \;, \label{eq:AMat22}
\end{align}
and the mass of the charged Higgs boson is
\begin{equation}
	m_{H^\pm}^2 = M_A^2 + m_W^2 - \lambda^2 v^2 \;.
\end{equation} 

The neutralino sector of the NMSSM is extended by the singlino $\widetilde{S}$ with respect to the MSSM. In the basis $\{\widetilde{B}, \widetilde{W}^3, \widetilde{H}_d^0, \widetilde{H}_u^0, \widetilde{S}\}$, where $\widetilde{B}$ and $\widetilde{W}^3$ are the bino and the neutral wino, respectively, and $\widetilde{H}_d^0$ and $\widetilde{H}_u^0$ are the neutral Higgsinos belonging to the respective doublet superfields, the symmetric tree-level neutralino mass matrix reads
\begin{equation} \label{eq:mneu}
	M_{\chi^0}= \begin{pmatrix} M_1 & 0 & -m_Z \sw c_\beta & m_Z \sw s_\beta & 0
	\\ & M_2 & m_Z \cw c_\beta & -m_Z \cw s_\beta & 0
	\\ & & 0 & -\mu & -\lambda v s_\beta
	\\ & & & 0 & -\lambda v c_\beta
	\\ & & & & 2\kappa \mu / \lambda \end{pmatrix},
\end{equation}
where $s_W \equiv \sin\theta_W,$ with $\theta_W$ the weak mixing angle. In this paper, we decouple the gauginos from the collider phenomenology by setting the bino and wino mass parameters $\{|M_1|, |M_2|\} \gg |\mu|$.

%*********************************************************
\subsection{Alignment}
%*********************************************************

The (neutral) interaction states of the Higgs basis mix into three CP-even and two CP-odd mass eigenstates. We denote the CP-even mass eigenstates $h_i = \{ h_{125}, H, h\}$,
\begin{equation}\label{eq:h_mix}
	h_i = S_{h_i}^{\rm SM} H^{\rm SM} + S_{h_i}^{\rm NSM} H^{\rm NSM} + S_{h_i}^{\rm S} H^{\rm S} \;,
\end{equation}
where $h_{125}$ is identified with the $m_{h_{125}} \approx 125\,$GeV SM-like state observed at the LHC, $H$ and $h$ are the new eigenstates ordered by masses, $m_H > m_h$, and $S_{h_i}^j$ denotes the $j = \{\rm SM, NSM, S\}$ component of the $h_i$ mass eigenstate. Likewise, we denote the two CP-odd mass eigenstates $a_i = \{A, a\}$,
\begin{equation} \label{eq:a_mix}
	a_i = P_{a_i}^{\rm NSM} A^{\rm NSM} + P_{a_i}^{\rm S} A^{\rm S} \;,
\end{equation}
where again $m_A > m_a$, and $P_{a_i}^j$ denotes the $j = \{\rm NSM, S\}$ component of the $a_i$ mass eigenstate. The $S_{h_i}^j$ and $P_{a_i}^j$ are obtained by diagonalizing the squared mass matrices for the CP-even states, Eqs.~\eqref{eq:NMSSM_MS11}--\eqref{eq:NMSSM_MS33}, and CP-odd states, Eqs.~\eqref{eq:AMat}--\eqref{eq:AMat22}, respectively.
 
The measured branching ratios of the 125\,GeV mass eigenstate observed at the LHC are compatible with those of a SM Higgs boson, although current experimental precision allows for $\mathcal{O}(10\,\%)$ deviations~\cite{Khachatryan:2016vau,CMS-PAS-HIG-16-042,CMS-PAS-HIG-17-031,Sirunyan:2017exp,ATLAS-CONF-2017-047}. Thus, in order to be compatible with the observed phenomenology, the $h_{125}$ eigenstate we identify with the observed Higgs boson must have a mass of $\sim 125\,$GeV and be dominantly composed of the interaction eigenstates $H^{\rm SM}$, whose couplings are identical to the SM Higgs boson's.

As is well known, the (squared) mass of the $h_{125}$ mass eigenstate receives an additional contribution $\lambda^2 v^2 s_{2\beta}^2$ relative to the MSSM case, and at tree-level is given by
\begin{equation}
	m_{h_{125}}^2 \simeq \mathcal{M}_{S,11}^2 = m_z^2 \cos^2(2\beta) + \lambda^2 v^2 \sin^2(2\beta)\,.
\end{equation} 
For small to moderate values of $\tan\beta$ the $\lambda^2$ contribution to the mass is sizable and allows for a tree-level mass of $h_{125}$ close to $125\,$GeV. This makes the low $\tan\beta$ region in the NMSSM particularly interesting because there the observed mass of the SM-like Higgs boson can be obtained without the need for large radiative corrections, as for example are required in the MSSM.

There are two possibilities to achieve approximate {\it alignment} of $H^{\rm SM}$ with $h_{125}$~\cite{Craig:2012vn,Craig:2013hca,Carena:2013ooa,Haber:2013mia,Das:2015mwa,Dev:2015bta,Bernon:2015qea,Carena:2015moc}: Either, the remaining mass eigenstates $H$ and $h$ are much heavier than 125\,GeV, the so-called {\it decoupling limit}, or, the parameters of the model conspire to (approximately) cancel the entries of the mass matrix corresponding to the mixing of $H^{\rm SM}$ with $H^{\rm NSM}$ and $H^{\rm S}$, the so-called {\it alignment (without decoupling) limit}~\cite{Carena:2015moc}. The latter option is of particular interest for LHC phenomenology since it allows the additional Higgs bosons to remain relatively light and thus accessible at the LHC. 

Including the dominant contributions from stop loops absorbed in the definition of $\mathcal{M}_{S,11}^2$~\cite{Carena:2015moc}
\begin{equation}
	\mathcal{M}_{S,12}^2 = \frac{1}{t_\beta} \left( \mathcal{M}_{S,11}^2 - m_Z^2 c_{2\beta} - 2 \lambda^2 v^2 s_\beta^2 \right) \;,
\end{equation}
and identifying $\mathcal{M}_{S,11} \simeq m_{h_{125}}^2$, we can write the conditions for alignment as
\begin{align} \label{eq:alignNMSSM21}
	\lambda^2 &= \frac{m_{h_{125}}^2 - m_Z^2 c_{2\beta}}{2 v^2 s_\beta^2} \;, \\
	\frac{\kappa}{\lambda} &= \left( \frac{2}{s_{2\beta}} - \frac{M_A^2}{2 \mu^2} s_{2\beta} \right) \;,\label{eq:alignNMSSM22}
\end{align} 
where the first condition ensures $\mathcal{M}_{S,12}^2 \to 0$, suppressing the mixing of $H^{\rm SM}$ with $H^{\rm NSM}$, and the condition in the second line ensures $\mathcal{M}_{S,13}^2 \to 0$, suppressing the mixing of $H^{\rm SM}$ with $H^{\rm S}$.

Close to the alignment limit, the CP-even mass matrix approximately reduces to a $2\times2$ system for $\{H^{\rm NSM}, H^{\rm S}\}$ which then form the mass eigenstates $\{H, h\}$. In this case, the mixing angle is simply given by
\begin{equation}\label{eq:MS23Align}
	\mathcal{M}_{S,23}^2 \approx -\frac{2\lambda v \mu}{ t_{2\beta}} = S_{H}^{\rm NSM} S_{H}^{\rm S} \left( m_H^2-m_h^2 \right).
\end{equation}
Eliminating the dependence on $M_A$ and $\mu$ using Eq.~\eqref{eq:alignNMSSM22}, the CP-odd mixing angle is instead given by
\begin{equation}\label{eq:MP12Align}
	\mathcal{M}_{P,12}^2 \approx \frac{2\lambda v \mu}{s_{2\beta}}\left(1-\frac{2\kappa}{\lambda} s_{2\beta} \right) = P_{A}^{\rm NSM} P_{A}^{\rm S} \left( m_A^2-m_a^2 \right).
\end{equation}
From the above, we see that if $|\kappa|/\lambda$ is small, $\mathcal{M}_{P,12}^2 \sim \mathcal{M}_{S,23}^2$. On the other hand, while the mixing in the CP-odd sector can be suppressed by judicious choices of larger values of $\kappa/\lambda$ consistent with the alignment limit, the $H^{\rm NSM} - H^{\rm S}$ mixing in the CP-even sector will usually remain sizable, and can only be eliminated for $\tan\beta = 1$. 

%*********************************************************
\subsection{Physical Re-Parameterization}\label{sec:CorrelConsequence}
%*********************************************************

The widely used description of the Higgs sector of the NMSSM in terms of the parameters appearing in the scalar potential listed in Eq.~\eqref{eq:NMSSMparams} does not reflect the correlations in the parameters due to the SM-like nature of $h_{125}$ in a transparent fashion. Instead, it is useful to re-parameterize the physically relevant region of parameter space by approximate alignment and the physical masses of the CP-odd Higgs bosons. The remaining freedom of the parameter space can be described by the mixing angle in the CP-odd sector $P_A^{\rm S}$ [Eq.~\eqref{eq:a_mix}] and the value of $\tan\beta$. Hence the basis
\begin{equation}\label{eq:Physparams}
	\{ \lambda,\; \kappa, \; \tan\beta,\; m_A,\; m_a,\; P_A^{\rm S}\} ,
\end{equation}
is physically much more intuitive than the usual parameterization in terms of the parameters appearing in the scalar potential, cf. Eq.~\eqref{eq:NMSSMparams}. 

While current experimental constraints allow for $\lambda$ and $\kappa$ to be slightly shifted from the values expected from perfect alignment, in practice, we can use the alignment conditions, Eqs.~\eqref{eq:alignNMSSM21} and~\eqref{eq:alignNMSSM22}, to fix $\lambda$ and $\kappa$ to a very good approximation~\cite{Carena:2015moc,Baum:2017gbj}. Further, it is well known (and easily seen from the mass matrices) that the precise value of $\tan\beta$ is a small effect in the low $\tan\beta$ regime, which is of most interest in the NMSSM. Hence, the phenomenology of the Higgs and neutralino sectors is, to a large degree, governed by the three parameters
\begin{equation}\label{eq:Relparams}
	\{ m_A,\; m_a,\; P_A^{\rm S}\} .
\end{equation}

Keeping the $\tan\beta$ dependence but assuming alignment, the NMSSM parameters listed in Eq.~\eqref{eq:NMSSMparams} can be obtained in terms of the more physical parameters listed in Eq.~\eqref{eq:Physparams} by using the elements of the CP-odd mass matrix Eqs.~\eqref{eq:AMat}-\eqref{eq:AMat22}. The value of $M_A^2$ can be obtained directly from the definition of the $\mathcal{M}_{P,11}^2$ matrix element
\begin{equation} \label{eq:getMA}
	M_A^2 = (P_A^{\rm NSM})^2 m_A^2 + (P_A^{\rm S})^2 m_a^2 \, .
\end{equation}
The value of $\mu$ can be obtained from $\mathcal{M}_{P,12}^2$ as given in Eq.~\eqref{eq:MP12Align}, and using the relationship for $\kappa/\lambda$ as dictated by alignment, Eq.~\eqref{eq:alignNMSSM22},
\begin{equation} \label{eq:getmu}
	\mu = - \frac{s_{2\beta}}{12\lambda v} P_A^{\rm NSM} P_A^{\rm S} \left( m_A^2 - m_a^2 \right) \left[ 1 \pm \sqrt{ 1 + \frac{48 \lambda^2 v^2 M_A^2}{(P_A^{\rm NSM})^2 (P_A^{\rm S})^2 \left( m_A^2 - m_a^2 \right)^2}}~ \right] ,
\end{equation}
with the corresponding
\begin{equation}\label{eq:getkappa}
	\frac{\kappa}{\lambda} = \frac{1}{2 s_{2\beta}} \left[1- \frac{P_A^{\rm NSM} P_A^{\rm S} \left( m_A^2 - m_a^2 \right) s_{2\beta}}{2\lambda v \mu} \right]~.
\end{equation}
Finally, from the matrix element $\mathcal{M}_{P,22}^2 =[(P_A^{\rm S})^2 m_A^2 + (P_A^{\rm NSM})^2 m_a^2]$ we obtain $A_\kappa$,
\begin{equation}\label{eq:getAkappa}
	A_\kappa = \frac{\lambda}{3\kappa\mu} \left[ \lambda^2 v^2 \left( 3 - \frac{M_A^2}{2\mu^2} s_{2\beta}^2 \right) - (P_A^{\rm S})^2 m_A^2 - (P_A^{\rm NSM})^2 m_a^2 \right] .
\end{equation}
where $M_A^2$ in Eqs.~\eqref{eq:getmu} and~\eqref{eq:getAkappa} is given by Eq.~\eqref{eq:getMA}, $\mu$ in Eqs.~\eqref{eq:getkappa} and ~\eqref{eq:getAkappa} by Eq.~\eqref{eq:getmu}, $P_A^{\rm NSM} = \sqrt{1-(P_A^{\rm S})^2}$, and the alignment relations have been assumed for $\kappa$ and $\lambda$. Note that for each set of input parameters $\{m_a, m_A, P_A^{\rm S}\}$, there are two sets of correlated solutions for $\mu$, $\kappa$ and $A_\kappa$. In our analytical formulae and figures, we will denote these by $\mu^\pm$. We also note that $P_A^{\rm S} \leftrightarrow (- P_A^{\rm S})$ corresponds to $\mu^\pm \leftrightarrow (- \mu^\mp)$. 

We stress that these relations define the masses as well as all couplings between the NMSSM Higgs bosons and between Higgs bosons, neutralinos\footnote{Here, we assume the bino and wino to be much heavier than the singlino and the Higgsinos. The mass of the bino and wino is specified by the additional parameters $M_1$ and $M_2$, respectively.}, and SM particles from the input parameters $\{\tan\beta, m_A, m_a, P_A^{\rm S}\}$, assuming (approximate) alignment as dictated by $h_{125}$ phenomenology.

A comment about radiative corrections is in order here. In general, sizable corrections are present in the NMSSM, in particular via stop loops due to the large top Yukawa couplings as well as via Higgs loops via potentially large quartic couplings between the Higgs bosons, see e.g. Refs.~\cite{Ellwanger:2009dp,Carena:2011jy,Cheung:2014lqa,Carena:2015moc}. Since our re-parameterization of the parameter space is obtained at tree level except for the first alignment condition, Eq.~\eqref{eq:alignNMSSM21}, not all such corrections are  explicitly included. This may somewhat cloud the relation of our parameter basis, which uses physical masses, a mixing angle, $\tan\beta$, and the couplings $\lambda$ and $\kappa$, with the usual parameterization of the Higgs sector in terms of the parameters appearing in the scalar potential. Our relations in Eqs.~\eqref{eq:getMA}--\eqref{eq:getAkappa} should strictly be understood as relations to obtain parameters shifted with respect to the bare parameters after absorbing relevant radiative corrections. Note also that we  did not include obtaining $m_{h_{125}} = 125\,$GeV as a condition on our parameter basis, rather, the required mass of the SM-like eigenstate should be understood as setting the size of the stop corrections. In the NMSSM, a 125\,GeV mass for the SM-like Higgs mass eigenstate can be obtained without large radiative corrections. Thus, the phenomenology of the Higgs and neutralino sector can be studied in a region of parameter space where the radiative corrections from the stops are small and the relation of the parameters obtained from our Eqs.~\eqref{eq:getMA}--\eqref{eq:getAkappa} with input parameters for numerical tools is rather direct. Thus, even though the parameters obtained from Eqs.~\eqref{eq:getMA}--\eqref{eq:getAkappa}  cannot generally be directly used as input in spectrum generators like \texttt{NMSSMTools}, \texttt{SOFTSUSY}, \texttt{NMSSMCALC}, etc, in practice, this is a minor problem as discussed further in the following section. Finally, the main advantage of our re-parameterization is that it allows for the  transparent understanding of the Higgs sector in the physically viable region of parameter space. While a precision study would require one to carefully incorporate radiative corrections, here we are interested in mapping the qualitative behavior and in identifying search strategies to cover as much of the NMSSM's parameter space as possible. Radiative corrections  shift the parameters, but do not impact the qualitative behavior of the NMSSM.

%*********************************************************
\subsection{Mass Correlations} \label{sec:paramCorrel}
%*********************************************************

In the previous section we re-parameterized the NMSSM parameters governing the Higgs and neutralino sectors of the NMSSM. We showed that in the alignment limit, only four free parameters remain in the Higgs sector. Most of the phenomenology is controlled by $\{m_a, m_A, P_A^{\rm S}\}$, while $\tan\beta$ plays a minor role. This leads to strong correlations between the masses in the Higgs sector as well as between the Higgs masses and the Higgsino and singlino parameters. 

From the form of the mass matrices, it is straightforward to see that the scale of all masses, except for the SM-like Higgs state, are controlled by the parameter $|\mu|$, which is in turn highly correlated with $M_A$ due to the requirement of approximate alignment~[see Eq.~\eqref{eq:getmu}]. Because of the large numerical factor in the square root in Eq.~\eqref{eq:getmu}, regardless of the mixing angles and the mass splitting in the CP-odd sector,
\begin{equation}\label{eq:muApprox}
	|\mu| \sim \frac{M_A s_{2\beta}}{2}.
\end{equation}
Combining this with the alignment condition given in Eq.~\eqref{eq:alignNMSSM22} dictates that $|\kappa|/\lambda$ should be small for most of the region under consideration, however it can be driven to larger values even for small deviations of $\mu$ from Eq.~\eqref{eq:muApprox},
\begin{equation}
	\epsilon \equiv 2|\mu|-M_A s_{2\beta}\;,
\end{equation}
due to the $1/s_{2\beta}$ dependence of the alignment condition. These quantities are most directly related to the neutralino sector. The mass of the Higgsinos is controlled by $\mu$, while the singlino mass is parameterized by $2\kappa \mu/\lambda$, cf. Eq.~\eqref{eq:mneu}. The mixing between the Higgsinos and the singlino is~\cite{Cheung:2014lqa}
\begin{equation}
	\frac{N_{i 3}^2+N_{i4}^2}{N_{i5}^2} = \frac{\lambda^2 v^2}{\left(\mu^2 - m_{\chi_i}^2 \right)^2} \left( m_{\chi_i}^2+\mu^2-2 \mu \,m_{\chi_i} s_{2\beta} \right),
\end{equation}
where the $\chi_i$ with $1 \leq i \leq 5$ are the neutralino mass eigenstates in ascending order of their masses $m_{\chi_i}$, and the $N_{ij}$ denote the interaction eigenstate components of the $\chi_i$ with $j = \{3,4,5\} = \{\widetilde{H}_d^0, \widetilde{H}_u^0, \widetilde{S}\}$. Since $\tan\beta$ is small, the neutralino sector is mostly controlled by the mass splitting between $m_{\chi_i}$ and $\mu$, and hence by the value of $\kappa/\lambda$ given by Eq.~\eqref{eq:getkappa}.\footnote{This holds under the assumption that the absolute value of the bino and wino mass parameters $|M_1|$ and $|M_2|$ are much larger than $|\mu|$. However, note that allowing the bino and the wino to be light does not add new parameters beyond $M_1$ and $M_2$ since the bino and wino mix only with the Higgsinos, and the mixing is controlled only by $\tan\beta$.} Thus, we expect the masses of the Higgsino-like neutralinos to be correlated with the mass of the doublet-like Higgs states, while the mass of the singlino-like state is much more strongly effected by $m_a$ and $P_A^{\rm S}$.
 
Considering the Higgs sector, we first note from Eq.~\eqref{eq:alignNMSSM22} that $2 s_{2\beta} |\kappa|/\lambda\propto \epsilon/|\mu| \ll 1$, and hence the mixing angles and the heavy masses in the CP-odd and even sectors are generally expected to be correlated, cf. Eqs.~\eqref{eq:MS23Align} and \eqref{eq:MP12Align}, with masses approximately given by $M_A$. The singlet-like states are less tightly correlated; taking into account first-order mixing effects, their masses can be approximated as~\cite{Carena:2015moc}~\footnote{For compactness, we denote the masses of the singlet-like CP-even and CP-odd mass eigenstate by $m_h$ and $m_a$ here, respectively. A priori, the singlet-like states can be heavier than the doublet-like states, in such a case the singlet-like states should be identified with $A$ or $H$ in our notation. It should be noted however that for typical choices of parameters the singlet-like masses are lighter than $M_A$, such that they comprise the lighter CP-odd and non SM-like CP-even mass eigenstates.}
\begin{align}
	m_h^2 &\simeq \frac{\kappa \mu}{\lambda} \left(A_\kappa + \frac{4 \kappa \mu}{\lambda} \right) + \lambda^2 v^2 s_{2\beta}^4 \frac{M_A^2}{4\mu^2} - \frac{\lambda \kappa v^2}{2} s_{2\beta} \left(1+2c_{2\beta}^2\right) - \kappa^2 v^2 \frac{\mu^2}{M_A^2} c_{2\beta}^2 \;, \\
	m_a^2 &\simeq 3 \kappa v^2 \left[ \frac{3}{2} \lambda s_{2\beta} - \left(\frac{1}{\lambda} \frac{\mu A_\kappa}{v^2} + 3 \kappa \frac{\mu^2}{M_A^2} \right) \right] .
\end{align} 
Note that the sum $(3 m_h^2 + m_a^2)$ is independent of $A_\kappa$~\cite{Carena:2015moc}. The opposite sign contribution from $A_\kappa$ to $m_h^2$ and $m_a^2$ induces an anti-correlation in their masses for fixed values of the remaining parameters. Further, compared to the CP-even state, the singlet-like CP-odd state receives a factor of 3 larger contribution from $A_\kappa$, and no large contribution from either $M_A$ or $\mu$. Thus, $m_a$ has the smallest correlation with $M_A$ (or $\mu$) of the non SM-like Higgs states, justifying our choice of parameterization for the Higgs sector in Eq.~\eqref{eq:Physparams}. On the other hand, apart from its anti-correlation with $m_a$, the mass of the singlet-like CP-even state receives large contributions proportional to $\kappa^2\mu^2/\lambda^2$, and is thus expected to be quite correlated with the masses of the doublet-like states, as well as the CP-odd mixing angle. It can be further shown that in the parameter region of interest, the maximal value for the CP-even state is obtained for the smallest values of $m_a$, and generally obeys $m_h \lesssim M_A/2$~\cite{Carena:2015moc}. 

\begin{table}
 	\begin{centering}
 		\begin{tabular}{ccc}
 		 	\hline\hline
 		 	& ``standard'' & ``light subset'' \\ \hline
 		 	$\tan\beta$ & $ \left[ 1; 5 \right] $ & $ \left[ 1; 5 \right] $ \\ 
 		 	$\lambda$ & $ \left[ 0.5; 2 \right] $ & $ \left[ 0.5; 1 \right] $ \\ 
 		 	$\kappa$ & $ \left[ -1; +1 \right] $ & $ \left[ -0.5; +0.5 \right] $ \\ 
 		 	$A_\lambda$ & $ \left[ -1; +1 \right] \,$TeV & $ \left[ -0.5; +0.5 \right] \,$TeV \\ 
 		 	$A_\kappa$ & $ \left[ -1; +1 \right] \,$TeV & $ \left[ -0.5; +0.5 \right] \,$TeV \\ 
 		 	$\mu$ & $ \left[-1; +1 \right] \,$TeV & $ \left[ -0.5; +0.5 \right] \,$TeV \\ 
 		 	$M_{Q_3}$ & $ \left[ 1; 10 \right] \,$TeV & $ \left[ 1; 10 \right] \,$TeV \\ \hline \hline
 		\end{tabular}
 		\caption{NMSSM parameter ranges used in \texttt{NMSSMTools} scans. We decouple the remaining supersymmetric partners by setting all sfermion mass parameters (except the stop parameters $M_{Q_3} = M_{U_3}$) to 3\,TeV, the bino and wino mass parameters to $M_1 = M_2 = 1\,$TeV, and the gluino mass to $2\,$TeV. The stop and sbottom mixing parameters are set to $X_t \equiv (A_t - \mu \cot\beta) = 0$ and $X_b \equiv (A_b - \mu\tan\beta) = 0$. See the text for a discussion of these choices as well as the ranges of the parameters we scan over.}
 		\label{tab:scan_param}
 	\end{centering}
\end{table}

\begin{figure}
 	\begin{center}
 		\includegraphics[width=0.49\linewidth]{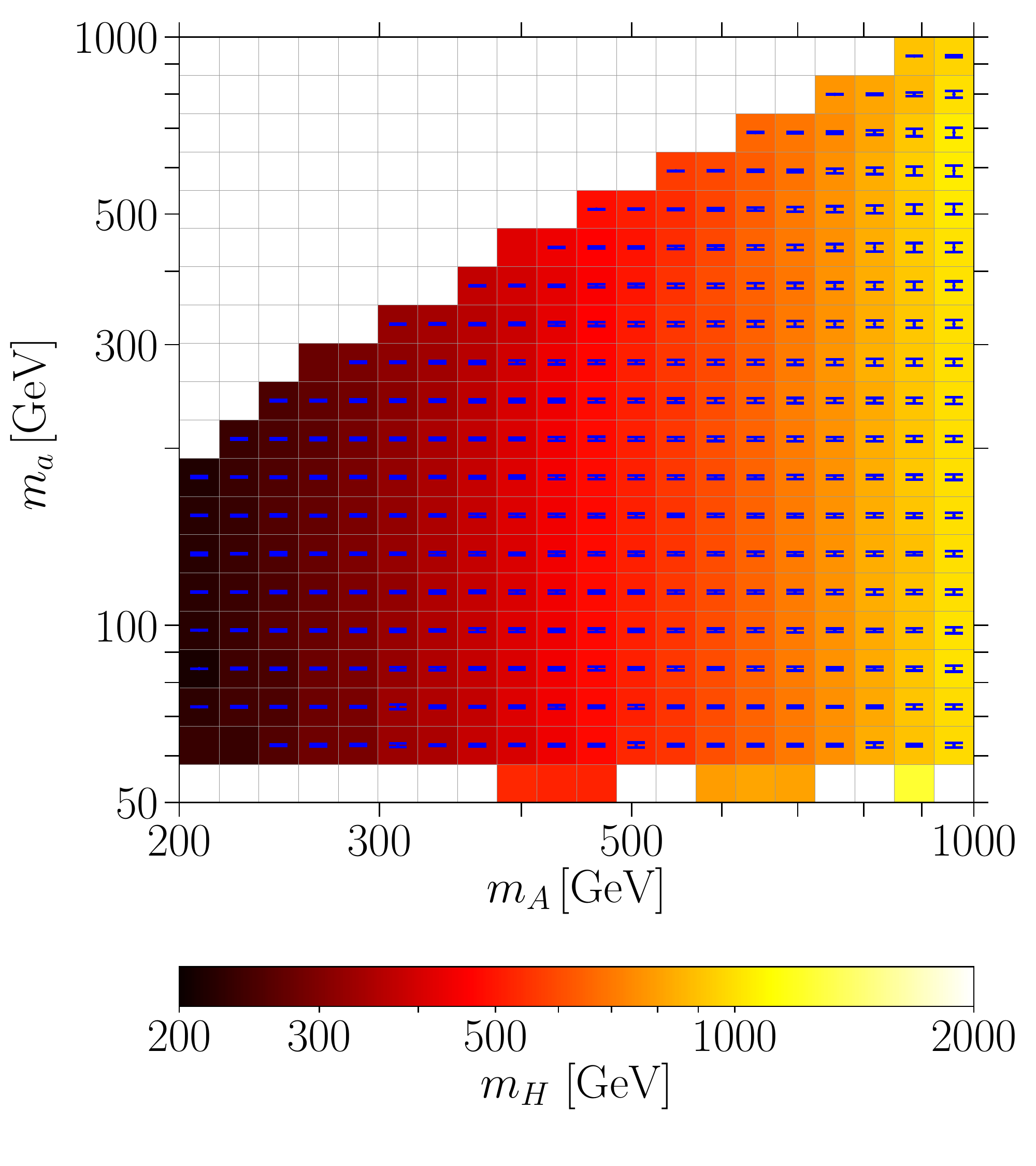}
 		\includegraphics[width=0.49\linewidth]{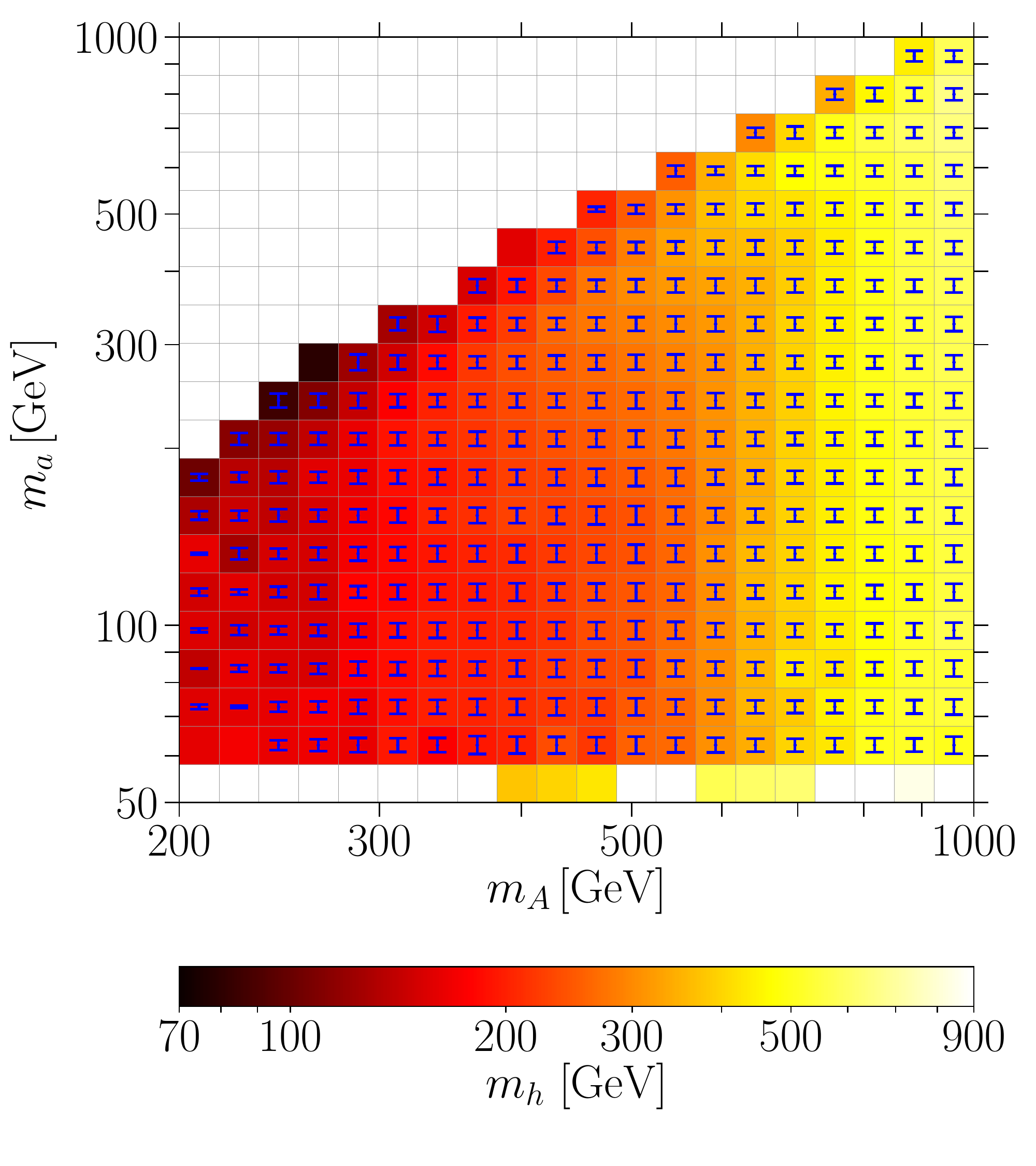}
 		\caption{The additional CP-even non SM-like Higgs boson masses $m_H$ and $m_h$ in the NMSSM parameter space. We show the statistical properties of points from our parameter scan as discussed in the text. The color scale in the left (right) shows the mean of the mass of $H$ ($h$) in the $m_A$--$m_a$ plane. Recall that the non SM-like states are defined by their mass ordering, $m_h < m_H$ and $m_a < m_A$. For the observed SM-like 125\,GeV Higgs state we reserve the notation $h_{125}$. As discussed in the text, usually the heavier states $H$ and $A$ are mostly composed of the non SM-like doublet interaction states, while the lighter states $h$ and $a$ are usually singlet-like. The blue error-bars show the standard deviation of the masses in the respective bin, normalized such that the error-bar would span the height of the bin if the standard deviation is equal to the mean. Note that the scale of the error-bar is linear with respect to the bin height, while the bin widths as well as the color scale are logarithmic. Note also the different color scaling for the masses in the left and right panels. Bins containing only 1 data point receive no error-bar, while bins without any data points are white.}
 		\label{fig:masscorrelations_Higgs}
 	\end{center}
\end{figure}

\begin{figure}
 	\begin{center}
		\includegraphics[width=0.49\linewidth]{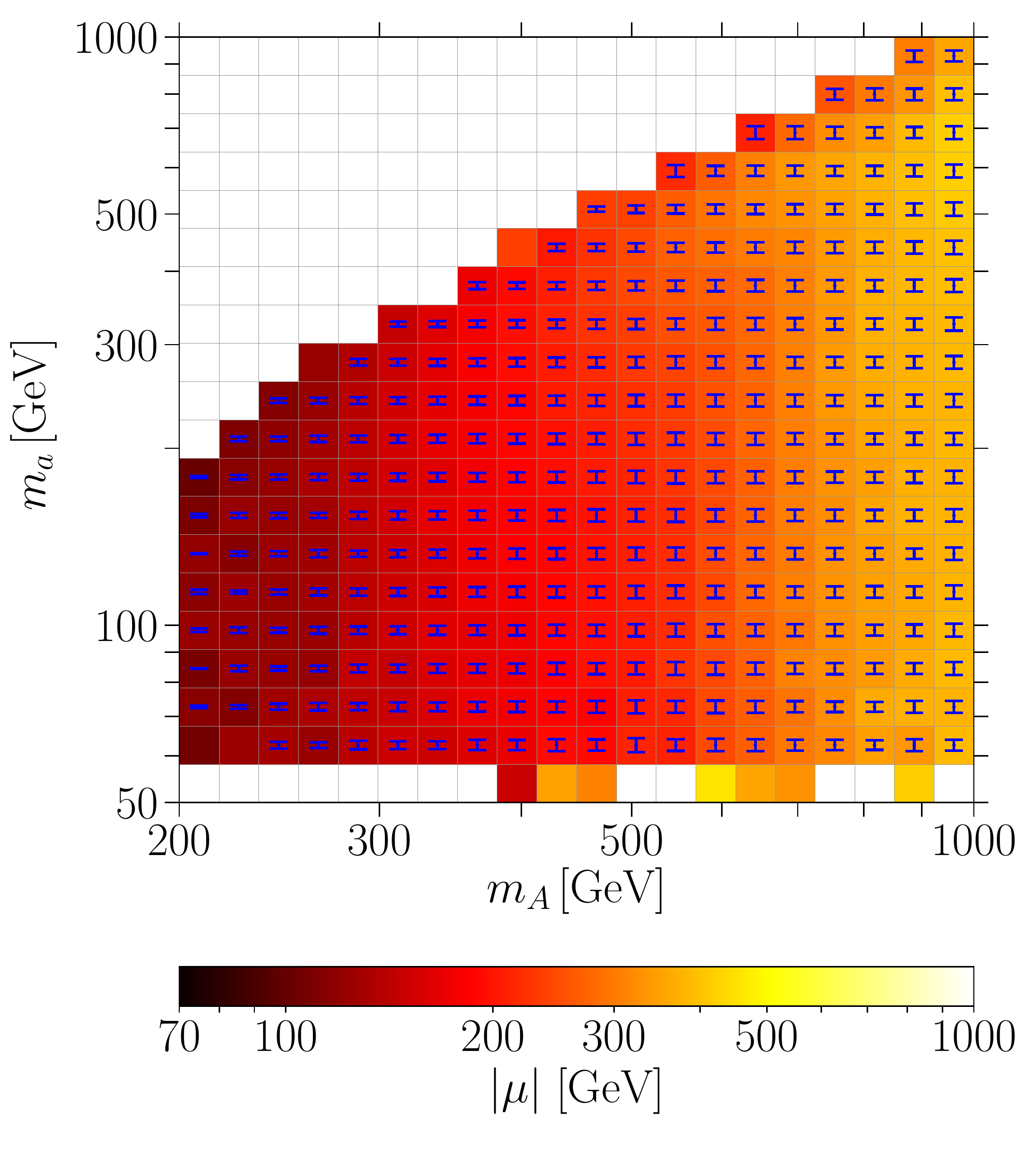}
		\includegraphics[width=0.49\linewidth]{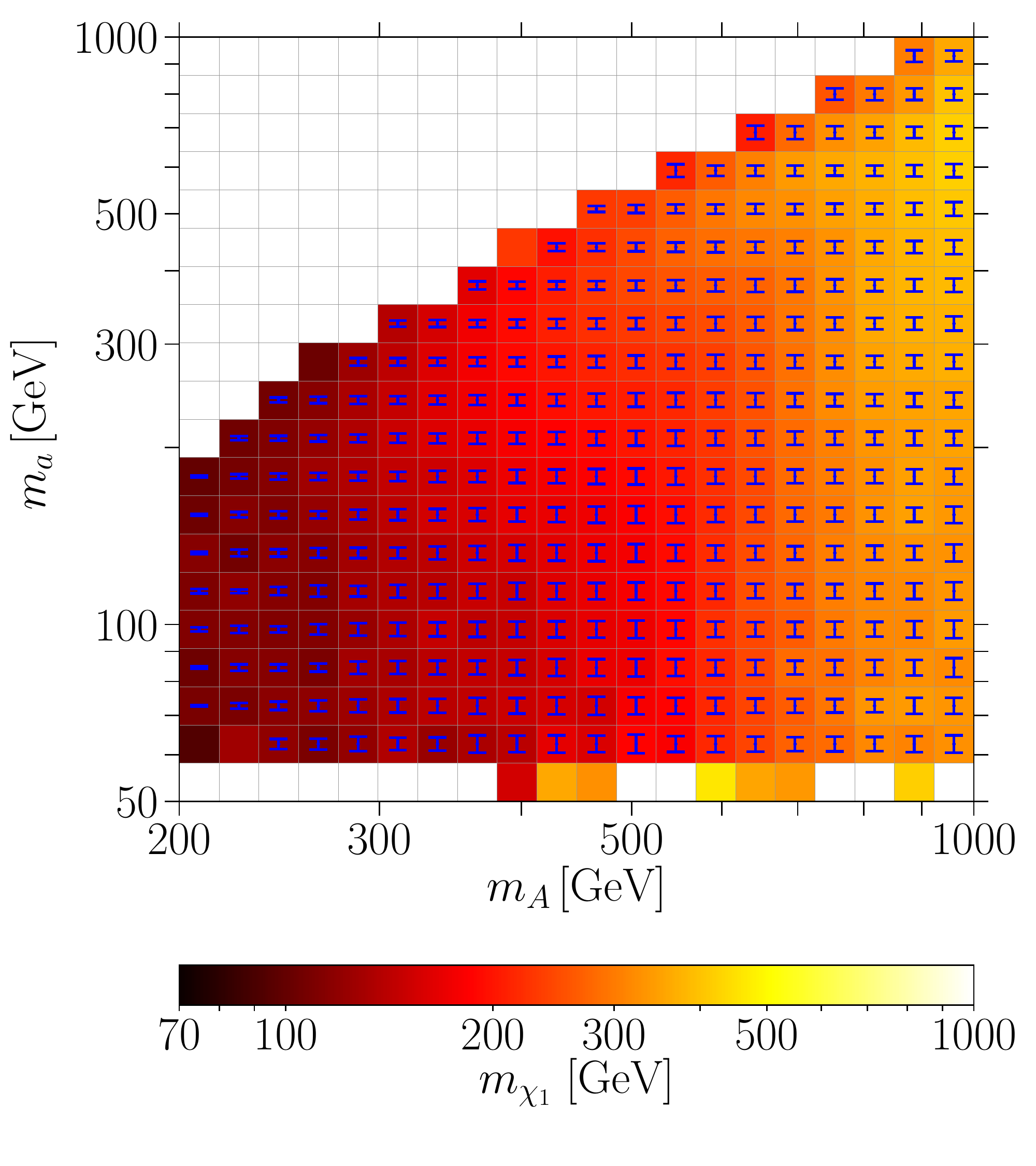}

		\includegraphics[width=0.49\linewidth]{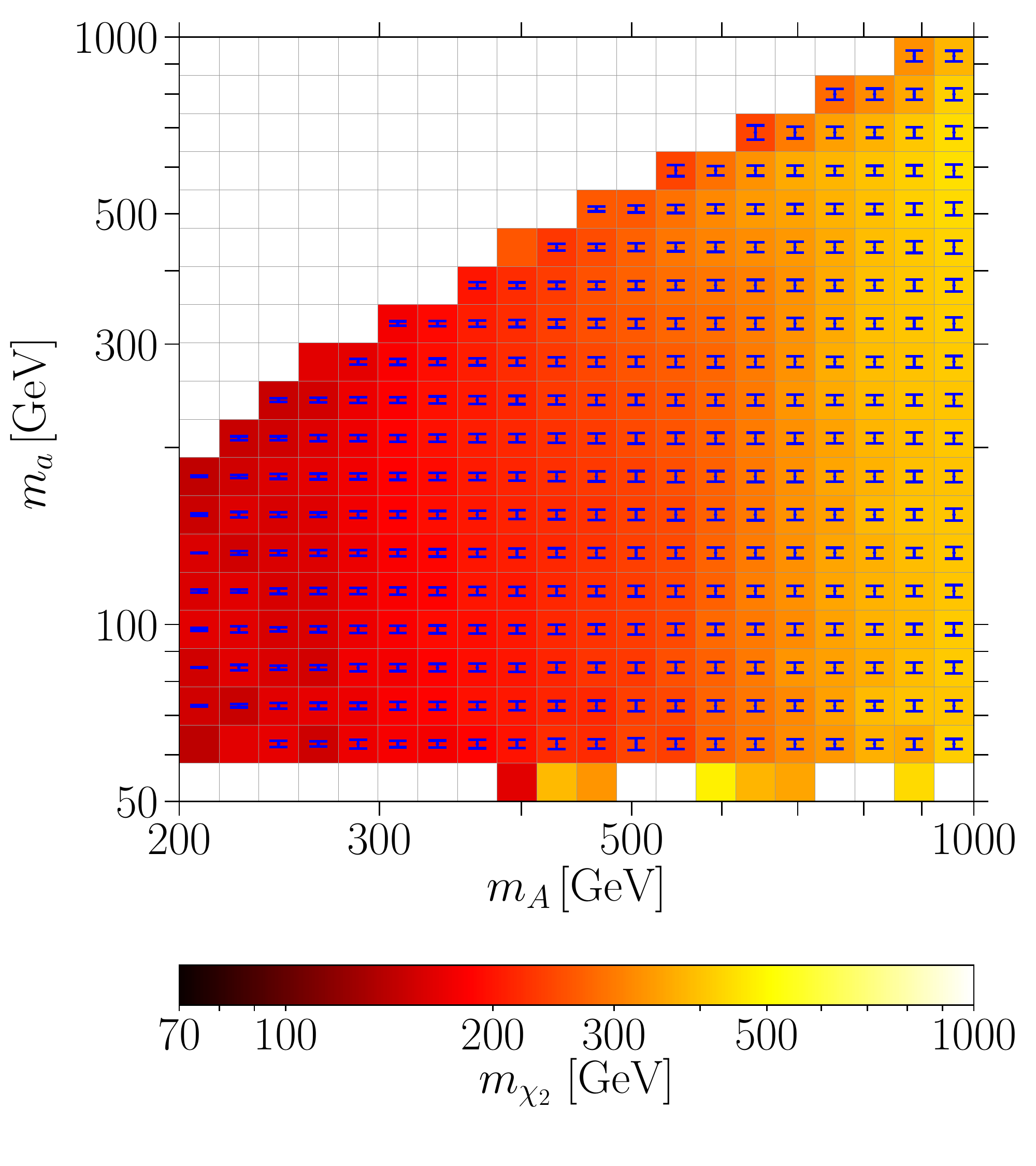}
		\includegraphics[width=0.49\linewidth]{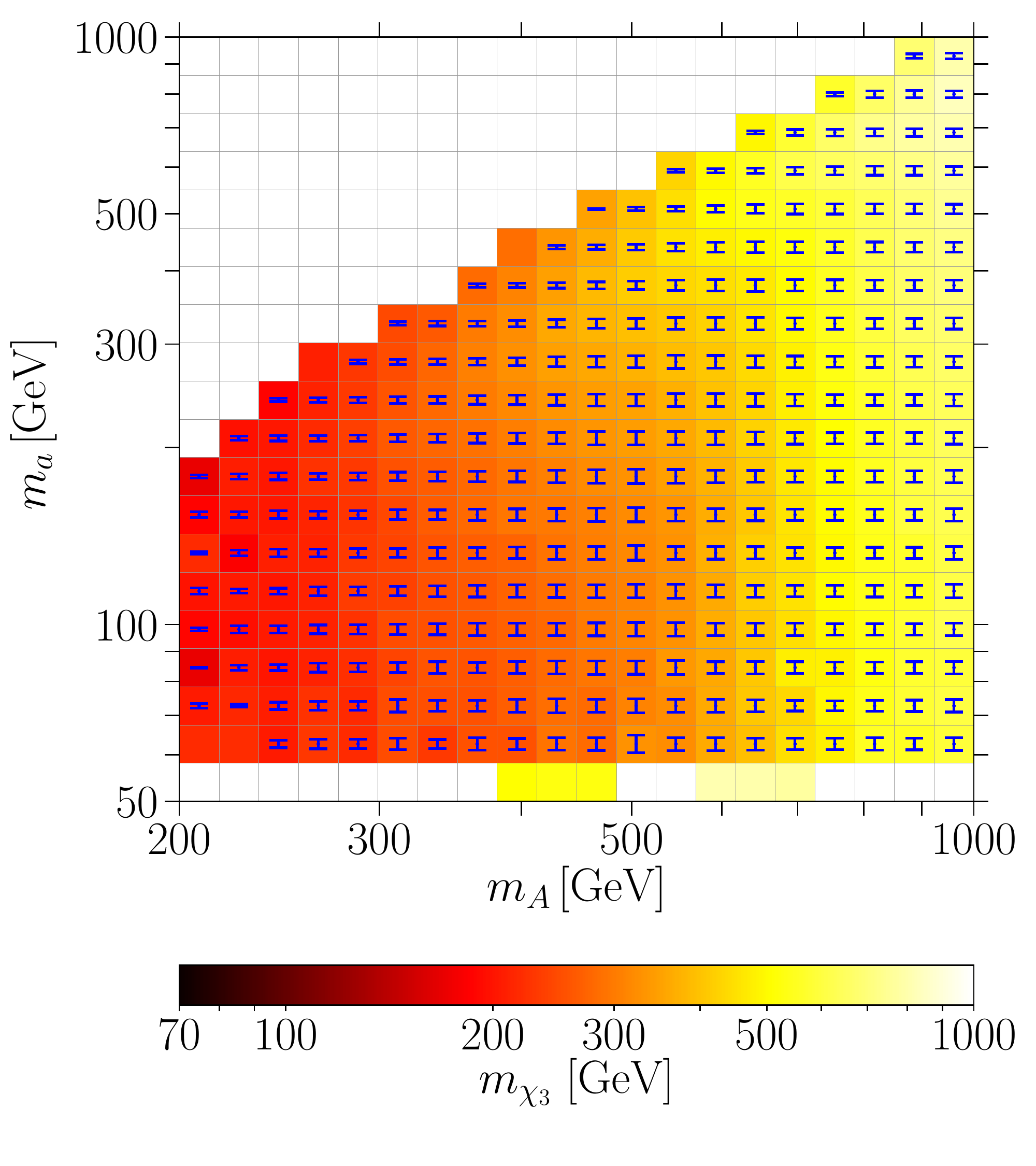}
		\caption{Same as Fig.~\ref{fig:masscorrelations_Higgs}, but the top left panel shows the $\mu$ parameter and the remaining three panels the masses of the 3 lightest neutralinos. Note that due to the choices of parameters in our scan, i.e. decoupling the bino and the wino by choosing their mass parameters $\{|M_1,|M_2|\} \gg |\mu|$, the 3 lightest neutralinos are dominantly composed of the Higgsinos and the singlino.}
		\label{fig:masscorrelations_neu}
	\end{center}
\end{figure}

In order to demonstrate these correlations, we will show the statistical properties of the masses of NMSSM spectra obtained from a parameter scan with \texttt{NMSSMTools\_4.9.3}~\cite{NMSSMTools,Ellwanger:2004xm, Ellwanger:2005dv,Das:2011dg, Muhlleitner:2003vg} in Figs.~\ref{fig:masscorrelations_Higgs} and~\ref{fig:masscorrelations_neu}. NMSSM parameters are drawn from linear flat distributions over the ranges listed in Tab.~\ref{tab:scan_param}. Note that for our numerical scans \texttt{NMSSMTools} requires us to use the parameters appearing in the scalar potential, $\{\tan\beta, \lambda, \kappa, A_\lambda, A_\kappa, \mu\}$, and the stop mass parameter, $M_{Q_3}$, as input, not the parameters of our more physical re-parameterization given in Eq.~\eqref{eq:Physparams}. In addition to the {\it standard} scan we also perform a scan over a narrower range of parameters focused on producing lighter Higgs spectra accessible at the LHC which we label the {\it light subset}. The range $1 \leq \tan\beta \leq 5$ is motivated by obtaining $m_{h_{125}} = 125\,$GeV without the need for large radiative corrections; recall that the contribution $\lambda^2 v^2 s_{2\beta}^2$ to $\mathcal{M}_{S,11}^2$ is suppressed for larger values of $\tan\beta$. We decouple the remaining supersymmetric partners from our study by setting all sfermion mass parameters (except the stop parameters) to 3\,TeV, the bino and wino mass parameters to $M_1 = M_2 = 1\,$TeV, and the gluino mass to $2\,$TeV. Since large third generation squark mixing is not necessary to obtain the correct SM-like Higgs mass in the NMSSM, we set the stop and sbottom mixing parameters $X_t \equiv (A_t - \mu \cot\beta) = 0$ and $X_b \equiv (A_b - \mu\tan\beta) = 0$. Parameter points from the scan are kept if they satisfy a subset of the standard constraints implemented in \texttt{NMSSMTools}, in particular, the Higgs spectra must contain a SM-like Higgs boson with mass and couplings compatible with the SM-like 125\,GeV state observed at the LHC, as well as evade constraints from searches for additional Higgs bosons and sparticles at the Large Electron-Positron Collider (LEP), the Tevatron, and the LHC. Furthermore, we require the lightest neutralino $\chi_1$ to be the lightest supersymmetric particle (LSP). Beyond the constraints implemented in \texttt{NMSSMTools}, we require compatibility with direct searches for Higgs bosons at the LHC listed in Tab.~\ref{tab:LHCSearches} located in the Appendix. As discussed in Ref.~\cite{Baum:2017gbj} we find that parameter points with a Higgs boson with mass and couplings compatible with the 125\,GeV state observed at the LHC approximately satisfy the alignment conditions, although our chosen parameter ranges do not a priori impose these conditions. 

Before we validate our analytical claims with numerics, let us highlight a few points regarding the coverage of the NMSSM parameters in terms of the physical basis we have chosen. First we note that the requirement of non-tachyonic $m_{h}^2$ means that not all values of $\{m_a, m_A, P_A^{\rm S} \}$ are physically allowed. Second, as discussed above, depending on the choice of the CP-odd mass parameters, the alignment conditions may lead to large values of $|\kappa|/\lambda$. However, perturbative consistency generally demands $|\kappa|/\lambda \lesssim 1/2$~\cite{Ellwanger:2009dp,Carena:2015moc}. Generically, this implies that for a fixed value of $m_A$, large mixing angles would demand too large values of $|\kappa|/\lambda$ for small values of $m_a$, whereas values of $m_a$ close to degeneracy with $m_A$ tend to drive $m_h$ tachyonic. We also note that large mixing angles for light $m_a$ can be in tension with direct searches for doublet-like scalars at the LHC, cf. Tab.~\ref{tab:LHCSearches}, further reducing the allowed range of mixing angles in such situations. Therefore, even though we started off with a linear flat distribution for our scans of the NMSSM parameter space, the resultant physically viable regions predominantly correspond to CP-odd masses with small mixing angles. The sum of these requirements leads to a rather small dependence of the mass spectra on the mixing angles beyond the ordering of the neutralino masses and the mass of $m_h$, despite our random scan a priori allowing for large mixing angles. Hence, we present our numerical results for the mass spectra in the $m_A - m_a$ plane. 

As commented at the end of the previous section, radiative corrections affect the relations between our parameter basis and the inputs used in the \texttt{NMSSMTools} scan. Since we set the stop and sbottom mixing parameters to zero in our numerical scan, the stop corrections to the Higgs mass matrices are relatively simple, cf. Refs.~\cite{Ellwanger:2009dp,Carena:2015moc} for the relevant expressions. However, depending on the value of the stop mass parameter $M_{Q_3}$ as well as the size of the quartic couplings between the Higgs bosons, sizable radiative corrections may still be present. Hence, care must be taken when performing precision studies of the NMSSM parameter space to ensure that radiative corrections are properly incorporated when comparing \texttt{NMSSMTools} numerical output to our analytical alignment conditions. The phenomenologically most interesting region of parameter space is where the additional Higgs bosons have masses below 1\,TeV and are hence accessible at the LHC. In this region, excellent agreement is obtained between analytic alignment conditions and the full numerical output from \texttt{NMSSMTools} as shown in Fig.~1 of Ref.~\cite{Baum:2017gbj}. We have further checked that our Eqs.~\eqref{eq:getMA}--\eqref{eq:getAkappa} yield broad agreement when comparing the \texttt{NMSSMTools} input parameters with the corresponding quantities obtained from these equations for the points in our numerical scan.

In Fig.~\ref{fig:masscorrelations_Higgs} we show the masses of the non SM-like CP-even states $H$ and $h$, and in Fig.~\ref{fig:masscorrelations_neu} the masses of the singlino- and Higgsino-like neutralinos $\chi_{i}$, $i=\{1,2,3\}$, together with the value of $|\mu|$ in the $m_A - m_a$ plane. In these figures, the color scale shows the mean of the respective mass (or $|\mu|$) binned in the $m_A - m_a$ plane. In addition, we show the standard deviation of the entries in each bin in units of the mean value with the blue error-bars. The error-bars are normalized such that the they would span the height of the bin if the standard deviation is equal to the mean. Note that the scale of the error-bar is linear with respect to the bin height, while the bin widths as well as the color scale are logarithmic.

From the left panel of Fig.~\ref{fig:masscorrelations_Higgs}, we see that as expected the heavy (usually doublet-like) CP-even state $H$ and the CP-odd state $A$ are approximately mass degenerate and tightly correlated. The mass of $H$ is virtually independent of $m_a$. From the right panel of Fig.~\ref{fig:masscorrelations_Higgs} we observe that the mass of the light (usually singlet-like) CP-even state $h$ is also correlated with the mass of the heavy doublet-like states, with larger $m_A$ leading to heavier $m_h$. Further, as expected the mass satisfies $m_h \lesssim m_A/2$. From the same panel, we also observe that the lightest $m_h$ is obtained for the heaviest $m_a$ and vice versa, showing the expected anti-correlation in their masses. The large standard deviation $\sim$50\,\% for most values of $\{m_a, m_A\}$ shows the weaker dependence on the mixing angle. 

Fig.~\ref{fig:masscorrelations_neu} shows the correlation of the $\mu$ parameter and the masses of the three lightest neutralinos with the Higgs masses. We first note that the value of $|\mu|$ is very tightly correlated with $m_A$, with correspondingly small error-bars, particularly in the low $m_A$ region. This is in agreement with Eq.~\eqref{eq:muApprox}, and stems from viable $h_{125}$ phenomenology requiring approximate alignment. We stress again that these parameter points are selected from a random parameter scan by requiring compatibility with the observed $h_{125}$ phenomenology without a priori imposing alignment conditions. Thus, these results justify our use of the alignment conditions to reduce the number of free parameters, which facilitates the analytic understanding of the parameter space. The $|\mu|$ parameter not only controls the Higgsino masses, but the singlino mass is also proportional to $\mu$. In our parameter scans we decoupled the bino and wino mass parameters $\{|M_1,|M_2|\} \gg |\mu|$, hence the three lightest neutralinos are dominantly composed of the Higgsinos and the singlino. Since the Higgsinos are mass degenerate before taking into account mixing effects, we expect either the lightest or the third-lightest neutralino to be singlino-like, while the remaining two of the three lightest neutralinos are Higgsino-like. The effect on the masses can be seen in Fig.~\ref{fig:masscorrelations_neu}: The second lightest neutralino is usually Higgsino-like and its mass thus quite tightly correlated with $\mu$. The mass-scales of the lightest and third-lightest neutralino on the other hand are also correlated with $\mu$, but we see both larger standard deviations as well as masses smaller than the mean of $|\mu|$ for $\chi_1$ and larger than $|\mu|$ for $\chi_3$. Both effects are due to either $\chi_1$ or $\chi_3$ being singlino-like.

In summary, we find that the mass spectra of the Higgs sector as well as the associated
neutralinos can be described by a simple parameterization in terms of four quantities: the two physical masses and the mixing angle in the CP-odd sector, and $\tan\beta$ (the latter plays a minor role). The presence of a SM-like state with a mass of 125 GeV requires approximate alignment for Higgs spectra accessible at the LHC, determining the preferred values of $\tan \beta$, $\lambda$ and $\kappa/\lambda$. The non SM-like CP-even doublet-like state and the CP-odd doublet-like state are approximately mass degenerate and heavier than the singlet-like states. We use the masses of the CP-odd states $m_a$ and $m_A$ as an input parameters. Together with the value of $\tan\beta$, one can then obtain the value of $\mu$ which controls the Higgsino masses. Since the value of $\kappa/\lambda$ is given by the alignment condition one also directly obtains the singlino mass. In the limit where the bino and the wino are heavy, $\{|M_1|, |M_2|\} \gg |\mu|$, we find that the second-lightest neutralino is Higgsino-like with mass given by $|\mu|$. Either the lightest (if $2\kappa/\lambda \lesssim 1$) or third lightest neutralino (if $2\kappa/\lambda \gtrsim 1$) is singlino-like with mass $\sim |2\kappa\mu/\lambda|$, and the remaining state is again Higgsino-like with its mass pushed away from $|\mu|$ due to the mixing effects with the singlino. The remaining Higgs states are the singlet-like CP-odd and CP-even states. The mass of the CP-even state is mostly governed by $m_A$ and satisfies $m_h \lesssim m_A/2$. The mass of the CP-odd singlet-like state is only weakly correlated with the mass of the remaining states. Most prominently, the mass of the singlet-like scalar is (weakly) anti-correlated with the mass of the singlet-like pseudo-scalar. 

%*********************************************************
\section{Higgs Decays}\label{sec:signals}
%*********************************************************

\begin{figure}
 	\begin{center}
 		\includegraphics[width=1\linewidth]{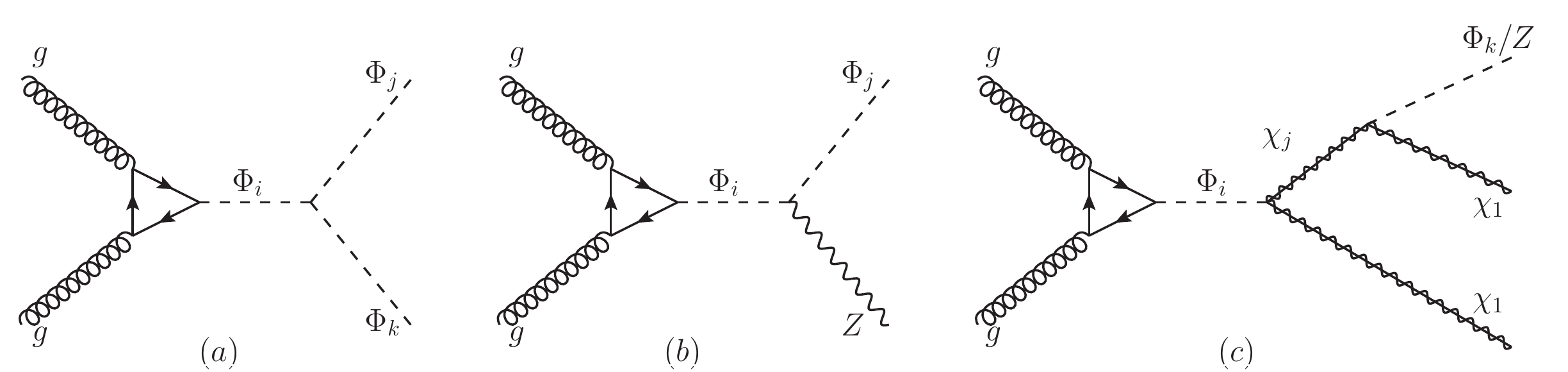}
 		\caption{Illustration of NMSSM-specific Higgs decay channels, where the $\Phi_{i,j,k}$ stand for one of the five NMSSM Higgs bosons. For channel $(a)$, either one or all three of the $\Phi_{i,j,k}$ must be CP-even. For channel $(b)$, if $\Phi_i$ is CP-even, $\Phi_j$ must be a CP-odd state, and vice-versa. For channel $(c)$, the final state can be $\chi_1 \chi_1 h_i$, $\chi_1 \chi_1 a_i$, or $\chi_1 \chi_1 Z$, and $\Phi_i$ can be CP-even or -odd. As discussed further in the text, the most important channels considered in this work are ($gg \to H \to h h_{125}$) and ($gg \to A \to a h_{125}$) through channel $(a)$, ($gg \to H \to Z a$) and ($gg \to A \to Z h$) through channel $(b)$, and ($gg \to \{H,A\} \to \chi_1 \chi_1 h_{125}$) and ($gg \to \{H,A\} \to \chi_1 \chi_1 Z$) through channel $(c)$.}
 		\label{fig:Hdiagrams}
 	\end{center}
\end{figure}

The Higgs bosons in the NMSSM can decay into a variety of final states, however the bulk of the experimental searches at the LHC have been focused on Higgs bosons decaying into pairs of SM particles, see Tab.~\ref{tab:LHCSearches}. The presence of the singlet-like states in the NMSSM both poses a challenge and offers new opportunities for Higgs searches at the LHC when compared to the MSSM. On the one hand, since the singlet does not directly couple to any SM particle, production cross sections of the NMSSM Higgs bosons at colliders are suppressed by the respective singlet component of the Higgs boson in question. On the other hand, the additional singlet-like states offer new decay modes for the doublet-like Higgs bosons, illustrated in Fig.~\ref{fig:Hdiagrams}. As discussed e.g. in Refs.~\cite{Carena:2015moc,Baum:2017gbj}, branching ratios into pairs of lighter Higgs bosons or a light Higgs and a $Z$ boson can be sizable and even compete with decays into pairs of top quarks. Note that decays into pairs of SM-like Higgs bosons or a SM-like Higgs and a $Z$ boson are suppressed, since the corresponding couplings vanish in the alignment limit. Therefore, of all the decays into bosons, the experimentally most promising channels are cascade decays into a SM-like Higgs and an additional non SM-like Higgs boson, or into a $Z$ boson and an additional non SM-like Higgs. The corresponding couplings are not suppressed by the presence of the SM-like $h_{125}$, and the $Z$ or $h_{125}$ in the final state allows for tagging of such events due to their known masses and branching ratios. We will discuss such decays in some detail below.

%*********************************************************
\subsection{Cascade Decays} \label{sec:ratios}
%*********************************************************

In order to study which of the different final states is most relevant for the different regions of NMSSM parameter space, it is useful to start by studying the ratios of $\sigma(gg \to \Phi_1 \to Z \Phi_2)$ and $\sigma(gg \to \Phi_1 \to h_{125} \Phi_2)$ at the LHC, where $\Phi_i$ stands for any of the non SM-like NMSSM Higgs mass eigenstates. The branching ratio ${\rm BR}(\Phi_1 \to h_{125} \Phi_2)$ in particular is intimately related to the couplings $\lambda$ and $\kappa$, while the $(\Phi_1 \to Z \Phi_2)$, $(\Phi \to {\rm SM}~{\rm SM})$ and $(\Phi \to \chi_i \chi_j)$ branching ratios depend mostly on the mass spectrum and the respective mixing angles. The dependence of these Higgs cascade decays on the relevant masses and mixing angles have been studied in great detail in the context of a generic 2HDM+singlet model in Ref.~\cite{Baum:2018zhf}. However, unlike the generic 2HDM+S model, in the NMSSM many parameters are correlated as discussed in the previous section. In the following we will discuss how these parameter correlations dictate the behavior of the Higgs cascade decays.

In terms of the mixing angles and masses and assuming alignment, the most relevant ratios can be written as~\cite{Baum:2018zhf}
\begin{equation} \begin{split} \label{eq:ratio_AZhS_AhaS}
	\frac{\sigma(gg \to A \to Z h)}{\sigma(gg \to A \to h_{125} a)} &= \left(\frac{S_H^{\rm S}}{P_A^{\rm S}}\right)^2 \frac{ \left( m_A^2 - m_h^2\right)^2 - 2\left( m_A^2 + m_h^2 \right) m_Z^2 + m_Z^4}{\left\{ \left[ 1 - 2(P_A^{\rm S})^2 \right] \left(m_A^2 - m_a^2 \right) + \sqrt{2} v \tilde{g}_{A} \right\}^2} \\
		&\qquad \times \sqrt{ \frac{ 1 - 2 \left(m_h^2 + m_Z^2\right)/m_A^2 + \left( m_h^2 - m_Z^2 \right)^2/m_A^4 }{ 1- 2 \left(m_a^2 + m_{h_{125}}^2\right)/m_A^2 + \left(m_a^2 - m_{h_{125}}^2\right)^2/m_A^4} } \;,
\end{split} \end{equation}

\begin{equation} \begin{split}\label{eq:ratio_HZaS_HhhS}
	\frac{\sigma(gg \to H \to Z a)}{\sigma(gg \to H \to h_{125} h)} &= \left( \frac{P_A^{\rm S}}{S_H^{\rm S}} \right)^2 \frac{ \left( m_H^2 - m_a^2 \right)^2 - 2\left( m_H^2 + m_a^2 \right) m_Z^2 + m_Z^4}{ \left\{ \left[ 1 - 2(S_H^{\rm S})^2 \right] \left( m_H^2 - m_h^2 \right) + \sqrt{2} v \tilde{g}_H \right\}^2 } \\
		&\qquad \times \sqrt{ \frac{ 1 - 2 \left(m_a^2 + m_Z^2\right)/m_H^2 + \left(m_a^2 - m_Z^2\right)^2/m_H^4 }{ 1 - 2 \left(m_h^2 + m_{h_{125}}^2\right)/m_H^2 + \left(m_h^2 - m_{h_{125}}^2\right)^2/m_H^4 } } \;, 
\end{split} \end{equation}

\begin{equation} \begin{split} \label{eq:ratio_AZhS_HhhS}
	\frac{\sigma(gg \to A \to Z h)}{\sigma(gg \to H \to h_{125} h)} &= \frac{\sigma_{ggh}(m_A)}{\sigma_{ggh}(m_H)} \left( \frac{\tau_A f(\tau_A)}{\tau_A - \left(\tau_A-1\right) f(\tau_A)} \right)^2 \left(\frac{ P_A^{\rm NSM}}{S_H^{\rm NSM}}\right)^4 \frac{m_H}{m_A} \frac{\Gamma_H}{\Gamma_A} \\
		& \qquad \times \frac{ \left( m_A^2 - m_h^2\right)^2 - 2\left( m_A^2 + m_h^2 \right) m_Z^2 + m_Z^4 }{ \left\{ \left[ 1 - 2(S_H^{\rm S})^2 \right] \left( m_H^2 - m_h^2 \right) + \sqrt{2} v \tilde{g}_H \right\}^2 } \\
		& \qquad \times \sqrt{ \frac{ 1 - 2 \left(m_{h}^2 + m_Z^2\right)/m_A^2 + \left( m_{h}^2 - m_Z^2 \right)^2/m_A^4 }{ 1- 2 \left(m_h^2 + m_{h_{125}}^2\right)/m_H^2 + \left(m_h^2 - m_{h_{125}}^2\right)^2/m_H^4 } } \;,
\end{split} \end{equation}

\begin{equation} \begin{split} \label{eq:ratio_HZaS_AhaS}
	\frac{\sigma(gg \to H \to Z a)}{\sigma(gg \to A \to h_{125} a)} &= \frac{\sigma_{ggh}(m_H)}{\sigma_{ggh}(m_A)} \left( \frac{1}{f(\tau_A)} + \frac{\tau_A-1}{\tau_A} \right)^2 \left(\frac{ S_H^{\rm NSM}}{P_A^{\rm NSM}}\right)^4 \frac{m_A}{m_H} \frac{\Gamma_A}{\Gamma_H} \\
		& \qquad \times \frac{ \left( m_H^2 - m_a^2 \right)^2 - 2\left( m_H^2 + m_a^2 \right) m_Z^2 + m_Z^4 }{ \left\{ \left[ 1 - 2(P_A^{\rm S})^2 \right] \left(m_A^2 - m_a^2 \right) + \sqrt{2} v \tilde{g}_A \right\}^2 } \\
		& \qquad \times \sqrt{ \frac{ 1 - 2 \left(m_a^2 + m_Z^2\right)/m_H^2 + \left( m_a^2 - m_Z^2 \right)^2/m_H^4 }{ 1- 2\left(m_a^2 + m_{h_{125}}^2\right)/m_A^2 + \left(m_a^2 - m_{h_{125}}^2\right)^2/m_A^4 } }\;, 
\end{split} \end{equation}
where $\sigma_{ggh}(m)$ is the gluon fusion production cross section of a SM Higgs boson of mass $m$, and the form factor is defined as 
\begin{equation}
	f(\tau) = \begin{cases}
		\arcsin^2\sqrt{\tau} &, \quad \tau \leq 1 \;,
		\\ -\frac{1}{4} \left[ \log\left(\frac{1+\sqrt{1-1/\tau}}{1-\sqrt{1-1/\tau}}\right) - i\pi \right]^2 &, \quad \tau >1 \;,
	\end{cases}
\end{equation}
with $\tau \equiv (m/2 m_t)^2$.

In addition to the masses and the mixing angles, these ratios depend on the decay widths of the parent state $\Gamma_\Phi$, and one combination of trilinear couplings between the involved states, which are given by
\begin{align}
	\tilde{g}_H \equiv g_{H^{\rm SM} H^{\rm S} H^{\rm S}} - g_{H^{\rm SM} H^{\rm NSM} H^{\rm NSM}} &= \frac{v}{\sqrt{2}} \left[ \lambda s_{2\beta} \left( 3\lambda s_{2\beta} - 2\kappa \right) - \frac{m_Z^2}{v^2} \left( 2 s_{2\beta}^2 - c_{2\beta}^2 \right) \right] \;, \\
	\tilde{g}_A \equiv g_{H^{\rm SM} A^{\rm S} A^{\rm S}} - g_{H^{\rm SM} A^{\rm NSM} A^{\rm NSM}} &= \frac{v}{\sqrt{2}} \left[ \lambda s_{2\beta} \left( \lambda s_{2\beta} + 2\kappa \right) + \frac{m_Z^2}{v^2} c_{2\beta}^2 \right] \;. 
\end{align}

For values in proximity of alignment $\lambda \sim 0.65$, moderate values of $|\kappa| < 1$, and low $\tan\beta$, these combinations of couplings are at most $\sim\mathcal{O}(v)$. Hence, unless the channel in the denominator of Eqs.~\eqref{eq:ratio_AZhS_AhaS}--\eqref{eq:ratio_HZaS_AhaS} is kinematically suppressed, or the relevant mixing angle takes values $(P_A^{\rm S})^2 \approx 0.5$ [$(S_H^{\rm S})^2 \approx 0.5$], these couplings play no important role for the ratios.
 
\begin{figure}
	\includegraphics[width=\linewidth]{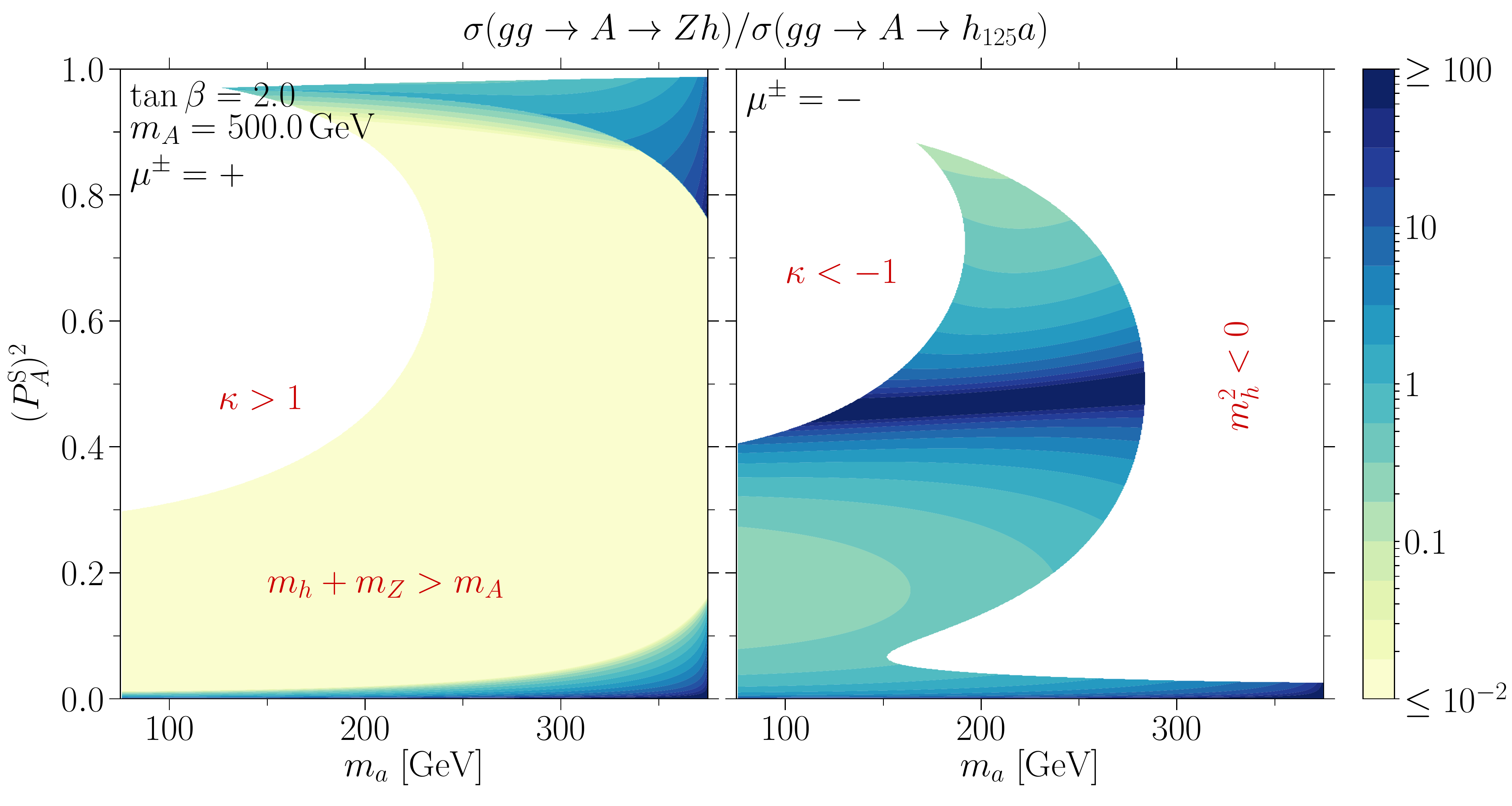}
	\includegraphics[width=\linewidth]{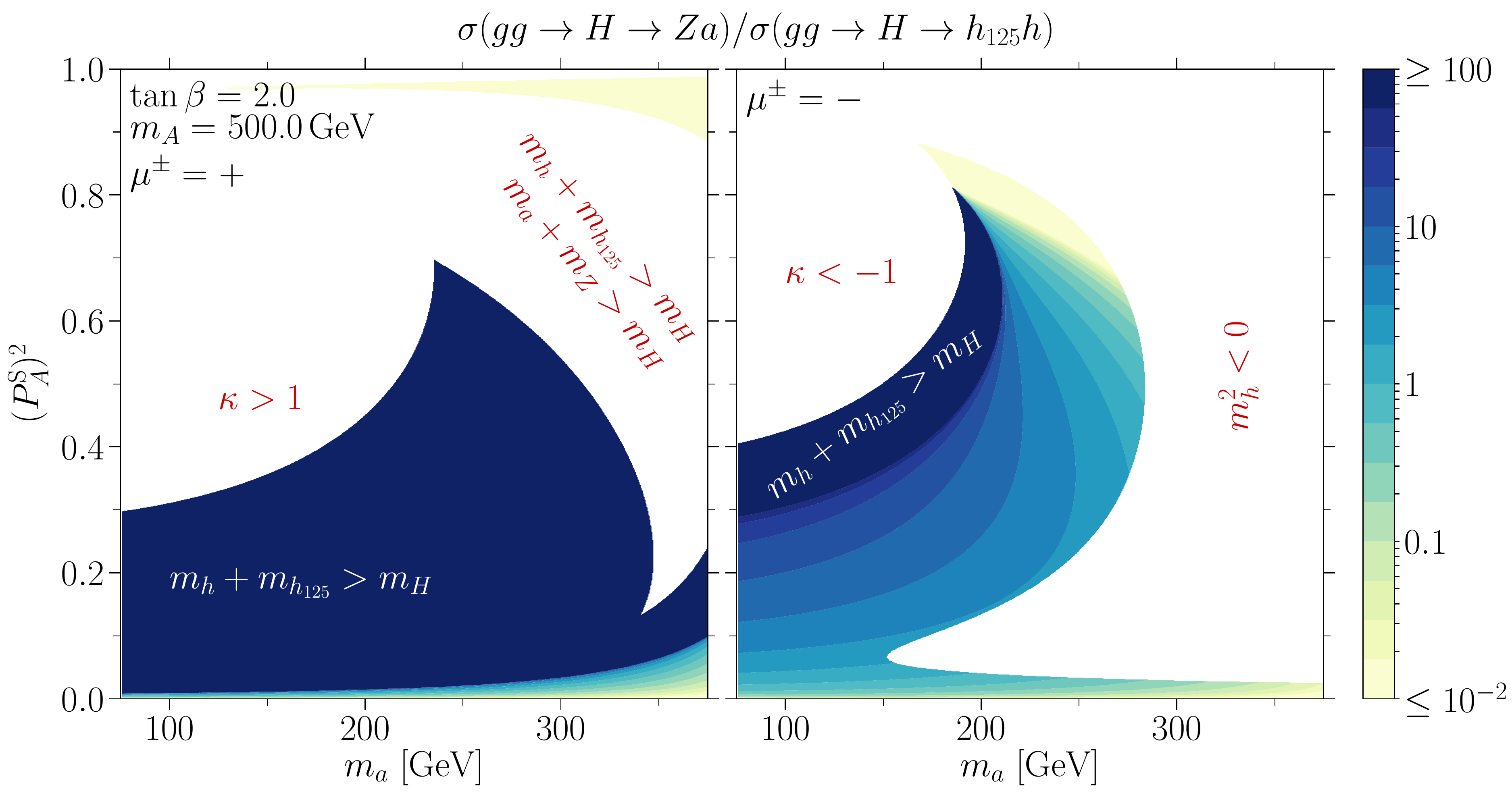}
	\caption{Ratios of various heavy Higgs cross sections given in Eq.~\eqref{eq:ratio_AZhS_AhaS} [\eqref{eq:ratio_HZaS_HhhS}] for the top [bottom] panel in the $m_a$--$(P_A^{\rm S})^2$ plane. Recall that $m_a$ is the mass of the lighter CP-odd mass eigenstate and $P_A^{\rm S}$ parameterizes the mixing angle in the CP-odd sector. Specifically, $(P_A^{\rm S})^2$ is the singlet fraction of the heavier CP-odd state $A$ and numerically $(P_A^{\rm S})^2 = (P_a^{\rm NSM})^2$, where $(P_a^{\rm NSM})^2$ is the (non SM-like) doublet component of the lighter CP-odd state $a$, cf. Eq.~\eqref{eq:a_mix}. The remaining parameters are fixed to $m_A = 500\,$GeV and $\tan\beta = 2$, and the left and right panels are for the two different $\mu^\pm$ solutions, cf. Eq.~\eqref{eq:getmu}. The red/white labels indicate regions of parameter space where one or more of the channels are kinematically forbidden, the lighter non SM-like CP-even state $m_h$ becomes tachyonic, or $\kappa$ takes large values $|\kappa| > 1$.}
	\label{fig:ratios1}
\end{figure}

\begin{figure}
	\includegraphics[width=\linewidth]{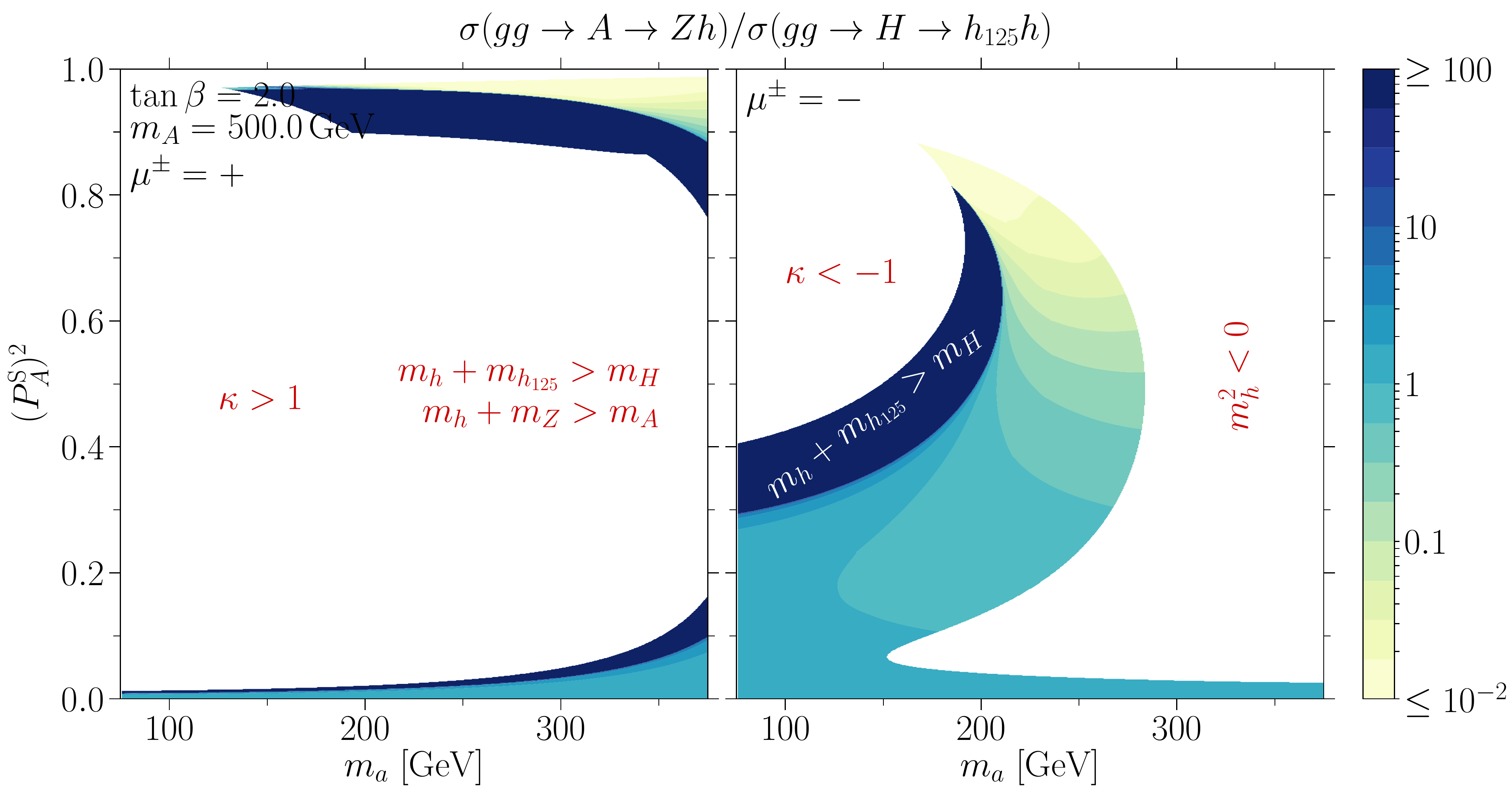}
	\includegraphics[width=\linewidth]{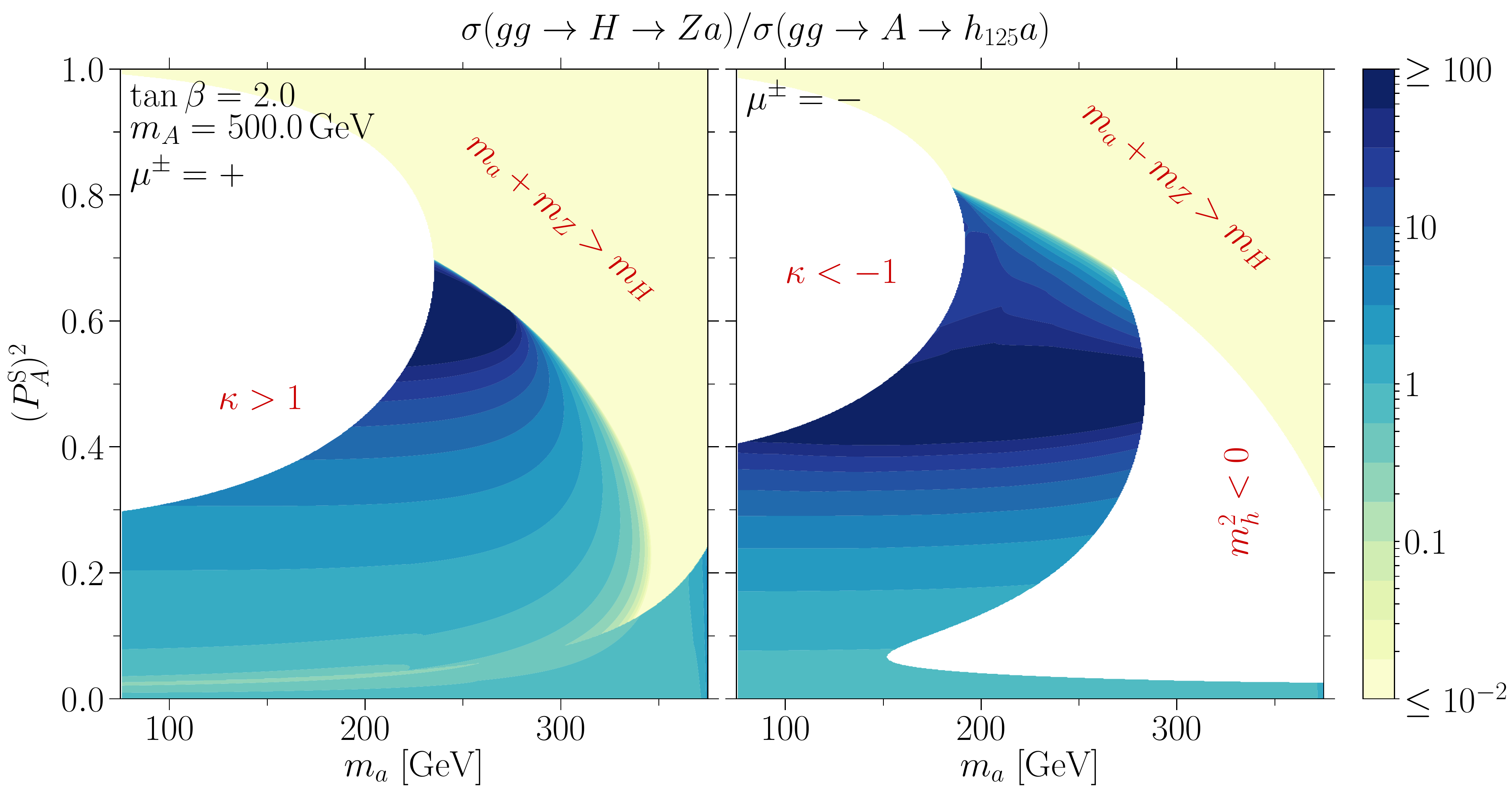}
	\caption{Same as Fig.~\ref{fig:ratios1} but for the ratios given in Eq.~\eqref{eq:ratio_AZhS_HhhS} [\eqref{eq:ratio_HZaS_AhaS}] for the top [bottom] panel.}
	\label{fig:ratios2}
\end{figure}

Note that the ratio [$\sigma(gg\to H\to Za)/\sigma(gg\to H\to h_{125}h)$] in Eq.~\eqref{eq:ratio_HZaS_HhhS} can be obtained from the ratio [$\sigma(gg\to A\to Zh)/\sigma(gg\to A\to h_{125}a)$] in Eq.~\eqref{eq:ratio_AZhS_AhaS} when exchanging all quantities referring to CP-even states to the corresponding quantity for CP-odd states, and vice versa. Likewise, the ratio [$\sigma(gg\to H\to Za)/\sigma(gg\to A\to h_{125}a)$] in Eq.~\eqref{eq:ratio_HZaS_AhaS} can be obtained from the ratio [$\sigma(gg\to A\to Zh)/\sigma(gg\to H\to h_{125}h)$] in Eq.~\eqref{eq:ratio_AZhS_HhhS} keeping in mind that exchanging the ratio of the gluon fusion production cross sections entails replacing
\begin{equation}
	\left( \frac{\tau_A f(\tau_A)}{\tau_A - \left(\tau_A-1\right) f(\tau_A)} \right)^2 \leftrightarrow \left( \frac{\tau_A f(\tau_A)}{\tau_A - \left(\tau_A-1\right) f(\tau_A)} \right)^{-2} = \left( \frac{1}{f(\tau_A)} + \frac{\tau_A-1}{\tau_A} \right)^2 \;.
\end{equation}

In Figs.~\ref{fig:ratios1} and~\ref{fig:ratios2} we show these ratios in the plane of the light CP-odd mass $m_a$ vs. its NSM fraction $(P_a^{\rm NSM})^2=(P_A^{\rm S})^2$ for fixed values of $m_A$ and $\tan\beta$. As discussed in section~\ref{sec:CorrelConsequence}, in the alignment limit all parameters controlling the NMSSM Higgs sector are fixed in terms of these inputs, except for the choice of the two different solutions for $\mu^\pm$, cf. Eq.~\eqref{eq:getmu}, corresponding to the two panels in each of the figures. As discussed in section~\ref{sec:CorrelConsequence}, due to the correlation of parameters, not all of the parameter region shown in Figs.~\ref{fig:ratios1} and~\ref{fig:ratios2} is allowed. In particular, $\kappa$ takes large absolute values for large $(P_A^{\rm S})^2 \gtrsim 0.3$ and small values of $m_a \lesssim 150\,$GeV. In order to prevent a Landau pole close to the SUSY breaking scale, we constrain $|\kappa| < 1$ in the figures. Furthermore, for one of the solutions for $\mu^\pm$, the light non SM-like CP-even Higgs state becomes tachyonic for large values of $m_a$ as indicated in the corresponding panels.

In some regions of parameter space only one of the decay modes appearing in the respective ratio is kinematically allowed. Recalling the correlation of the masses discussed previously, this is in particular due to the fact that the mass of the singlet-like CP-odd state can be tuned quite independently, while the masses of the doublet-like mass eigenstates and the singlet-like CP-even state are more tightly correlated.

In the region of parameter space where both channels appearing in the respective ratios are allowed, we find the two cross sections to generally be of the same order of magnitude. However, in particular regions of parameter space one of the channels can dominate, even far from regions where kinematic suppression is effective. This occurs for example in the top panels of Fig.~\ref{fig:ratios1} and the bottom panels of Fig.~\ref{fig:ratios2} for $(P_A^{\rm S})^2 \sim 0.5$; for this value the cross section in the denominator $\sigma(gg \to A \to h_{125} a)$ is strongly suppressed, cf. Eqs.~\eqref{eq:ratio_AZhS_AhaS} and~\eqref{eq:ratio_HZaS_AhaS}. Therefore, since no single decay channel is dominant throughout parameter space, it is important to consider all of them in order to fully cover the parameter space of the NMSSM at the LHC. We have verified the analytical results presented in this section by computing and comparing these ratios from the output of our \texttt{NMSSMTools} scan.

The final decay products of the daughter Higgs and $Z$ bosons produced from the decay of the heavy parent Higgs in the Higgs cascade decays discussed above will dictate the sensitivity of the LHC to such channels. Higgs and $Z$ bosons decay into pairs of SM particles such as $\tau^+ \tau^-$, $b\bar{b}$, $ZZ$, or $W^+ W^-$, or, if kinematically accessible, they might also decay into pairs of neutralinos. In general, in the low $\tan\beta$ regime, the branching ratios of the light non SM-like Higgs bosons are similar to those of a SM Higgs boson of the same mass, with the exception that they can have sizable branching ratios into pairs of neutralinos if kinematically accessible, and that CP-odd Higgs bosons do not decay into pairs of gauge bosons at tree-level. Note that the decays into pairs of SM fermions are controlled by the ($\tan\beta$ suppressed/enhanced) Yukawa couplings. The decays into pairs of Higgsino/singlino-like neutralinos are controlled by $\lambda$ and $\kappa$ instead. In the alignment limit we find $\lambda \sim 0.65$, much larger than all Yukawa couplings but the top Yukawa. Thus, if kinematically accessible, NMSSM Higgs bosons typically have large branching ratios into pairs of neutralinos below the top threshold, i.e. $m_{\Phi_i}\lesssim 2m_t \sim 350\,$GeV. This qualitative behavior is unchanged when allowing for light binos or winos. The couplings of the doublet-like Higgs bosons to binos and winos are controlled by the $U(1)_Y$ and $SU(2)_L$ gauge couplings, respectively, which are again larger than all Yukawa couplings but the top Yukawa. Expressions for the couplings of the NMSSM Higgs bosons to pairs of SM particles as well as pairs of neutralinos can e.g. be found in Refs.~\cite{Ellwanger:2009dp,Cheung:2014lqa,Baum:2017enm}, and a more detailed discussion of the branching ratios can be found in Refs.~\cite{Baum:2017gbj, Baum:2018zhf}.

Incorporating the final state decays of the non SM-like daughter Higgs bosons, we classify the Higgs cascade decay channels leading to different final states as follows: If the heavy Higgs boson decays into a SM-like Higgs and a light Higgs [Fig.~\ref{fig:Hdiagrams} (a)], one obtains 
\begin{itemize}
	\item {\it Higgs+visible} final states if the additional light Higgs decays into a pair of SM particles visible in the detector, or 
	\item {\it Mono-Higgs} signatures if the additional light Higgs decays into a pair of neutralinos, leading to a boosted SM-like Higgs and missing transverse energy (\MET) in the detector. 
\end{itemize}
Likewise, decays of the heavy Higgs bosons into a $Z$ and a light Higgs boson [Fig.~\ref{fig:Hdiagrams} (b)] yield 
\begin{itemize}
	\item {\it $Z$+visible} final states if the additional Higgs boson decays into pairs of SM particles, or 
	\item {\it Mono-$Z$} signatures if the additional Higgs decays into neutralinos. 
\end{itemize}

Mono-Higgs or mono-$Z$ signatures can also arise if the heavy Higgs decays directly into neutralinos where one of the neutralinos is not the lightest one, and subsequently decays into the lightest neutralino and a SM-like Higgs or a $Z$ boson [Fig.~\ref{fig:Hdiagrams} (c)]. However, as discussed in Refs.~\cite{Baum:2017gbj, Baum:2018zhf}, such decays are kinematically unfavorable for collider searches since the neutralinos might conspire to be produced approximately back-to-back in the transversal plane yielding small \MET.

Note that the categorization above misses some final states, such as when both the $h_{125}$ (or the $Z$) and the light additional Higgs boson decay into invisible states, or if the heavy Higgs decays to two heavier neutralinos which subsequently decay into the lightest neutralino and additional particles. The former type of decay channels may e.g. be probed via monojet-type searches. The latter decay channel may in principle be probed with strategies similar to what is discussed here, although it will in general be more challenging since they yield softer final states.

%*********************************************************
\subsection{LHC Prospects for Cascade Decays}
%*********************************************************

Not all final states are equal - the sensitivity of the LHC is very channel dependent. To determine the coverage of the NMSSM parameter space at the LHC we need to compare the cross sections for each channel to the sensitivity of the LHC. To this end we compare the cross sections for our NMSSM parameter scan to the projected sensitivity in the different channels at the 13\,TeV LHC assuming $\mathcal{L} = 3000\,{\rm fb}^{-1}$ of data. For the first time, we exploit all of the mono-$Z$, mono-Higgs, $Z$+visible, and Higgs+visible classes of final states for probing the NMSSM at the LHC, whereas the previous literature considered one class of final states at a time.

The sensitivity of the mono-Higgs and mono-$Z$ channels has been extensively discussed in Refs.~\cite{Baum:2017gbj, Baum:2018zhf}. The sensitivity of Higgs+visible channels in the $b\bar{b}b\bar{b}$ and $b\bar{b}\gamma\gamma$ final states has been discussed in Ref.~\cite{Ellwanger:2017skc}. The importance of the $Z$+visible channel has been discussed in~\cite{Carena:2015moc,Baum:2017gbj}, but to date no estimate of the sensitivity at the 13\,TeV LHC is available. For the purposes of this work, we extrapolate the sensitivity at the 13\,TeV LHC for $\mathcal{L} = 3000\,{\rm fb}^{-1}$ of data, $\sigma_{Z+{\rm vis}}^{13\,{\rm TeV};\,3000\,{\rm fb}^{-1}}$ from the limit set by the CMS collaboration at the 8~TeV LHC with $\mathcal{L} = 19.8\,{\rm fb}^{-1}$ of data in the [$(Z \to \ell^+ \ell^-) + (\Phi \to \tau^+ \tau^-)$] final state~\cite{Khachatryan:2016are}. We rescale the reported limit $\sigma_{Z+{\rm vis}}^{8\,{\rm TeV};\,19.8\,{\rm fb}^{-1}}$ with the number of events as
\begin{equation}
	\sigma_{Z+{\rm vis.}}^{13\,{\rm TeV};\;3000\,{\rm fb}^{-1}} (m_{\Phi_i}, m_{\Phi_j}) = \sqrt{\frac{\sigma_{ggh}^{8\,{\rm TeV}}(m_{\Phi_i})}{\sigma_{ggh}^{13\,{\rm TeV}}(m_{\Phi_i})} \times \frac{19.8\,{\rm fb}^{-1}}{3000\,{\rm fb}^{-1}}} ~\sigma_{Z+{\rm vis.}}^{8\,{\rm TeV};\;19.8\,{\rm fb}^{-1}} (m_{\Phi_i}, m_{\Phi_j}) \;,
\end{equation}
where $\sigma_{ggh}^{\sqrt{s}}(m)$ is the gluon fusion production cross section of a SM Higgs boson with mass $m$ at the LHC with center-of-mass energy $\sqrt{s}$. Note that this is a conservative extrapolation of the sensitivity relying purely on the increased statistics, while the ATLAS and CMS collaborations have demonstrated significant improvements in background rejection as well as increased control of the systematic errors when updating searches in the past. 

\begin{figure}
	\begin{center}
		\includegraphics[width=0.48\linewidth,trim={0cm, 3.8cm, 0cm, 0cm},clip=True]{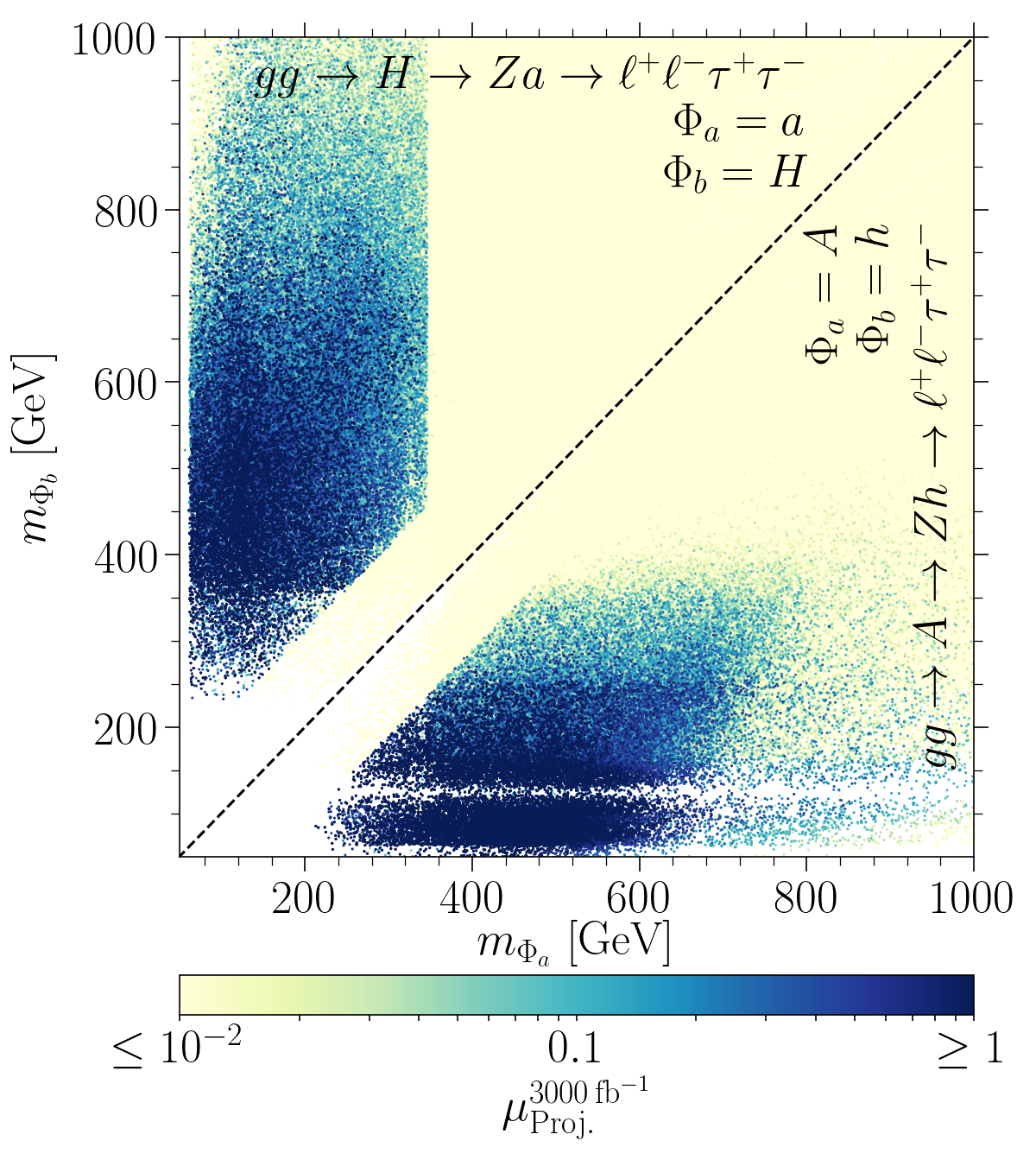}
		\includegraphics[width=0.48\linewidth,trim={0cm, 3.8cm, 0cm, 0cm},clip=True]{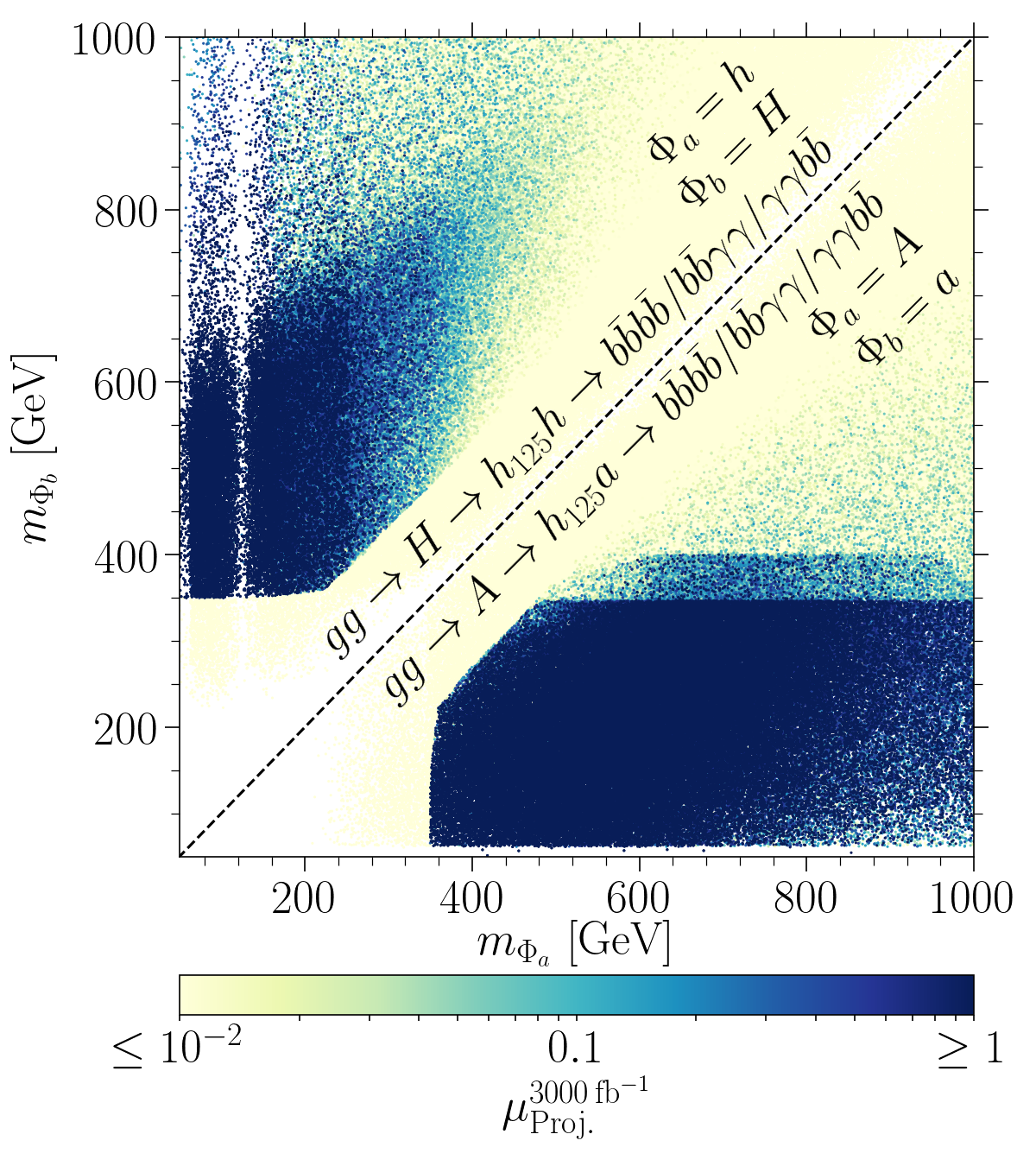}

		\includegraphics[width=0.48\linewidth]{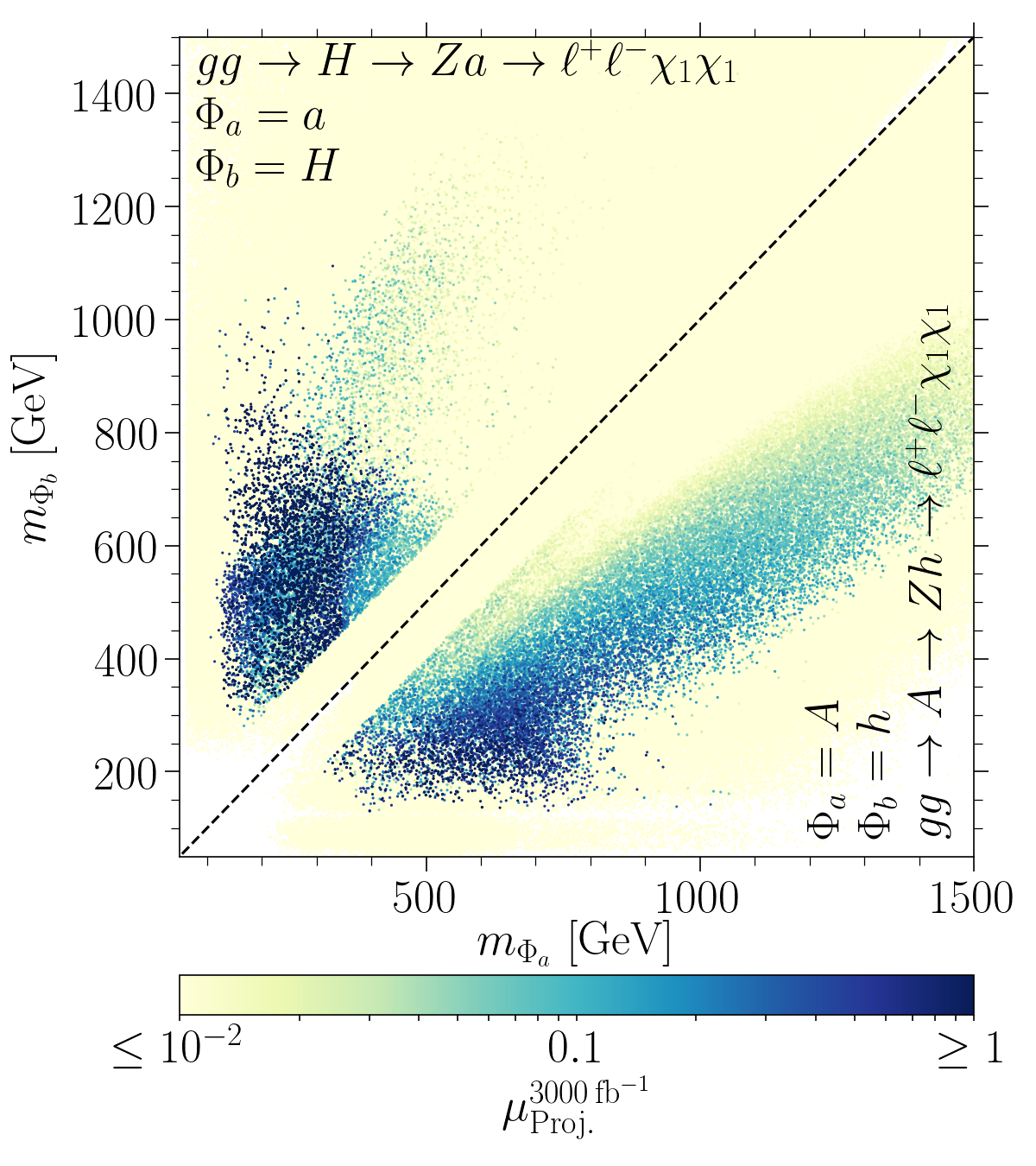}
		\includegraphics[width=0.48\linewidth]{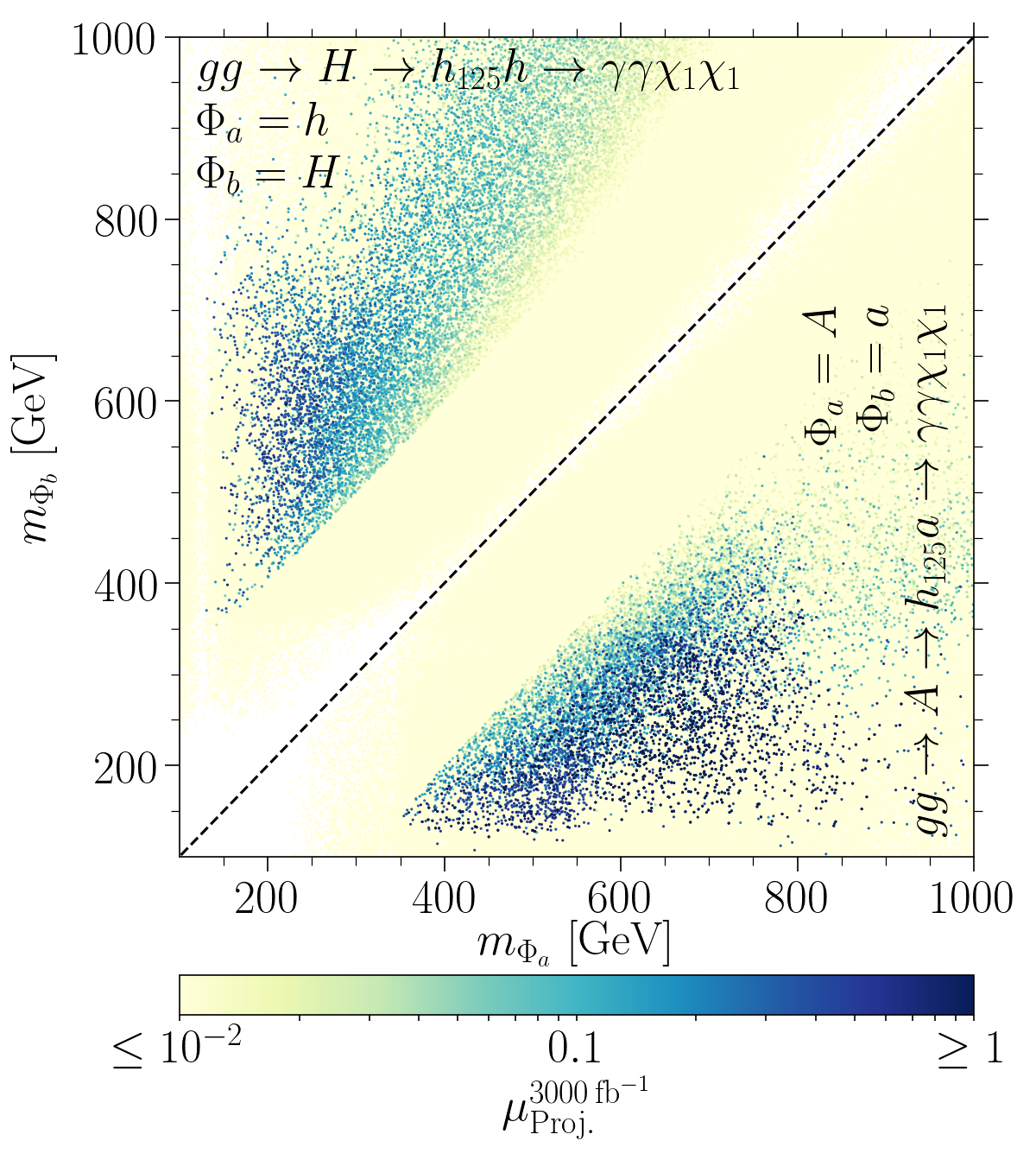}

		\caption{Signal strength $\mu_{\rm Proj.}^{3000\,{\rm fb}^{-1}} \equiv \sigma / \sigma^{3000\,{\rm fb^{-1}}}_{\rm Proj.}$ as indicated by the color bar, where $\sigma$ is the cross section of the parameter point and $\sigma^{3000\,{\rm fb^{-1}}}_{\rm Proj.}$ the estimated sensitivity for $\mathcal{L}=3000\,{\rm fb}^{-1}$ of data in the respective channel. The top panels are for $Z$+visible (left) and Higgs+visible (right). The bottom panels show the sensitivity in the mono-Higgs (left) and mono-$Z$ final states (right). In each panel, the two triangles separated by the dashed line correspond to a parent CP-even state $H$ (upper left triangle) or a parent CP-odd state $A$ (lower right triangle). For the left panels, the $x$-axis corresponds to $m_a$ ($m_A$) and the $y$-axis to $m_H$ ($m_h$) for the upper left (lower right) triangle. For the right panels, the $x$-axis corresponds to $m_h$ ($m_A$) and the $y$-axis to $m_H$ ($m_a$). The hard cutoff at masses of the parent state below $\sim 350\,$GeV in the top right panel (Higgs+visible) is due to the mass ranges for which the sensitivity in these finals states is available in Ref.~\cite{Ellwanger:2017skc}. Note also that in the bottom left panel the mass range extends up to $1.5\,$TeV, while in the other panels we show only the mass range up to $1\,$TeV.}
	\label{fig:variousreach1}
	\end{center}
\end{figure}

In Fig.~\ref{fig:variousreach1} we show the signal strength for our NMSSM parameter scan 
\begin{equation}
	\mu_{\rm Proj.}^{3000\,{\rm fb}^{-1}} \equiv \sigma / \sigma^{3000\,{\rm fb^{-1}}}_{\rm Proj.} \;,
\end{equation}
where $\sigma$ is the cross section of the parameter point and $\sigma^{3000\,{\rm fb^{-1}}}_{\rm Proj.}$ is the cross section expected to be probed with $\mathcal{L}=3000\,{\rm fb}^{-1}$ of data in the respective channel. The different panels in Fig.~\ref{fig:variousreach1} correspond to the different Higgs cascade channels (mono-Higgs, Higgs+visible, mono-$Z$ and $Z$+visible) arising from the decay of a heavy Higgs into a pair of lighter Higgses or a light Higgs and a $Z$ boson, corresponding to the diagrams (a) and (b) in Fig.~\ref{fig:Hdiagrams}. The signal strength for mono-$Z$ and mono-Higgs final states arising via decays of a heavy Higgs into neutralinos, where one of the neutralinos in turn radiates off a $Z$ or a Higgs boson, cf. diagram (c) in Fig.~\ref{fig:Hdiagrams}, is shown in Fig.~\ref{fig:variousreach2} located in Appendix~\ref{app:ratios}.\footnote{Note that the sensitivity of the mono-$Z$ channel via processes shown in diagram (c) in Fig.~\ref{fig:Hdiagrams} may be enhanced with respect to what is shown here by using recently proposed kinematic variables, cf. Ref.~\cite{Gori:2018pmk}.} For completeness, we present four benchmark points in Appendix~\ref{app:BM}, chosen to have large signal cross sections in the mono-Higgs, Higgs+visible, mono-$Z$, and $Z$+visible classes of Higgs cascades, respectively.

All four search channels shown in Fig.~\ref{fig:variousreach1} are able to probe sizable regions of the NMSSM parameter space. Note that the comparison between different channels should not be taken at face value but as a qualitative comparison, since the extrapolations of the future LHC sensitivity assume different systematic uncertainties for each channel; in particular, the assumptions in Ref.~\cite{Ellwanger:2017skc} for the Higgs+visible channel are more optimistic than those in the extrapolation of the remaining channels. Most notably, all searches maintain sensitivity in the $\{m_H, m_A\} \gtrsim 350\,$GeV region which is difficult to probe with conventional Higgs searches. For $\{m_H, m_A\} \gtrsim 350\,$GeV, decays into top quarks $\{H,A\} \to t\bar{t}$ dominate the decays of heavy Higgs bosons into pairs of SM particles; this decay channel is very challenging to probe at the LHC due to interference with the QCD background~\cite{Dicus:1994bm,Barcelo:2010bm,Barger:2011pu,Bai:2014fkl,Jung:2015gta,Craig:2015jba,Gori:2016zto,Carena:2016npr}. 

Comparing various final states arising in individual triangles portrayed in each panel of Fig.~\ref{fig:variousreach1}, we clearly find the effects of the correlation of masses discussed in section~\ref{sec:paramCorrel}. For the mono-$Z$ and $Z$+visible (mono-Higgs and Higgs+visible) final states, the decay chains induced by the parent CP-odd state $A$ (CP-even state $H$) contain only the light CP-even state $h$ and neutralinos with tightly correlated masses. This leads to the behavior of the cross sections of the respective channels with respect to the extrapolated LHC sensitivity being relatively uniform, cf. the bottom right triangles in the left panels and the top left triangles in the right panels of Fig.~\ref{fig:variousreach1} respectively. On the other hand, mono-$Z$ and $Z$+visible (mono-Higgs and Higgs+visible) final states induced by the parent CP-even state $H$ (CP-odd state $A$) involve the light pseudo-scalar $a$, whose mass is much less tightly correlated with the remaining Higgs states. This leads to somewhat less regular behavior, as can be seen from top left triangles in left panels and bottom right triangles in right panels of Fig.~\ref{fig:variousreach1} respectively. In particular, while one may find mass spectra with large mass gaps leading to readily observable final states, one can also easily end up in situations where the $a$ is too heavy such that this decay of the heavy parent state is kinematically suppressed or may not yield sufficiently hard decay products required for large \MET \;in the mono-$Z$ and mono-Higgs final states. On the other hand, one may also end up in the situation where $a$ is lighter than the pair of lightest neutralinos, such that $a \to \chi_1 \chi_1$ decays required for mono-$Z$ and mono-Higgs final states are kinematically forbidden.

\begin{figure}
	\begin{center}
		\includegraphics[width=0.8\linewidth]{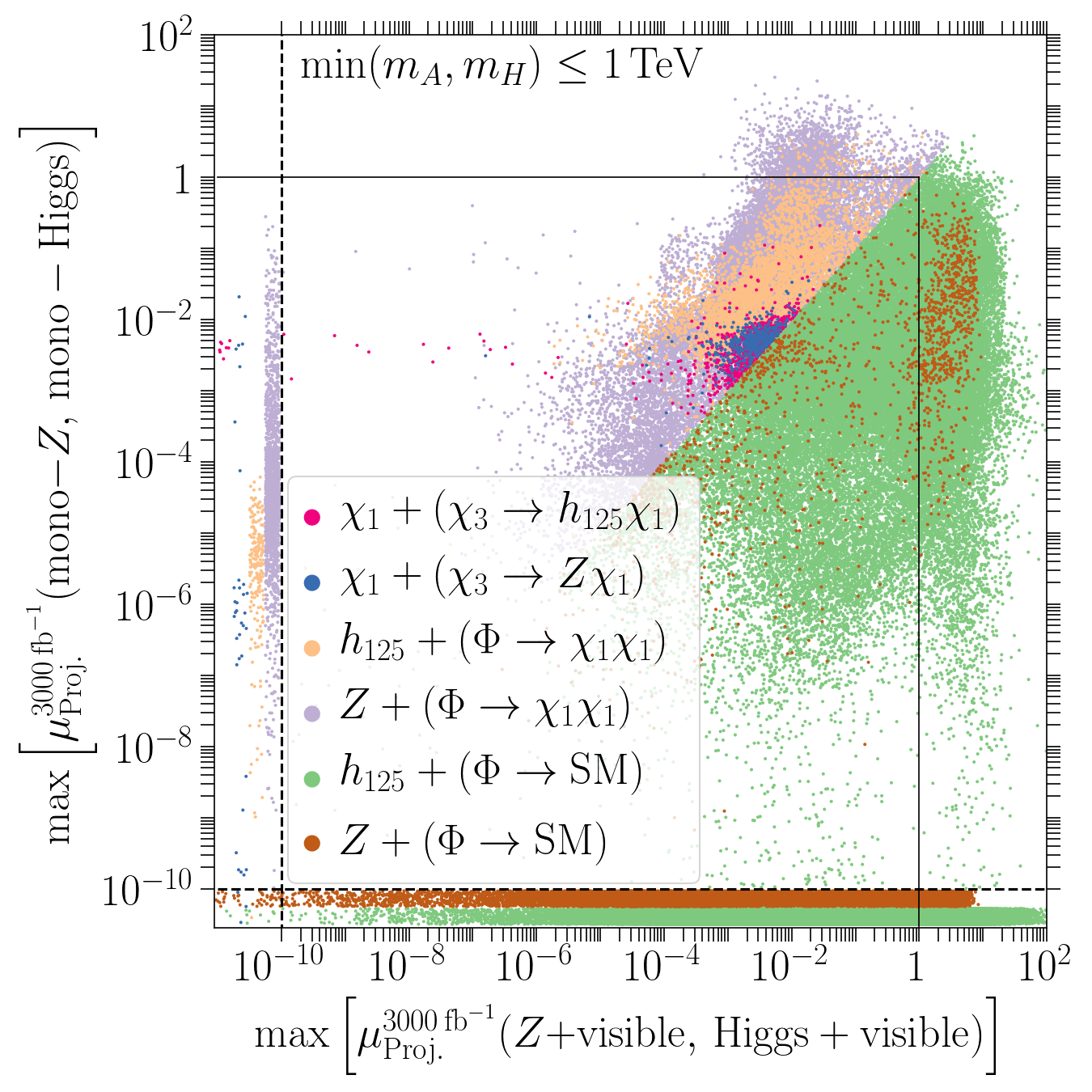}
		\caption{Largest $Z$+visible or Higgs+visible signal strength ($x$-axis) vs. the largest mono-$Z$ or mono-Higgs signal strength ($y$-axis) for points from our NMSSM parameter scan. The color coding denotes the Higgs cascade channel with the largest signal strength as indicated in the legend. The $\Phi$ in the legend can be either the light non SM-like CP-even state $h$ or the light CP-odd state $a$, depending on the CP properties of the parent state in the cascade, and whether it is accompanied by an $h_{125}$ or a $Z$ boson, cf. Fig.~\ref{fig:Hdiagrams}. The displayed parameter points satisfy all current constraints from conventional searches at the LHC as listed in Tab.~\ref{tab:LHCSearches} and feature at least one of the heavy Higgs bosons $H$ or $A$ lighter than 1\,TeV. Points in the L-shaped region either above or to the right of the solid lines, indicating a signal strength $\mu_{\rm Proj.}^{3000\,{\rm fb}^{-1}} = 1$, are within our projected sensitivity of the LHC with 3000\,fb$^{-1}$ of data. The different Higgs cascade channels are clearly complimentary such that one must employ all of them in order to probe as large a portion of the parameter space as possible. Note that as discussed further in the text, the LHC collaborations may improve the sensitivities by at least one order of magnitude compared to our estimates. Points below (to the left of) the dashed lines have signal strengths $\mu_{\rm Proj.}^{3000\,{\rm fb}^{-1}} < 10^{-10}$ in the mono-$Z$ and mono-Higgs ($Z$+visible and Higgs+visible) channels, rendering such points difficult to detect in the respective channels even at the high energy LHC~\cite{Cepeda:2650162}, and may require the 100 TeV collider~\cite{Arkani-Hamed:2015vfh}.}
		\label{fig:visiblevinvisible}
	\end{center}
\end{figure}

Comparing the $Z$+visible and Higgs+visible to the mono-$Z$ and mono-Higgs final states, we find that the $Z$+visible and Higgs+visible channels are usually more effective as long as the light Higgs involved in the decay chain is below the top threshold, $m_{h/a} \lesssim 350\,$GeV. Once the light state is allowed to decay to a pair of top quarks, such decays will usually dominate, rendering searches in ($h/a \to b\bar{b}/\tau^+ \tau^-/\gamma\gamma$) final states less effective. This effect is particularly visible in the ($gg \to H \to Za$) and ($gg \to A \to h_{125} a$) decay channels, while it is somewhat less pronounced in the ($gg \to A \to Zh$) and ($gg \to H \to h_{125} h$) channels because the kinematic cutoff for $a \to t\bar{t}$ decays is much harder than for $h \to t\bar{t}$ decays and because decays of CP-odd Higgs bosons into pairs of vector bosons are forbidden at tree-level. Mono-$Z$ and mono-Higgs final states remain sensitive above the top threshold for the light Higgs; as discussed above, the branching ratios of the light Higgs bosons into pairs of neutralinos can be comparable with the branching ratios into pairs of top quarks. Thus, mono-$Z$ and mono-Higgs final states are particularly powerful in the parameter region hard to probe with conventional searches where all of the non SM-like Higgs bosons are above the top threshold. This region may also be accessible with $Z$+visible or Higgs+visible final states when using final states arising from decays of the light Higgs bosons into top quarks or $W$ or $Z$ bosons. Sizable cross sections into such final states have been demonstrated in Ref.~\cite{Baum:2017gbj}, however, no estimate of the LHC sensitivity for such final states exists to date.

Finally, in Fig.~\ref{fig:visiblevinvisible} we compare the signal strengths of the mono-$Z$ and mono-Higgs channels to the Higgs+visible and $Z$+visible channels for all points from our NMSSM parameter scans passing all constraints, in particular evading all current bounds from conventional searches at the LHC as listed in Tab.~\ref{tab:LHCSearches}. The different Higgs cascade channels are clearly complimentary such that one must employ all of them in order to cover as large a portion of the parameter space as possible. Recall that the comparison of the different channels should be understood qualitatively and not be taken at face value, since the extrapolation of the sensitivity for the different channels assume e.g. different systematic errors of the background. 

We stress that the sensitivity extrapolations we have used are somewhat conservative, in particular in the mono-Higgs, mono-$Z$, and $Z$+visible final states, where systematic errors may be reduced significantly by the experimental collaborations compared to what was assumed when estimating the sensitivity. Hence we expect that the true sensitivity of LHC searches may be up to approximately one order of magnitude better than what is shown in Figs.~\ref{fig:variousreach1} and \ref{fig:visiblevinvisible}. This renders in particular the mono-$Z$ final state very promising, allowing the LHC to probe heavy Higgs boson with masses larger than $1\,$TeV.

%*********************************************************
\section{Combining Searches to Cover the NMSSM Parameter Space} \label{sec:reach}
%*********************************************************

In the previous section we discussed the sensitivity of the LHC for the different final states arising from Higgs cascade decays. In this section we will demonstrate how by combining these searches with conventional searches (utilizing direct decays of the heavy Higgs bosons into pairs of SM particles), significant progress towards coverage of the NMSSM parameter space can be made. We find that the NMSSM parameter space which realizes non SM-like Higgs bosons lighter than $\sim 1\,$TeV could be almost completely probed by the 13 TeV LHC with 3000 fb$^{-1}$ of data.

\begin{figure}
	\begin{center}
		\includegraphics[width=0.49\linewidth]{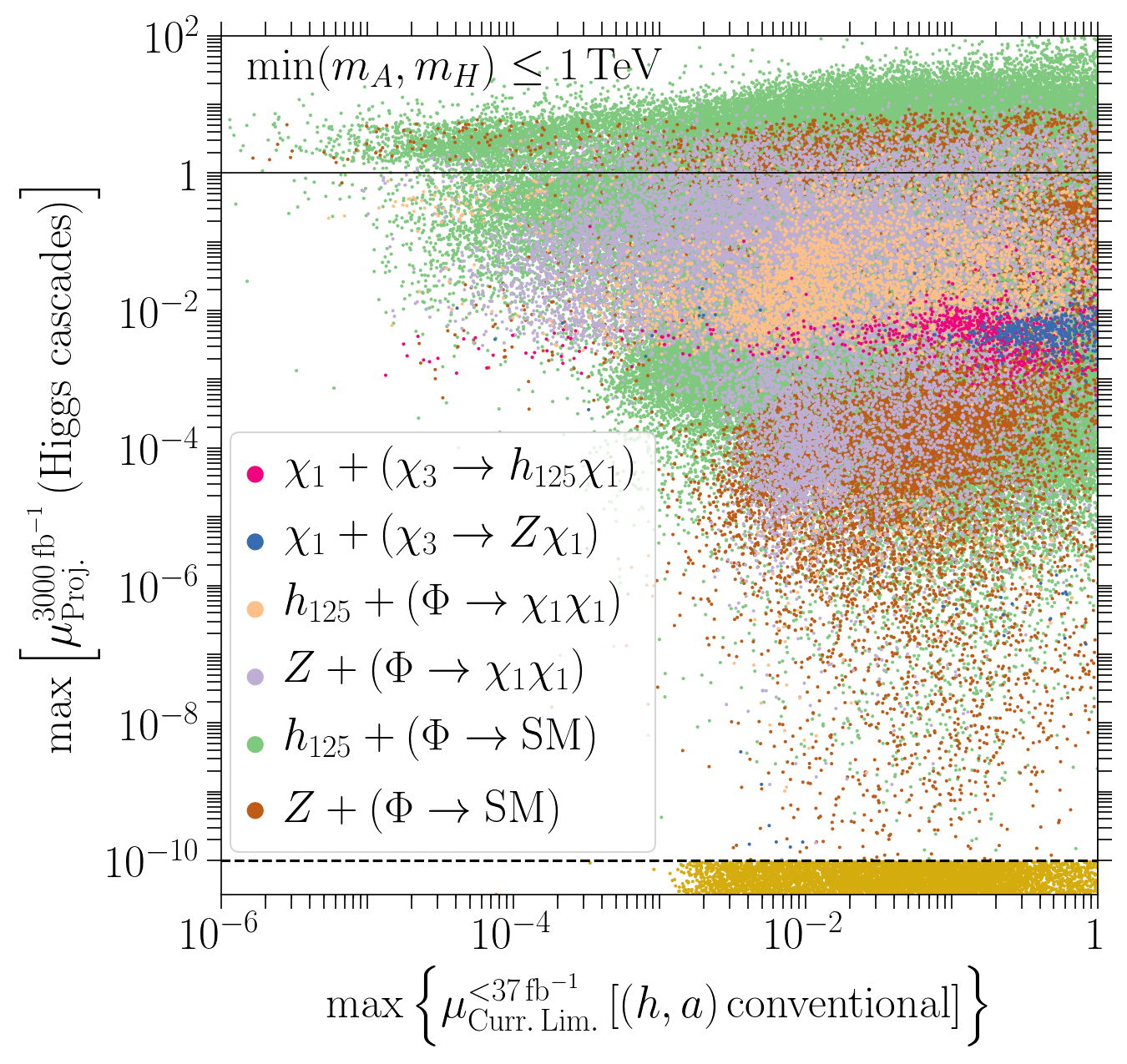}			\includegraphics[width=0.49\linewidth]{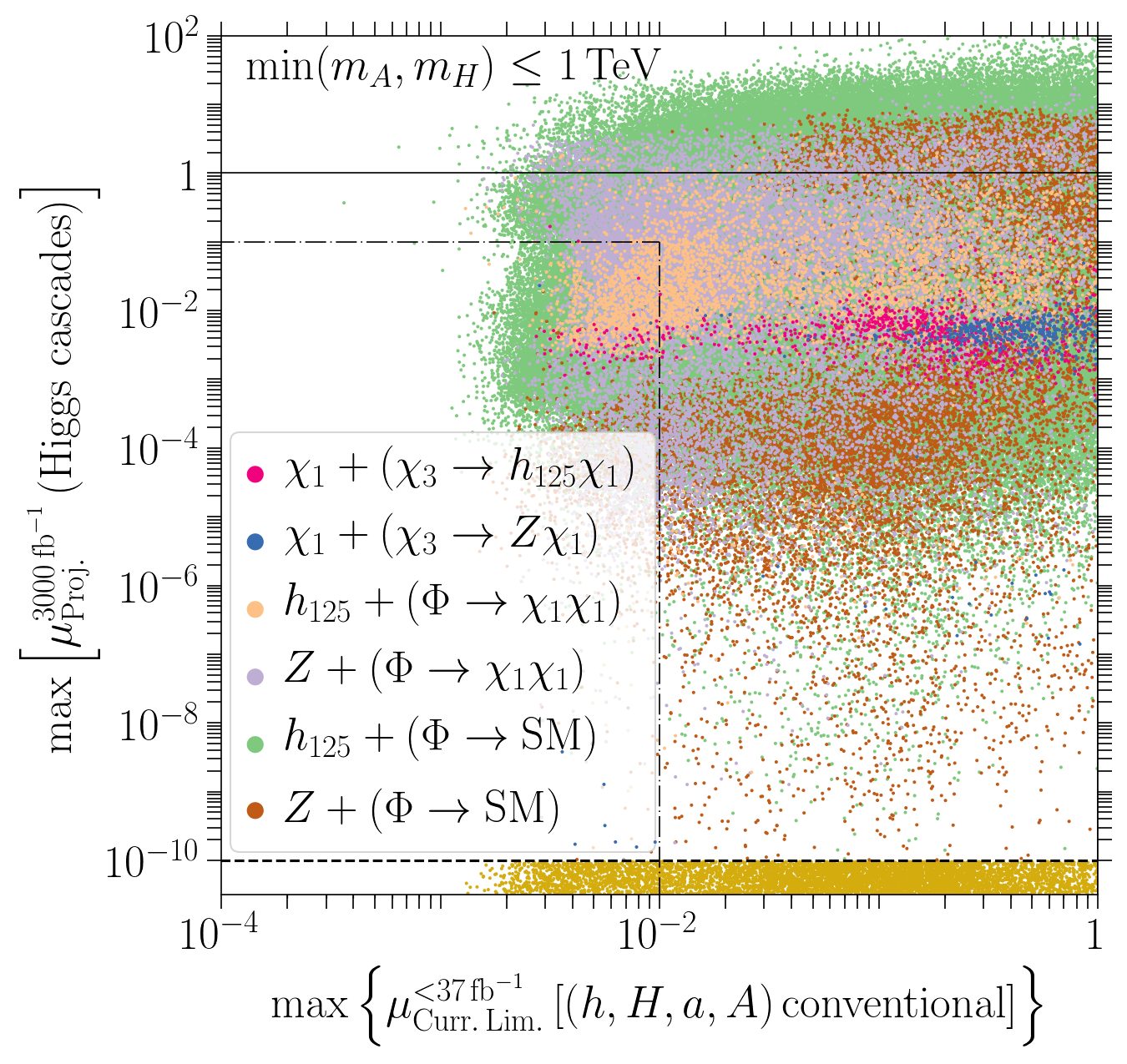}
		\caption{{\it Left:} Same as Fig.~\ref{fig:visiblevinvisible} but that the $x$-axis shows the largest signal strength of all conventional Higgs searches listed in Tab.~\ref{tab:LHCSearches} arising through the production of the light states $h$ and $a$, and the $y$-axis shows the largest signal strength of all the Higgs cascade searches. Note that for the Higgs cascades modes we use the projected sensitivity for $\mathcal{L}=3000\,{\rm fb}^{-1}$ of data while for the conventional searches we use the best current limit. We cut off the $x$-axis at $\mu_{\rm Curr.\,Lim.}^{\rm <37\,fb^{-1}} = 1$ since points to the right of that are already excluded. {\it Right:} Same as the left panel, but the $x$-axis shows the best signal strength of the conventional channels including those arising via the direct production of the heavy non SM-like CP-even and CP-odd states $H$ and $A$. In Sec.~\ref{sec:reach}, we entertain the scenario that the LHC collaborations will be able to improve their sensitivities by one order of magnitude for the Higgs cascade decays compared to our projections (i.e. to $\mu_{\rm Proj}^{\rm 3000\,fb^{-1}}=0.1$ on the $y$-axis) and two orders of magnitude compared to current conventional limits (i.e. to $\mu_{\rm Curr.\,Lim.}^{\rm <37\,fb^{-1}}=10^{-2}$ on the $x$-axis). Then, all points except those in the bottom-left quadrangle bounded by the dash-dotted lines may be probed at the LHC with 3000\,fb$^{-1}$ of data. This quadrangle encloses only $\approx 10\,\%$ of the points shown - thus, such an improvement would allow future runs of the LHC to cover almost all ($\approx 90\,\%$) of the phenomenologically viable NMSSM parameter space containing additional Higgs bosons with masses below 1\,TeV. Note that the scales of the $x$-axes differ between the panels.}
		\label{fig:standardvcascades}
	\end{center}
\end{figure}

In Fig.~\ref{fig:standardvcascades} we compare the projected signal strength of the Higgs cascade channels, $\mu_{\rm Proj.}^{\rm 3000~fb^{-1}}$, with the current signal strength for the conventional Higgs searches, $\mu_{\rm Curr.\,Lim.}^{\rm <37\,fb^{-1}}$. Once more we note the complementarity of the different channels. In particular, when considering the detectability of the lighter states $h$ and $a$ via conventional searches, cf. the left panel, we find that for parameter points where one class of searches becomes ineffective, the other one usually fares well. If the lighter states $h$ and $a$ evade constraints from conventional Higgs searches, they are usually quite singlet-like such that their production cross section at the LHC is suppressed. However, mostly singlet-like light states are readily produced via Higgs cascade decays: For example, $(\Phi_1 \to h_{125} \Phi_2)$ decays, where $\Phi_i$ stands for a non SM-like Higgs mass eigenstate, are mostly controlled by the coupling $\lambda$ if $\Phi_2$ is singlet-like, and $\Phi_1$ doublet-like. Since $\lambda$ takes large values $\lambda \sim 0.65$ in the alignment limit, the corresponding branching ratios are large such that searches utilizing Higgs cascades remain sensitive. From the right panel of Fig.~\ref{fig:standardvcascades}, we see that direct searches for the heavy states $H$ and $A$ provide an additional handle for the Higgs cascade decays. Combining Higgs cascade decays with (conventional) direct searches for all the NMSSM Higgs bosons, the entire parameter space of the NMSSM with heavy Higgs bosons $H$ and $A$ lighter than $\sim 1\,$TeV is at most $2-3$ orders of magnitude below current limits. Note the different scale for the $x$-axes between the left and right panels in Fig.~\ref{fig:standardvcascades}. 

In Fig.~\ref{fig:standardvcascades} we used our projected sensitivity for the 13\,TeV LHC with $3000\,{\rm fb}^{-1}$ of data for the Higgs cascade channels, while we used current limits based, depending on the channel, on at most $37\,{\rm fb}^{-1}$ of data for the conventional Higgs searches. The increased statistical power of the future $3000\,{\rm fb}^{-1}$ data set should allow the bounds in the conventional searches to improve by approximately one order of magnitude. This would allow $\approx 50\,\%$ of our parameter points with masses of the additional Higgs bosons below 1\,TeV to be probed by the LHC. Note that all of these parameter points satisfy  current constraints. Hence, by combining all search channels, the LHC can make significant progress towards complete coverage of the NMSSM parameter space.

In the remainder of this section, we entertain the scenario that the LHC collaborations will be able to improve the sensitivity of their searches in the Higgs cascade channels by one order of magnitude compared to our projections (i.e. $\mu_{\rm Proj}^{3000\,{\rm fb}^{-1}} = 0.1$) and two orders of magnitude compared to current limits in the conventional channels (i.e. $\mu_{\rm Curr.Lim.}^{, 37\,{\rm fb}^{-1}} = 0.01$). For the Higgs cascade decay based searches, such improvements could be realized by a combination of better rejection of reducible backgrounds and reduced systematic uncertainty of the remaining backgrounds. Note that improvements of comparable size have been demonstrated by both the ATLAS and the CMS collaboration in the past when updating analyses with increased statistics. For the conventional searches, we estimate that increasing the luminosity from the current $\mathcal{L} = \mathcal{O}(30)\,{\rm fb}^{-1}$ of data to $\mathcal{L} = 3000\,{\rm fb}^{-1}$ could yield one order of magnitude better sensitivity as discussed above, while another order of magnitude of improvement may be possible by improved background rejection/systematics and search strategies. While this is an optimistic scenario, it presents a clear target for the experimental collaborations which would allow the LHC to probe almost all of the remaining phenomenologically viable NMSSM parameter space featuring additional Higgs bosons with masses below 1\,TeV.

\begin{figure}
	\includegraphics[width=.49\linewidth]{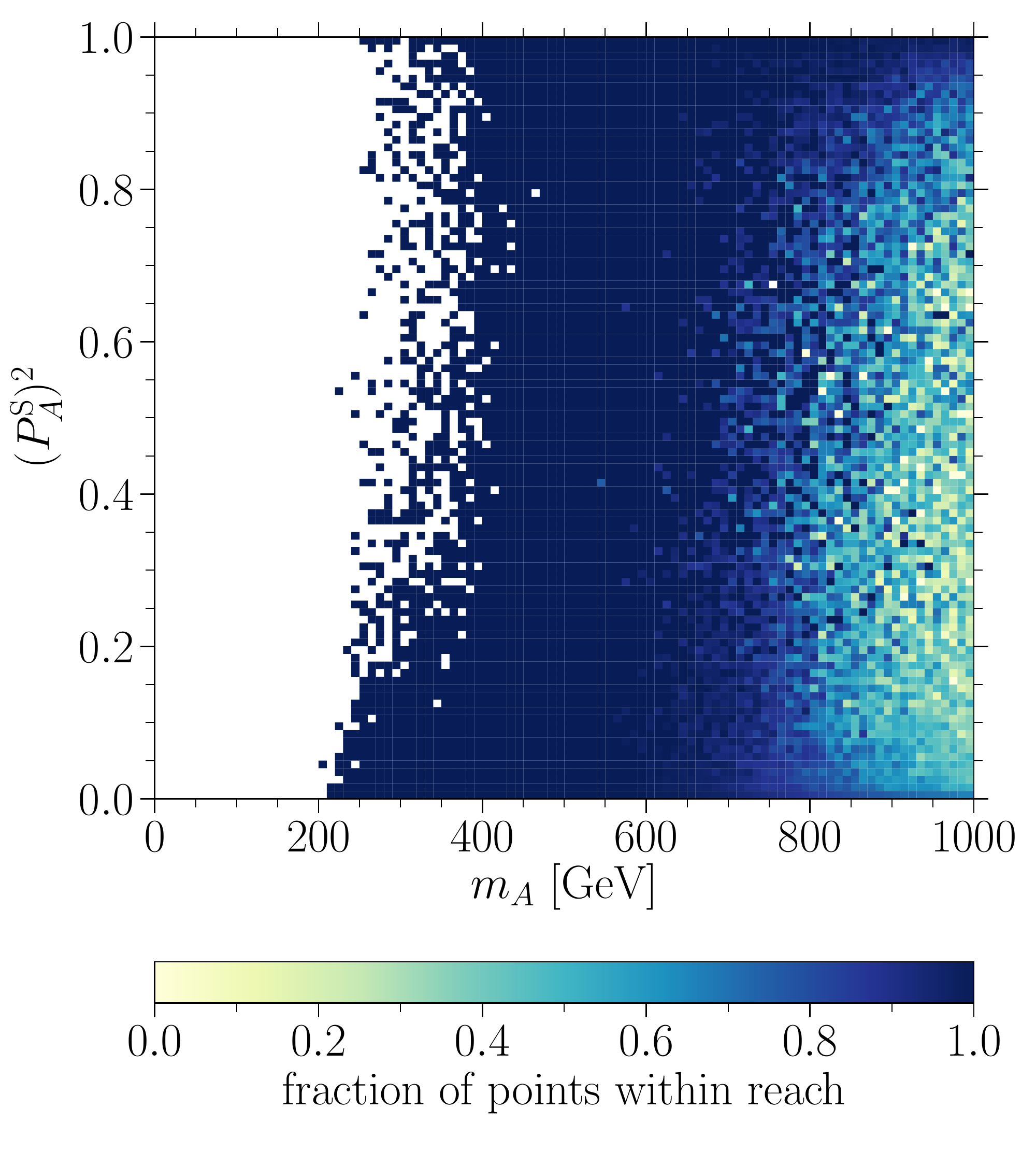}
	\includegraphics[width=.49\linewidth]{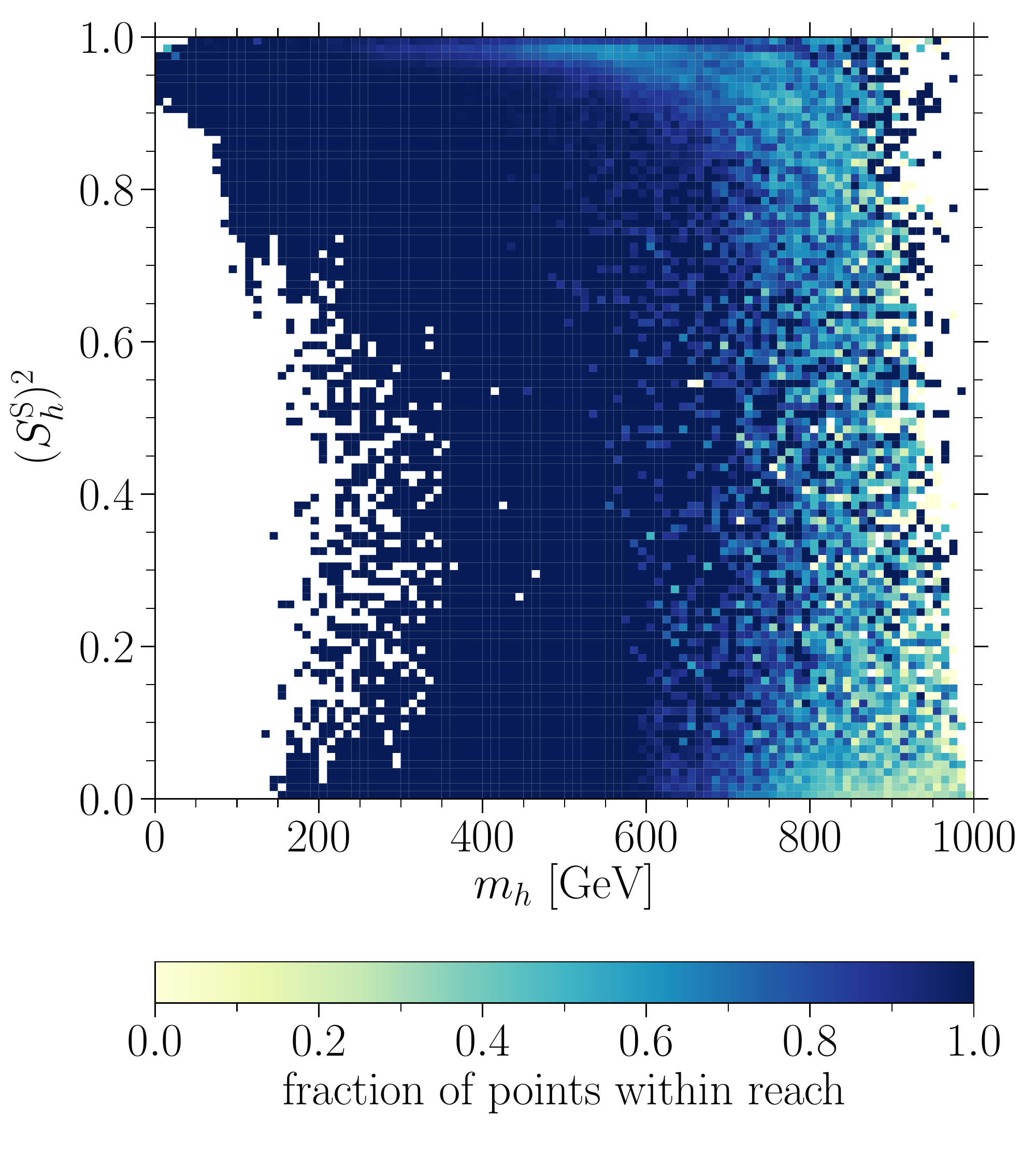} 
	\caption{The left [right] panel shows the distribution of LHC sensitivity in the plane of the mass $m_A$~[$m_h$] and the singlet fraction $(P_A^{\rm S})^2$ [$(S_h^{\rm S})^2$], cf. Eq.~\eqref{eq:a_mix}~[Eq.~\eqref{eq:h_mix}], of the heavy CP-odd state $A$ [the light non SM-like CP-even state $h$]. The color coding denotes the fraction of points in each bin which will be probed at the LHC using a combination of conventional Higgs searches and searches utilizing Higgs cascade decays. Here, we make somewhat more optimistic assumptions on the sensitivity of the LHC than before. As discussed further in the text we consider points to be within reach if ($\max[\mu_{\rm Curr.\,Lim.}^{\rm <37\,fb^{-1}}({\rm conventional})] > 0.01$ or $\max[\mu_{\rm Proj.}^{3000\,{\rm fb}^{-1}}({\rm Higgs~cascades})] > 0.1$). We consider only points where at least one of the heavy states $H$ or $A$ has a mass below 1\,TeV, as in Figs.~\ref{fig:visiblevinvisible} and~\ref{fig:standardvcascades}. White regions do not contain any parameter points.}
	\label{fig:combinedreach}
\end{figure}

\begin{figure}
	\includegraphics[width=.49\linewidth]{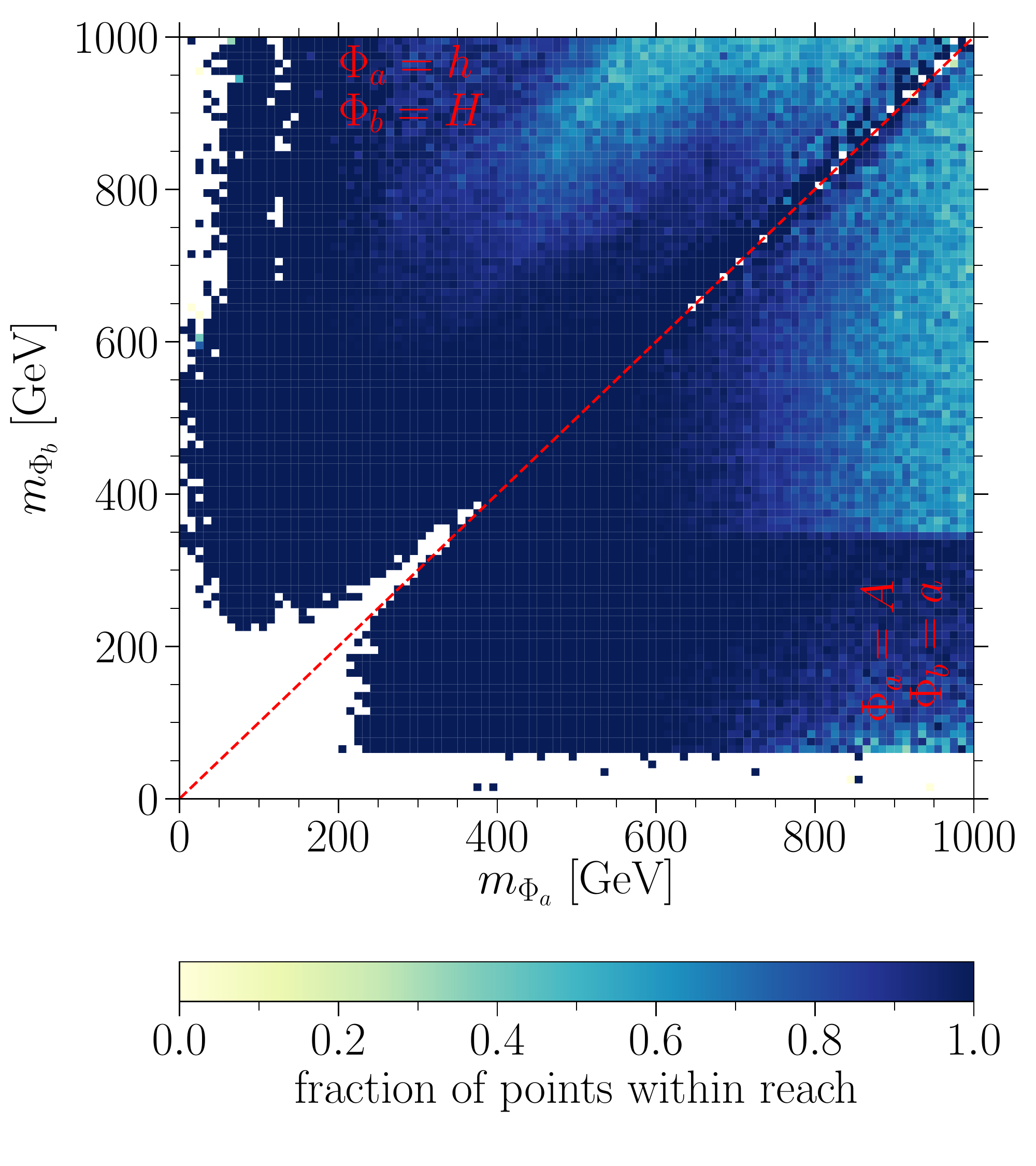} 
	\includegraphics[width=.49\linewidth]{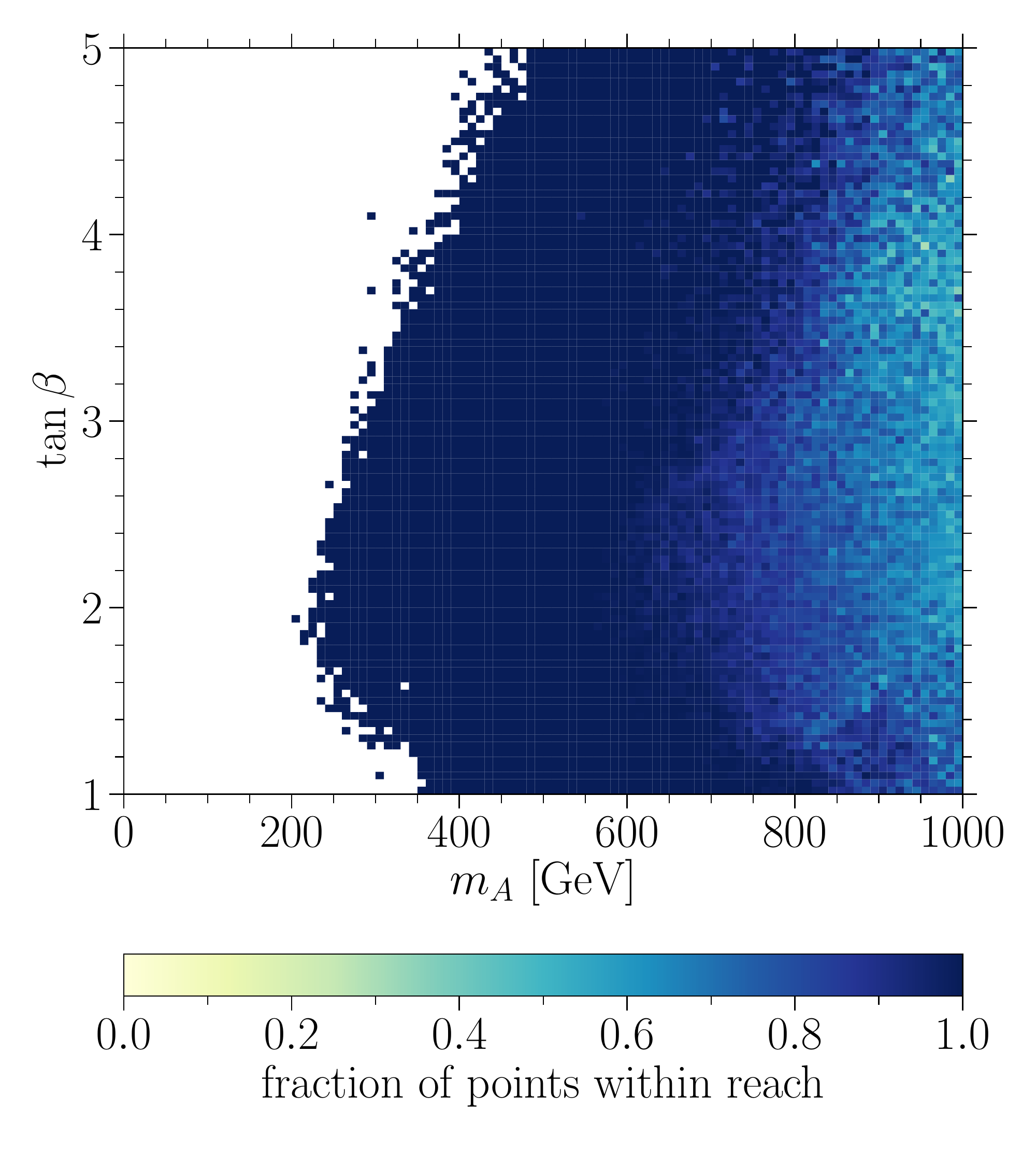}
	\caption{Same as Fig.~\ref{fig:combinedreach} but the left panel shows the distribution of LHC sensitivity for all the masses: non SM-like CP-even masses $m_H$ vs. $m_h$ in the top left triangle, and CP-odd masses $m_A$ vs. $m_a$ in the lower right triangle. The right panel instead shows the sensitivity in the conventional $m_A$-$\tan\beta$ plane predominantly used when presenting results in the MSSM.}
	\label{fig:combinedreach2}
\end{figure}

In Figs.~\ref{fig:combinedreach} and \ref{fig:combinedreach2} we show the coverage of the parameter space when combining conventional searches with searches utilizing Higgs cascades under the assumptions that the sensitivity of the searches will be improved to $\left[\mu_{\rm Proj.}^{3000\,{\rm fb}^{-1}}({\rm Higgs\,cascades}) = 0.1\right]$ and $\left[\mu_{\rm Curr.\,Lim.}^{<37\,{\rm fb}^{-1}}({\rm conventional)} = 0.01\right]$ as discussed in the previous paragraph. These figures demonstrate that such a combination will allow the LHC to probe most of the NMSSM parameter space where at least one of the heavy Higgs states $H$ or $A$ has a mass below 1\,TeV. The left panel of Fig.~\ref{fig:combinedreach} presents the fraction of points scanned in each bin which will be probed at the LHC with 3000 fb$^{-1}$ of data in the $m_A$ vs. $(P_A^{\rm S})^2$ plane. The left panel of Fig.~\ref{fig:combinedreach} and Fig.~\ref{fig:combinedreach2} present the same information in the $m_h$ vs. $(S_h^{\rm S})^2$ plane, $m_H$ vs. $m_h$~(top triangle left panel) and $m_A$ vs. $m_a$~(bottom triangle left panel) plane, and the $m_A$ vs. $\tan\beta$ plane, respectively. In these figures, the white regions for the lighter masses are excluded due to current direct search bounds. In the regions still allowed, future searches start to lose sensitivity when the mass of the heavy CP-odd state approaches 1\,TeV, unless the lighter CP-odd state $a$ is mostly doublet-like (corresponding to the $m_A \sim 1\,$TeV, $(P_A^{\rm S})^2 \sim 1$ region in the left panel of Fig.~\ref{fig:combinedreach}); in this region of parameter space $a$ retains relatively large production cross sections at the LHC such that it can be searched for with conventional Higgs searches. The region with the least sensitivity in the left panel of Fig.~\ref{fig:combinedreach}~(heavy $m_A$ with sizable mixing $(P_A^{\rm S})^2$, colored pale yellow) is correlated with the heavy $m_h$ region in the right panel, particularly clustered around small $(P_h^{\rm S})^2 \ll 1$, implying a dominantly doublet-like $m_h$ and a mostly singlet-like $m_H$. Further, it can be seen from the left panel of Fig.~\ref{fig:combinedreach2} that this region corresponds to $m_a\gtrsim 350$ GeV, where because of sizable mixing~(as seen from the left panel of Fig.~\ref{fig:combinedreach}), the $a$ is expected to have large branching ratios into pairs of top quarks, degrading search sensitivities. We again note the hard cut-off in the sensitivity for $a$ at the top threshold visible in the lower triangle in the left panel of Fig.~\ref{fig:combinedreach2}. As noted previously, while suppressed by alignment, $m_h$ can decay to pairs of gauge bosons, whereas such decays are forbidden for the CP-odd states at tree-level.

In the left panel of Fig.~\ref{fig:combinedreach2} we can most clearly see the complementarity between conventional searches and the Higgs cascade channels: If the non SM-like CP-even states ($h$ and $H$) and the CP-odd states ($a$ and $A$) are approximately mass degenerate, i.e. close to the diagonal, conventional searches are most powerful since all of the Higgs bosons could be heavily mixed and thus be copiously produced and decay to pairs of SM particles. On the other hand, if the Higgs bosons are not mass degenerate, the Higgs cascade channels are most powerful, as we have already seen in Fig.~\ref{fig:variousreach1}. In the right panel of Fig.~\ref{fig:combinedreach2} we shows these sensitivities in the conventional $m_A$-$\tan\beta$ plane for comparison with results generically presented in the MSSM. We see that the possibility of Higgs cascade decays in the NMSSM will allow the LHC to probe the low $\tan\beta$ region up to $m_A\sim 1$ TeV. 

We note again that the distribution of points portrayed in Figs.~\ref{fig:combinedreach} and~\ref{fig:combinedreach2} does not reflect the actual density of points scanned. As mentioned in section~\ref{sec:paramCorrel}, the viable NMSSM parameter space we have analyzed corresponds to predominantly doublet-like $m_A$ and singlet-like $m_a$, with $\tan\beta$ heavily clustered around $\sim$ 2.5. Combining conventional and Higgs cascade search channels, we expect ${\approx 50\,\%}$ of the points from our NMSSM parameter scan consistent with $h_{125}$ phenomenology, current direct search constraints, and featuring spectra with the additional Higgs bosons $\min(m_A,m_H) \leq 1\,$TeV to be probed by the LHC with $3000\,{\rm fb}^{-1}$ of data. Under the optimistic assumption that the LHC collaborations are able to improve their reach in the Higgs cascade channels by one order of magnitude with respect to our projections, and two order of magnitude in the conventional search channels with respect to current limits, almost all ($\approx 90\,\%$) of the points could be probed at the future LHC.

%*********************************************************
\section{Conclusions} \label{sec:conclusion}
%*********************************************************

In this work, we have studied the collider phenomenology of the $\mathbb{Z}_3$-invariant NMSSM. We have focused on the Higgs and neutralino sector of the model, which is usually described in terms of the parameters appearing in the scalar potential. However, in order to be compatible with the observed Higgs phenomenology, the model must contain a 125\,GeV Higgs mass eigenstate with SM-like couplings. This leads to strong correlations between the physical parameters, in particular the masses of the additional Higgs bosons and their supersymmetric partners, which are part of the neutralino sector. We have demonstrated that the Higgs and neutralino sector of the NMSSM can be effectively described by four physically intuitive parameters: the physical masses of the two CP-odd Higgs bosons, the mixing angle in the CP-odd Higgs sector, and $\tan\beta$, all of which are quite transparently connected to the couplings of the physical Higgs and neutralino states. This reduction in parameters due to $h_{125}$ phenomenology, and the induced correlation in the physical masses and couplings, makes the NMSSM much more tractable than previously thought. We stress that we verified our conclusions with intensive numerics using \texttt{NMSSMTools}. Without implementing alignment a priori as an input to our scans, the correlated parameters and masses were obtained as an output. 

Most search efforts for an extended Higgs sector at the LHC have been focused on direct searches, looking for resonant decays of heavy scalars/pseudo-scalars into SM particles as is generically predicted in supersymmetric models like the MSSM or generic 2HDMs. While such strategies have led to the exclusion of large mass ranges for the heavy doublet-like non SM-like states for large values of $\tan\beta$, the low $\tan\beta$ region is challenging to probe for masses of the non SM-like Higgs bosons above $\sim 350\,$GeV where decays into pairs of top quarks are kinematically allowed. The presence of the singlet sector in the NMSSM allows for Higgs cascade decays, where a heavy Higgs boson decays into two lighter Higgs bosons or one light Higgs and a $Z$ boson. As has been previously discussed e.g. in Refs.~\cite{Carena:2015moc,Baum:2017gbj,Ellwanger:2017skc}, these Higgs cascades can play an important role in the phenomenology of the NMSSM and provide a promising means to probe the low $\tan\beta$ regime. Of such decays, the branching ratios into pairs of SM-like Higgs boson or one SM-like Higgs and a $Z$ boson are suppressed by the presence of the SM-like 125\,GeV state observed at the LHC. Thus, the most relevant Higgs cascade modes are those into one SM-like Higgs and one non SM-like (light) Higgs boson, or into a $Z$ boson and one non SM-like (light) Higgs boson; the corresponding branching ratios are not suppressed and the presence of the SM-like Higgs or the $Z$ boson allows for the tagging of such processes at the LHC. If the additional Higgs bosons decay dominantly into pairs of SM particles, such Higgs cascades lead to Higgs+visible and $Z$+visible final states. If the dominant decay mode is into neutralinos, then the Higgs cascades lead to mono-Higgs and mono-$Z$ final states.

Previously, the potential of such Higgs cascade decays modes for probing the NMSSM parameter space has been studied on a channel-by-channel basis in the literature. Here, we have provided a systematic comparison of the different Higgs cascade modes, gaining analytical understanding of the phenomenology due to our re-parameterization of the NMSSM parameters in terms of the physical parameters of the CP-odd sector. Most importantly, we have demonstrated that it is crucial to use as many different final states arising through Higgs cascades as possible, since no single class of final states dominates throughout the NMSSM parameter space. Further, we note that Higgs cascade modes may play a crucial role for differentiating models, e.g. the MSSM from the NMSSM, if additional Higgs bosons are discovered at the LHC. Higgs cascade decays usually involve the singlet-like states characteristic of the NMSSM, which are challenging to probe via conventional Higgs searches at the LHC since their production cross sections are suppressed with respect to doublet-like Higgs bosons. 

In closing, we have demonstrated that the combination of Higgs cascade searches with conventional strategies to search for additional Higgs bosons via their decay into pairs of SM particles will allow $\approx 50$\% of the phenomenologically viable NMSSM parameter space with masses $\lesssim 1\,$TeV to be probed by the upcoming runs of the LHC. Under the optimistic assumption that the LHC collaborations are able to improve their reach in the Higgs cascade channels by one order of magnitude over our projections, and in the conventional search channels by two orders of magnitude with respect to current limits, $\approx 90\,\%$ of this NMSSM parameter space may be accessible to the LHC.  While this is an optimistic scenario, it sets a target for the sensitivity required to probe most of the remaining interesting parameter space of the NMSSM. 

%*********************************************************
\acknowledgments
%*********************************************************

We are indebted to Bibhushan Shakya for early collaboration in this project. We would also like to thank Carlos Wagner for interesting discussions. 
SB would like to thank the LCTP and the University of Michigan as well as Wayne State University, where part of this work was carried out, for hospitality. 
SB and KF acknowledge support by the Vetenskapsr\r{a}det (Swedish Research Council) through contract No. 638-2013-8993 and the Oskar Klein Centre for Cosmoparticle Physics. KF acknowledges support from DoE grant DE-SC007859 and the LCTP at the University of Michigan. NRS is supported by DoE grant DE-SC0007983 and Wayne State University. The work of NRS was partially performed at the Aspen Center for Physics, which is supported by National Science Foundation grant PHY-1607611.

%*********************************************************
\appendix
%*********************************************************

\newpage 
%*********************************************************
\section{LHC searches used to constrain the dataset}\label{app:LHCsearch}
%*********************************************************
\begin{table}[h!]
	\begin{center}
	\begin{tabular}{cccc}
		\hline\hline
		decay channel & NMSSM Higgs & Reference & Reference \\ & tested & $\sqrt{s}=8\,$TeV & $\sqrt{s}=13\,$TeV \\ \hline
		$\Phi_i \rightarrow \tau^+\tau^-$ & $h, H, a, A$ & \cite{Khachatryan:2014wca, CMS-PAS-HIG-14-029, Aad:2014vgg} & \cite{ATLAS-CONF-2016-085, CMS-PAS-HIG-16-037, CMS-PAS-HIG-17-020, Aaboud:2017sjh} \\
		$\Phi_i \rightarrow b\bar{b}$ & $h, H, a, A$ & -- & \cite{CMS-PAS-HIG-16-025} \\
		$\Phi_i \rightarrow \gamma\gamma$ & $h, H, a, A$ & \cite{Khachatryan:2015qba, CMS-PAS-HIG-14-037, Aad:2014ioa} & \cite{ATLAS-CONF-2016-018, CMS-PAS-EXO-16-027, ATLAS-CONF-2016-059, Aaboud:2017yyg} \\
		$\Phi_i \rightarrow ZZ$ & $h, H$ & \cite{Aad:2015kna} & \cite{CMS-PAS-HIG-16-001, ATLAS-CONF-2016-012, ATLAS-CONF-2016-016, ATLAS-CONF-2015-071, CMS-PAS-HIG-16-033, ATLAS-CONF-2016-056, ATLAS-CONF-2016-079, CMS-PAS-HIG-17-012, Aaboud:2017rel} \\
		$\Phi_i \rightarrow WW$ & $h, H$ & \cite{Aad:2015agg, CMS-PAS-HIG-13-027, ATLAS-CONF-2013-067} & \cite{ATLAS-CONF-2016-021, CMS-PAS-HIG-16-023, ATLAS-CONF-2016-074, ATLAS-CONF-2016-062, Aaboud:2017gsl, Aaboud:2017fgj} \\
		$\Phi_i \rightarrow h_{125}h_{125} \rightarrow b\bar{b}\tau^+\tau^-$ & $h, H$ & \cite{Khachatryan:2015tha, CMS-PAS-HIG-15-013, Aad:2015xja} & \cite{CMS-PAS-HIG-16-013, CMS-PAS-HIG-16-029, Sirunyan:2017djm} \\
		$\Phi_i \rightarrow h_{125}h_{125} \rightarrow b\bar{b} \ell \nu_\ell \ell \nu_\ell $ & $h, H$ & -- & \cite{CMS-PAS-HIG-16-011, Sirunyan:2017guj} \\
		$\Phi_i \rightarrow h_{125}h_{125} \rightarrow b\bar{b}b\bar{b} $ & $h, H$ & \cite{Khachatryan:2015yea, Aad:2015uka} & \cite{CMS-PAS-HIG-16-002, ATLAS-CONF-2016-017, ATLAS-CONF-2016-049, CMS-PAS-HIG-17-009} \\
		$\Phi_i \rightarrow h_{125}h_{125} \rightarrow b\bar{b} \gamma\gamma $ & $h, H$ & \cite{Khachatryan:2016sey, Aad:2014yja} & \cite{CMS-PAS-HIG-16-032, ATLAS-CONF-2016-004, CMS-PAS-HIG-17-008} \\
		$\Phi_i \rightarrow Z h_{125} \to Z b\bar{b} $ & $a, A$ & \cite{Aad:2015wra, Khachatryan:2015lba} & \cite{ATLAS-CONF-2016-015, Aaboud:2017cxo} \\
		$\Phi_i \rightarrow Zh_{125} \rightarrow Z \tau^+\tau^-$ & $a, A$ & \cite{Khachatryan:2015tha, Aad:2015wra} & -- \\
		$h_{125} \to a_i a_i \to \tau^+ \tau^- \tau^+ \tau^-$ & $a, A$ & \cite{Khachatryan:2240709} & --\\
		$h_{125} \to a_i a_i \to \mu^+ \mu^- b \bar{b}$ & $a, A$ & \cite{Khachatryan:2240709} & -- \\
		$h_{125} \to a_i a_i \to \mu^+ \mu^- \tau^+ \tau^-$ & $a, A$ & \cite{Khachatryan:2240709} & -- \\
		$h_{125} \to a_i a_i \to \mu^+ \mu^- \mu^+ \mu^-$ & $a, A$ & -- & \cite{CMS-PAS-HIG-16-035} \\
		$\Phi_i \to Z \Phi_j$ & $(A,h)$, $(H,a)$ & \cite{Khachatryan:2016are} & -- \\ \hline\hline 
	\end{tabular}
	\caption{Direct Higgs searches at the LHC used for this work. The second column indicates the NMSSM Higgs bosons which can take the place of the generic scalar $\Phi_i$ in the first column, recall that $H/h$ ($A/a$) are the heavier/lighter non SM-like CP-even (CP-odd) states and $h_{125}$ is the observed SM-like 125\,GeV Higgs boson. In the last row, the second column indicates possible pairs of $(\Phi_i, \Phi_j)$ in the corresponding process in the first column.}
	\label{tab:LHCSearches}
	\end{center}
\end{table}

%*********************************************************
\section{Benchmark Points}\label{app:BM}
%*********************************************************
In Tab.~\ref{tab:BP_params_masses} we present the NMSSM parameters and mass spectra, in Tab.~\ref{tab:BP_signal_strength} the signal strengths, and in Tab.~\ref{tab:BP_xsec_BR} the most relevant production cross sections and branching ratios for four Benchmark Points BP1$-$BP4. A description of the most important feature of the benchmark points can be found below. The benchmark points are chosen as examples of points which are simultaneously  within the projected reach of Higgs cascade search channels and difficult to detect with conventional search strategies. We categorize them according to the Higgs cascade channel corresponding to $\max\left[\mu_{\rm Proj.}^{3000\,{\rm fb}^{-1}}({\rm Higgs\;cascades})\right]$ with $\mathcal{L} = 3000\,{\rm fb}^{-1}$ of data as listed below:
\begin{itemize}
\item BP1: Mono-Higgs
\item BP2: Higgs+visible
\item BP3: Mono-$Z$
\item BP4: $Z$+visible
\end{itemize}

Note that since these benchmark points are obtained with \texttt{NMSSMTools}, we use the conventional set of NMSSM parameters as inputs and not our re-parameterization in terms of $\{m_a, m_A, P_A^{\rm S}, \tan\beta\}$ discussed in Section~\ref{sec:paramCorrel}. The parameters $\{\lambda, \kappa, \tan\beta, \mu, A_\lambda, A_\kappa\}$ are those appearing in the scalar potential, cf. Eq.~\eqref{eq:NMSSMparams}, and $M_{Q_3} = M_{U_3}$ is the stop mass parameter, and the remaining NMSSM parameters are fixed as detailed in section~\ref{sec:paramCorrel} and in the caption of Tab.~\ref{tab:BP_params_masses}.

All benchmark points presented here feature Higgs mass eigenstates approximately aligned with the Higgs basis interaction eigenstates. In particular, they all show very small doublet-doublet mixing $|S_{h_{125}}^{\rm NSM}| < 0.01$ as required by the observed phenomenology of the 125\,GeV SM-like state. The doublet-singlet mixing can take somewhat larger values; among the benchmark points we find the largest mixing angle for BP1 ($S_{h_{125}}^{\rm S} = 0.117$) and the smallest mixing angle for BP4 ($S_{h_{125}}^{\rm S} = -0.0487$). This proximity to the alignment limit is ensured by the values of $\lambda$ and $\kappa/\lambda$ close to what is dictated by the alignment conditions, and is found to be a generic feature of the allowed NMSSM parameter space we scanned. 
Note also that for all of the benchmark points, all non-SM states have masses larger than $m_{h_{125}}/2$. Therefore, $h_{125}$ can only decay into pairs of SM particles. Together with the approximate alignment of $h_{125}$ with $H^{\rm SM}$, this ensures compatibility of the $h_{125}$ phenomenology with LHC observations.

For all benchmark points, the lighter non SM-like CP-even state $h$ and the lighter CP-odd state $a$ are mostly singlet-like, while the heavier states $H$ and $A$ are dominantly composed of the non SM-like doublet interaction states $H^{\rm NSM}$ and $A^{\rm NSM}$, respectively. The only benchmark point featuring a sizable singlet component of one of the heavy doublet-like states $H$ or $A$ is BP4, with a mixing angle in the odd sector of $P_A^{\rm S} = 0.550$. The remaining benchmark points have mixing angles of $P_A^{\rm S} = -0.243$ for BP1, $P_A^{\rm S} = 0.0627$ for BP2, and $P_A^{\rm S} = 0.268$ for BP4. 

The mass spectra for all benchmark points are chosen such that the non SM-like doublet-like states $H$ and $A$ are heavy enough to be difficult to detect in conventional searches ($\{m_A, m_H\} \gtrsim 350\,$GeV) but light enough such that they are readily produced at the LHC. BP1, BP2, and BP3 feature masses of the doublet-like states of $m_A \sim m_H \sim 700\,$GeV. BP4 features somewhat lighter doublet-like states, $m_A \sim m_H \sim 500\,$GeV. In order to allow for sufficiently large mass gaps necessary for Higgs cascade decays, the mass of the singlet-like pseudo-scalar states has been chosen considerably lighter than the mass of the doublet-like states, $m_a \sim 200\,$GeV for BP1, $m_a \sim 160\,$GeV for BP2 and BP4, and $m_a \sim 290\,$GeV for BP3. Further, while BP1 features similar singlet masses $h$ and $a$, BP2, BP3 and BP4 have much larger mass splittings. The corresponding singlet-like scalar masses are $m_h \sim 165\,$GeV for BP1, $m_h \sim 560\,$GeV for BP2, $m_h \sim 70\,$GeV for BP3, and $m_h \sim 300\,$GeV for BP4. Regarding the lightest neutralino, BP1, BP3 and BP4 features $m_{\chi_1}\sim 100$ GeV, whereas BP2 features a much heavier neutralino $m_{\chi_1}\sim 500$ GeV.

\begin{table}
	\begin{center}
	\begin{tabular}{c|cccc}
		\hline\hline 
		& BP1 & BP2 & BP3 & BP4 \\
		& Mono-Higgs & Higgs+visible & Mono-$Z$ & $Z$+visible \\
		\hline
		$\lambda$ & $0.602$ & $0.602$ & $0.634$ & $0.668$ \\
		$\kappa$ & $-0.281$ & $0.347$ & $-0.203$ & $-0.734$ \\
		$\tan\beta$ & $2.73$ & $1.40$ & $2.09$ & $2.27$ \\
		$\mu$ [GeV] & $-193$ & $-466$ & $251$ & $141$ \\
		$A_\lambda$ [GeV] & $-784$ & $-270$ & $860$ & $741$ \\
		$A_\kappa$ [GeV] & $-200$ & $26.3$ & $470$ & $223$ \\
		$M_A$ [GeV] & $639$ & $732$ & $686$ & $472$ \\
		$M_{Q_3}$ [TeV] & $7.66$ & $7.78$ & $1.21$ & $3.85$ \\
		\hline
		$m_{h_{125}}$ [GeV] & $127$ & $128$ & $123$ & $126$ \\
		$m_h$ [GeV] & $165$ & $561$ & $66.8$ & $298$ \\
		$m_H$ [GeV] & $648$ & $750$ & $678$ & $460$ \\
		$m_a$ [GeV] & $205$ & $168$ & $290$ & $157$ \\
		$m_A$ [GeV] & $662$ & $749$ & $696$ & $533$ \\
		\hline
		$(S_{h_{125}}^{\rm S})^2$ & $1.34 \times 10^{-2}$ & $3.96 \times 10^{-3}$ & $6.90 \times 10^{-3}$ & $2.37 \times 10^{-3}$ \\
		$(S_h^{\rm S})^2$ & $0.972$ & $0.986$ & $0.984$ & $0.942$ \\
		$(S_H^{\rm S})^2$ & $1.41 \times 10^{-2}$ & $9.78 \times 10^{-3}$ & $8.84 \times 10^{-3}$ & $5.56 \times 10^{-2}$ \\
		$(P_A^{\rm S})^2$ & $5.92 \times 10^{-2}$ & $3.93 \times 10^{-3}$ & $7.17 \times 10^{-2}$ & $0.302$ \\
		\hline
		$m_{\chi_1}$ [GeV] & $102$ & $486$ & $97.6$ & $96.6$ \\
		$m_{\chi_2}$ [GeV] & $212$ & $494$ & $248$ & $142$ \\
		$m_{\chi_3}$ [GeV] & $292$ & $572$ & $323$ & $376$ \\
		%\hline
		%$\max\left[\mu_{\rm Proj.}^{3000\,{\rm fb}^{-1}}({\rm Higgs\;cascades})\right]$ & $2.36$ & $1.22$ & $2.14$ & $3.95$ \\
		%$\max\left[\mu_{\rm Curr.\,Lim.}^{<37\,{\rm fb}^{-1}}({\rm conventional})\right]$ & $8.76 \times 10^{-3}$ & $8.03\times 10^{-3}$ & $9.39 \times 10^{-3}$ & $0.117$ \\
		\hline\hline
		\end{tabular}
	\caption{NMSSM parameters and mass spectra for our benchmark points categorized according to $\max\left[\mu_{\rm Proj.}^{3000\,{\rm fb}^{-1}}({\rm Higgs\;cascades})\right]$, BP1: Mono-Higgs, BP2: Higgs+visible, BP3: Mono-$Z$, and BP: $Z$+visibles. The first block from the top shows the parameters used as input parameters in \texttt{NMSSMTools} $\{\lambda, \kappa, \tan\beta, \mu, A_\lambda, A_\kappa, M_{Q_3}\}$ where the first 6 parameters are those appearing in the scalar potential, cf. Eq.~\eqref{eq:NMSSMparams}, and $M_{Q_3} = M_{U_3}$ is the stop mass parameter which controls the radiative corrections to the scalar mass matrices. For the convenience of the reader we also record the value of $M_A$, defined in Eq.~\eqref{eq:MA}. As noted in Section~\ref{sec:paramCorrel}, the remaining parameters are fixed to $M_1 = M_2 = 1\,$TeV, $M_3 = 2\,$TeV, $A_t = \mu \cot\beta$, $A_b = \mu \tan\beta$, and all sfermion mass parameters (except $M_{Q_3} = M_{U_3}$) are fixed to 3\,TeV. The second block shows the mass spectrum of the Higgs sector and the third block  values of the singlet components of the non SM-like Higgs bosons. In particular, these blocks contain the masses of the CP-odd states $a$ and $A$ and the mixing angle in the CP-odd sector $P_A^{\rm S}$. Recall that these three quantities were used in our physical re-parameterization of the NMSSM (cf. Section~\ref{sec:CorrelConsequence}) together with the value of $\tan\beta$, which plays a minor role, and the values of $\lambda$ and $\kappa$, which are approximately fixed by alignment. In the fourth block we record the masses of the three lightest neutralinos. Since we set the bino and wino mass parameters to $M_1 = M_2 = 1\,$TeV, the two heaviest neutralinos $\chi_4$ and $\chi_5$ are bino- and wino-like with masses $m_{\chi_4} \approx m_{\chi_5} \approx 1\,$TeV, while the three lightest neutralinos, $\chi_1$, $\chi_2$, and $\chi_3$, are Higgsino- and singlino-like.}
	\label{tab:BP_params_masses}
	\end{center}
\end{table}

\begin{landscape}
\begin{table}
	\begin{center}
	\begin{tabular}{c|cccc}
		\hline\hline 
		& BP1 & BP2 & BP3 & BP4 \\
		& Mono-Higgs & Higgs+visible & Mono-$Z$ & $Z$+visible \\
		\hline
		$\max\left[\mu_{\rm Proj.}^{3000\,{\rm fb}^{-1}}({\rm Higgs\;cascades})\right]$ & $2.36$ & $1.22$ & $2.14$ & $3.95$ \\
		$\max\left[\mu_{\rm Curr.\,Lim.}^{<37\,{\rm fb}^{-1}}({\rm conventional})\right]$ & $8.76 \times 10^{-3}$ & $8.03\times 10^{-3}$ & $9.39 \times 10^{-3}$ & $0.117$ \\
		\hline
		Mono-Higgs Channels &&&&\\
		$\mu_{\rm Proj.}^{3000\,{\rm fb}^{-1}}(gg \to H \to h_{125} h \to \gamma\gamma \chi_1 \chi_1)$ & -- & -- & -- & -- \\
		$\mu_{\rm Proj.}^{3000\,{\rm fb}^{-1}}(gg \to A \to h_{125} a \to \gamma\gamma \chi_1 \chi_1)$ & $2.36$ & -- & $1.77$ & -- \\
		\hline
		Higgs+visible Channels &&&&\\
		$\mu_{\rm Proj.}^{3000\,{\rm fb}^{-1}}(gg \to H \to h_{125} h \to b\bar{b}b\bar{b})$ & $0.270$ & $1.60 \times 10^{-5}$ & $1.64$ & $1.61 \times 10^{-3}$ \\
		$\mu_{\rm Proj.}^{3000\,{\rm fb}^{-1}}(gg \to H \to h_{125} h \to b\bar{b} \gamma\gamma)$ & $8.37 \times 10^{-3}$ & -- & $1.25 \times 10^{-4}$ & $2.33 \times 10^{-4}$ \\
		$\mu_{\rm Proj.}^{3000\,{\rm fb}^{-1}}(gg \to H \to h_{125} h \to \gamma\gamma b\bar{b})$ & $0.125$ & -- & $0.383$ & $1.35 \times 10^{-4}$ \\
		$\mu_{\rm Proj.}^{3000\,{\rm fb}^{-1}}(gg \to A \to h_{125} a \to b\bar{b}b\bar{b})$ & $7.53 \times 10^{-2}$ & $1.22$ & $1.03 \times 10^{-2}$ & $3.68$ \\
		$\mu_{\rm Proj.}^{3000\,{\rm fb}^{-1}}(gg \to A \to h_{125} a \to b\bar{b} \gamma\gamma)$ & $5.75 \times 10^{-3}$ & $0.728$ & $6.21 \times 10^{-4}$ & $3.21 \times 10^{-2}$ \\
		$\mu_{\rm Proj.}^{3000\,{\rm fb}^{-1}}(gg \to A \to h_{125} a \to \gamma\gamma b\bar{b})$ & $3.80 \times 10^{-2}$ & $0.389$ & $3.46 \times 10^{-3}$ & $2.17$ \\
		\hline
		Mono-$Z$ Channels &&&&\\
		$\mu_{\rm Proj.}^{3000\,{\rm fb}^{-1}}(gg \to H \to Z a \to \ell^+ \ell^- \chi_1 \chi_1)$ & $1.73$ & -- & $2.14$ & -- \\
		$\mu_{\rm Proj.}^{3000\,{\rm fb}^{-1}}(gg \to A \to Z h \to \ell^+ \ell^- \chi_1 \chi_1)$ & -- & -- & -- & $0.189$ \\
		\hline
		$Z$+visible Channels &&&&\\
		$\mu_{\rm Proj.}^{3000\,{\rm fb}^{-1}}(gg \to H \to Z a \to \ell^+ \ell^- \tau^+ \tau^-)$ & $1.11 \times 10^{-2}$ & $0.136$ & $2.34 \times 10^{-3}$ & $3.95$ \\
		$\mu_{\rm Proj.}^{3000\,{\rm fb}^{-1}}(gg \to A \to Z h \to \ell^+ \ell^- \tau^+ \tau^-)$ & $4.24 \times 10^{-2}$ & $2.96 \times 10^{-6}$ & $0.131$ & $5.91 \times 10^{-4}$ \\
		\hline\hline
		\end{tabular}
	\caption{LHC signal strengths for the benchmark points BP1$-$BP4 defined in Tab.~\ref{tab:BP_params_masses}. In the first two rows we record the signal strength projected at the LHC for $\mathcal{L} = 3000\,{\rm fb}^{-1}$ of data in the dominant Higgs cascade channel, $\max\left[\mu_{\rm Proj.}^{3000\,{\rm fb}^{-1}}({\rm Higgs\;cascades})\right]$, and the largest signal strength in the conventional channels listed in Tab.~\ref{tab:LHCSearches}, $\max\left[\mu_{\rm Curr.\,Lim.}^{<37\,{\rm fb}^{-1}}({\rm conventional})\right]$, cf. the discussion in Sections~\ref{sec:reach}. In the remaining rows we record the projected signal strength at the LHC for $\mathcal{L} = 3000\,{\rm fb}^{-1}$ of data in the final states arising through Higgs cascades considered in this work.}
	\label{tab:BP_signal_strength}
	\end{center}
\end{table}
\end{landscape}

\begin{table}[h!]
	\begin{center}
	{\fontsize{11pt}{12.3pt}\selectfont
	\begin{tabular}{c|cccc}
		\hline\hline 
		& BP1 & BP2 & BP3 & BP4 \\
		& Mono-Higgs & Higgs+visible & Mono-$Z$ & $Z$+visible \\
		\hline
		$\sigma(gg \to h)$ [pb] & $7.10 \times 10^{-2}$ & $1.70 \times 10^{-4}$ & $0.484$ & $3.23\times 10^{-2}$ \\
		${\rm BR}(h \to \tau^+ \tau^-)$ & $1.74 \times 10^{-2}$ & $3.06 \times 10^{-5}$ & $8.78 \times 10^{-2}$ & $2.23 \times 10^{-4}$ \\
		${\rm BR}(h \to b\bar{b})$ & $0.151$ & $2.15 \times 10^{-4}$ & $0.909$ & $1.73 \times 10^{-3}$ \\
		${\rm BR}(h \to t\bar{t})$ & -- & $9.34 \times 10^{-4}$ & -- & -- \\
		${\rm BR}(h \to \gamma\gamma)$ & $4.32 \times 10^{-5}$ & $1.31 \times 10^{-6}$ & $1.17 \times 10^{-6}$ & $6.86 \times 10^{-6}$ \\
		${\rm BR}(h \to ZZ)$ & $1.77 \times 10^{-2}$ & $6.16 \times 10^{-2}$ & -- & $6.14 \times 10^{-3}$ \\
		${\rm BR}(h \to W^+ W^-)$ & $0.812$ & $0.128$ & -- & $1.38 \times 10^{-2}$ \\
		${\rm BR}(h \to \chi_1 \chi_1)$ & -- & -- & -- & $0.680$ \\
		\hline
		$\sigma(gg \to H)$ [pb] & $0.134$ & $0.239$ & $0.181$ & $0.907$ \\
		${\rm BR}(H \to \tau^+ \tau^-)$ & $8.66 \times 10^{-4}$ & $1.82 \times 10^{-4}$ & $4.19 \times 10^{-4}$ & $4.94 \times 10^{-4}$ \\
		${\rm BR}(H \to b\bar{b})$ & $6.02 \times 10^{-3}$ & $1.41 \times 10^{-3}$ & $2.92 \times 10^{-3}$ & $3.64 \times 10^{-3}$ \\ 
		${\rm BR}(H \to t\bar{t})$ & $0.281$ & $0.961$ & $0.405$ & $0.196$ \\
		${\rm BR}(H \to \gamma\gamma)$ & $2.29 \times 10^{-6}$ & $6.60 \times 10^{-6}$ & $3.94 \times 10^{-6}$ & $4.11 \times 10^{-6}$ \\
		${\rm BR}(H \to ZZ)$ & $7.31 \times 10^{-5}$ & $6.51 \times 10^{-4}$ & $1.70 \times 10^{-5}$ & $1.05 \times 10^{-4}$ \\
		${\rm BR}(H \to W^+ W^-)$ & $1.50 \times 10^{-4}$ & $1.33 \times 10^{-3}$ & $3.49 \times 10^{-5}$ & $2.21 \times 10^{-4}$ \\
		${\rm BR}(H \to \chi_1 \chi_1)$ & $6.66 \times 10^{-2}$ & -- & $4.47 \times 10^{-2}$ & $6.99 \times 10^{-2}$ \\
		${\rm BR}(H \to \chi_1 \chi_2)$ & $0.107$ & -- & $0.168$ & $4.03 \times 10^{-2}$ \\
		% ${\rm BR}(H \to \chi_1 \chi_3)$ & $2.28 \times 10^{-3}$ & -- & $1.44 \times 10^{-3}$ & -- \\
		% ${\rm BR}(H \to \chi_2 \chi_2)$ & $6.64 \times 10^{-3}$ & -- & $6.25 \times 10^{-3}$ & $2.23 \times 10^{-2}$ \\
		${\rm BR}(H \to \chi_2 \chi_3)$ & $0.110$ & -- & $4.00 \times 10^{-2}$ & -- \\
		% ${\rm BR}(H \to \chi_3 \chi_3)$ & $1.09 \times 10^{-3}$ & -- & $1.31 \times 10^{-4}$ & -- \\
		${\rm BR}(H \to hh)$ & $2.46 \times 10^{-3}$ & -- & $4.69 \times 10^{-3}$ & -- \\
		${\rm BR}(H \to hh_{125})$ & $0.102$ & $6.08 \times 10^{-3}$ & $7.12 \times 10^{-2}$ & $4.11 \times 10^{-2}$ \\
		${\rm BR}(H \to h_{125}h_{125})$ & $1.73 \times 10^{-3}$ & $8.56 \times 10^{-4}$ & $1.88 \times 10^{-3}$ & $8.56 \times 10^{-4}$ \\
		${\rm BR}(H \to aa)$ & $2.47 \times 10^{-3}$ & $1.01 \times 10^{-4}$ & $6.21 \times 10^{-4}$ & $3.53 \times 10^{-2}$ \\
		${\rm BR}(H \to Za)$ & $0.308$ & $2.69 \times 10^{-2}$ & $0.249$ & $0.569$ \\
		\hline
		$\sigma(gg \to a)$ [pb] & $0.195$ & $8.36 \times 10^{-2}$ & $0.335$ & $2.29$ \\
		${\rm BR}(a \to \tau^+ \tau^-)$ & $1.58 \times 10^{-3}$ & $9.05 \times 10^{-2}$ & $2.49 \times 10^{-4}$ & $0.101$ \\
		${\rm BR}(a \to b\bar{b})$ & $1.32 \times 10^{-2}$ & $0.797$ & $1.98 \times 10^{-3}$ & $0.885$ \\
		% ${\rm BR}(a \to t\bar{t})$ & -- & -- & -- & -- \\
		${\rm BR}(a \to \gamma\gamma)$ & $6.20 \times 10^{-6}$ & $5.60 \times 10^{-3}$ & $7.46 \times 10^{-7}$ & $5.55 \times 10^{-5}$ \\
		${\rm BR}(a \to \chi_1 \chi_1)$ & $0.985$ & -- & $0.994$ & -- \\
		\hline
		$\sigma(gg \to A)$ [pb] & $0.175$ & $0.336$ & $0.217$ & $0.619$ \\
		${\rm BR}(A \to \tau^+ \tau^-)$ & $8.37 \times 10^{-4}$ & $1.60 \times 10^{-4}$ & $3.88 \times 10^{-4}$ & $4.91 \times 10^{-4}$ \\
		${\rm BR}(A \to b\bar{b})$ & $5.84 \times 10^{-3}$ & $1.18 \times 10^{-3}$ & $2.76 \times 10^{-3}$ & $3.62 \times 10^{-3}$ \\
		${\rm BR}(A \to t\bar{t})$ & $0.350$ & $0.973$ & $0.478$ & $0.417$ \\
		${\rm BR}(A \to \gamma\gamma)$ & $4.39 \times 10^{-6}$ & $7.95 \times 10^{-6}$ & $8.31 \times 10^{-6}$ & $1.43 \times 10^{-5}$ \\
		${\rm BR}(A \to \chi_1 \chi_1)$ & $0.102$ & -- & $7.88 \times 10^{-2}$ & $8.81 \times 10^{-4}$ \\
		% ${\rm BR}(A \to \chi_1 \chi_2)$ & $1.53 \times 10^{-2}$ & -- & $1.35 \times 10^{-2}$ & $1.12 \times 10^{-2}$ \\
		% ${\rm BR}(A \to \chi_1 \chi_3)$ & $2.46 \times 10^{-5}$ & -- & $1.89 \times 10^{-2}$ & $1.41 \times 10^{-2}$ \\
		% ${\rm BR}(A \to \chi_2 \chi_2)$ & $4.05 \times 10^{-4}$ & -- & $1.72 \times 10^{-2}$ & $8.62 \times 10^{-2}$ \\
		% ${\rm BR}(A \to \chi_2 \chi_3)$ & $1.58 \times 10^{-2}$ & -- & $4.43 \times 10^{-3}$ & $1.25 \times 10^{-3}$ \\
		${\rm BR}(A \to \chi_3 \chi_3)$ & $0.112$ & -- & $7.62 \times 10^{-2}$ & -- \\
		${\rm BR}(A \to ha)$ & $3.31 \times 10^{-3}$ & $2.69 \times 10^{-4}$ & $2.93 \times 10^{-4}$ & $6.99 \times 10^{-2}$\\
		${\rm BR}(A \to h_{125}a)$ & $0.304$ & $1.88 \times 10^{-2}$ & $0.212$ & $0.111$ \\
		${\rm BR}(A \to Zh)$ & $8.40 \times 10^{-2}$ & $4.00 \times 10^{-3}$ & $5.56 \times 10^{-2}$ & $6.00 \times 10^{-2}$ \\
		${\rm BR}(A \to Zh_{125})$ & $5.61 \times 10^{-4}$ & $2.86 \times 10^{-4}$ & $1.79 \times 10^{-4}$ & $2.55 \times 10^{-5}$ \\
		\hline\hline
	\end{tabular}
	}
	\caption{Gluon fusion production cross sections at the $\sqrt{s} = 13\,$TeV LHC, $\sigma(gg \to \Phi)$, as well as the most relevant branching ratios for the non SM-like Higgs bosons $\Phi = \{h, H, a, A\}$ for the benchmark points BP1$-$BP4 defined in Tab.~\ref{tab:BP_params_masses}.}
	\label{tab:BP_xsec_BR}
	\end{center}
\end{table}

Regarding the branching ratios important for Higgs cascade decays, we first note that the branching ratio of heavy Higgs bosons into pairs of SM-like Higgs bosons or a SM-like Higgs and a $Z$ boson is suppressed due to the proximity to alignment as discussed in Section~\ref{sec:signals}, see also Refs.~\cite{Carena:2015moc, Baum:2017gbj, Baum:2018zhf}. For all benchmark points, we find
\begin{align*}
	{\rm BR}(H \to h_{125} h_{125}) &\ll \{ {\rm BR}(H \to h_{125} h), {\rm BR}(A \to h_{125} a)\} , \\
	{\rm BR}(A \to Z h_{125}) &\ll \{ {\rm BR}(A \to Z h), {\rm BR}(H \to Z a)\} .
\end{align*}

Additionally we note that in agreement with our discussions in Sec.~\ref{sec:ratios}, branching ratios of the heavy non-SM like doublets into either $h_{125}$ or $Z$ and an additional singlet like state are generally comparable. This leads to multiple channels that may be probed at the LHC for each benchmark point, as discussed in detail below.

\subsubsection*{BP1 - Mono-Higgs}

This benchmark point features a Higgs spectrum with comparable masses of the singlet-like states $a$ and $h$, $m_a = 205\,$GeV and $m_h = 165\,$GeV. The heavier states $A$ and $H$ are mostly composed of $A^{\rm NSM}$ and $H^{\rm NSM}$, respectively, and are approximately mass degenerate with $m_A \approx m_H \approx 650\,$GeV. The Higgsino mass parameter has a value of $\mu = -193\,$GeV, and $\kappa = -0.281$, leading to $2|\kappa|/\lambda = 0.93$. Thus, the lightest neutralino $\chi_1$ is mostly singlino-like but has sizable Higgsino components. Its mass is $m_{\chi_1} = 102\,$GeV, allowing for ($a \to \chi_1 \chi_1$) decays but not for ($h \to \chi_1 \chi_1$) decays. The second-lightest neutralino $\chi_2$ is dominantly Higgsino-like with a mass of $m_{\chi_2} = 212\,{\rm GeV} \approx |\mu|$, while $\chi_3$ is mostly Higgsino-like but has a sizable singlino component and a mass of $m_{\chi_3} = 292\,$GeV. 

Due to their singlet-like nature, the direct production cross sections of $a$ and $h$ are much smaller than those of a SM Higgs boson of the same mass, rendering them beyond the reach of conventional search channels at the LHC which rely on direct production of $a$ or $h$. The dominant decay modes of $h$ are into pairs of $b$-quarks, and, facilitated by its (small) doublet component, into pairs of $W$-bosons. The singlet-like pseudo-scalar on the other hand is kinematically allowed to decay into pairs of neutralinos ($a \to \chi_1 \chi_1$). Because $\chi_1$ has sizable singlino as well as Higgsino components, such decays proceed via both of the NMSSM's large couplings $\lambda$ and $\kappa$, rendering the corresponding branching ratio large, ${\rm BR}(a \to \chi_1 \chi_1) = 0.985$.

The heavier (doublet-like) CP-even state $H$ mostly decays into pairs of top quarks, neutralinos, and, most relevant for Higgs cascade channels, via ($H \to h h_{125}$) and ($H \to Z a$). The cross section ($gg \to H \to h h_{125}$) is not large enough for it to be within reach of the Higgs+visible search modes. However, facilitated by the sizable branching ratios of $(H \to Z a)$ and ($a \to \chi_1 \chi_1$), this benchmark point is within the projected reach of mono-$Z$ searches, $\mu_{\rm Proj.}^{3000\,{\rm fb}^{-1}}(gg \to H \to Z a \to \ell^+ \ell^- \chi_1 \chi_1) = 1.73$.

The heavier (doublet-like) CP-odd state $A$ mostly decays into pairs of top quarks, neutralinos, and through the ($A \to h_{125} a$) channel. The sizable branching ratio of the latter decay mode, ${\rm BR}(A \to Z h_{125}) = 0.304$, together with the large branching ratio corresponding to $(a \to \chi_1 \chi_1)$ decays leads to a large projected signal strength in mono-Higgs searches via the corresponding decay chain, $\mu_{\rm Proj.}^{3000\,{\rm fb}^{-1}}(gg \to A \to h_{125} a \to \gamma\gamma \chi_1 \chi_1) = 2.36$. 

Neither $H$ nor $A$ have large branching ratios into pairs of SM states except into pairs of top quarks, rendering them very difficult to detect by conventional searches. Thus, the best chances to detect BP1 would be in mono-$Z$ searches via ($gg \to H \to Z a$) (the projected signal strength in this channel is $\mu_{\rm Proj.}^{3000\,{\rm fb}^{-1}} = 1.73$) and particularly in mono-Higgs searches via ($gg \to A \to h_{125} a$) with a projected signal strength $\mu_{\rm Proj.}^{3000\,{\rm fb}^{-1}} = 2.36$.

\subsubsection*{BP2 - Higgs+visible}
Benchmark point BP2 features a Higgs spectrum with a large split between the masses of the singlet-like states $a$ and $h$, $m_a = 168\,$GeV and $m_h = 561\,$GeV. The heavier doublet-like states $A$ and $H$ are almost mass degenerate, $m_A \approx m_H \approx 750\,$GeV. The Higgsino mass parameter takes much larger absolute value than for BP1, $\mu = -466\,$GeV. Further, $\kappa$ also has a larger absolute value than for BP1, $\kappa = 0.347$, leading to $2|\kappa|/\lambda = 1.15$. Thus, the two lightest neutralinos, $\chi_1$ and $\chi_2$, are mostly Higgsino like and approximately mass degenerate, $m_{\chi_1} = 486\,$GeV and $m_{\chi_2} = 494\,$GeV. The third-lightest neutralino, $\chi_3$, is mostly composed of the singlino and has a mass of $m_{\chi_3} = 572\,$GeV. Note that because $|2|\kappa|/\lambda-1|$ is larger than for BP1, the Higgsino and singlino mass parameters are split further for BP2 than for BP1, leading to much smaller singlino-Higgsino mixing. Further, because of the relatively large masses of the neutralinos, none of the Higgs bosons are kinematically allowed to decay into pairs of neutralinos.

Similar to BP1, the large singlet components of $a$ and $h$ lead to direct production cross sections at the LHC much smaller than those of SM Higgs bosons of the same mass. Thus, they are out of reach of conventional search strategies. The dominant decay mode of the CP-even state $h$ is into pairs of $W$-bosons and pairs of the much lighter singlet-like CP-odd state, ${\rm BR}(h \to aa) = 0.740$. The CP-odd state $a$ decays mostly into pairs of $b$-quarks with a branching ratio of ${\rm BR}(a \to b\bar{b}) = 0.797$. It also has a sizable branching ratio into $\tau$-leptons, ${\rm BR}(a \to \tau^+ \tau^-) = 0.0905$.

The heavier (doublet-like) CP-even state $H$ predominantly decays into pairs of top quarks. Because of the small value of $\tan\beta$ compared to BP1, ($H \to h_{125} h$) decays, which are mostly controlled by the ($H^{\rm SM} H^{\rm NSM} H^{\rm S}$) coupling given in Tab.~\ref{tab:tricoup_NMSSM}, are suppressed. The largest branching ratio of $H$ relevant for Higgs cascade searches is ${\rm BR}(H \to Z a) = 0.0269$. However, this branching ratio is not sufficiently large to put BP2 within reach of $Z$+visible searches where the projected signal strength is only $\mu_{\rm Proj.}^{3000\,{\rm fb}^{-1}}(gg \to H \to Z a \to \ell^+ \ell^- \chi_1 \chi_1) = 0.136$.

Similar to $H$, the CP-odd doublet-like state $A$ mostly decays into pairs of top quarks. The largest branching ratio relevant for Higgs cascade searches is ${\rm BR}(A \to h_{125} a) = 0.0188$. This decay mode is mostly controlled by the $(H^{\rm SM} A^{\rm NSM} A^{\rm S})$ coupling, which becomes largest for values of $\tan\beta = 1$, but is suppressed for ${\rm sgn}(\kappa) = +1$ see Tab.~\ref{tab:tricoup_NMSSM}. Nonetheless, together with the large branching ratio of $a$ into pairs of $b$-quarks, this branching ratio is sufficiently large to render BP2 within reach of Higgs+visible searches with a projected signal strength $\mu_{\rm Proj.}^{3000\,{\rm fb}^{-1}}(gg \to A \to h_{125} a \to b\bar{b}b\bar{b}) =1.22$.

Since both $H$ and $A$ decay dominantly into pairs of top quarks, BP2 is very challenging to discover with conventional search strategies. The most promising channel to discover this benchmark point at the LHC are Higgs+visible Higgs cascade searches, particularly in the $b\bar{b}b\bar{b}$ final state with a projected signal strength of $\mu_{\rm Proj.}^{3000\,{\rm fb}^{-1}}(gg \to A \to h_{125} a \to b\bar{b}b\bar{b}) =1.22$. If the sensitivity of the LHC can be improved by an order of magnitude over our projections, BP2 could also be probed via $Z$+visible searches through the ($gg \to H \to Z a$) channel, $\mu_{\rm Proj.}^{3000\,{\rm fb}^{-1}}(gg \to H \to Z a \to \ell^+ \ell^- \chi_1 \chi_1) = 0.136$.

\subsubsection*{BP3 - Mono-$Z$}

Similar to BP2, the benchmark points BP3 also features a Higgs mass spectrum with a sizable split between the masses of the singlet-like states $a$ and $h$, $m_a = 290\,$GeV and $m_h = 66.8\,$GeV. However, note that we have the inverted hierarchy for BP3 compared to BP2: for BP3 the CP-even state $h$ is much lighter than the CP-odd state $a$, while for BP2 $h$ is much lighter than $a$. The doublet like states are $A$ and $H$ are again approximately mass degenerate, $m_A = 696\,$GeV and $m_H = 678\,$GeV. The Higgsino mass parameter $\mu$ takes a moderate absolute value, $\mu=251\,$GeV, and $\kappa = -0.203$ takes somewhat smaller absolute values than for BP1 and particularly BP2. The value of $\kappa$ implies $2|\kappa|/\lambda = 0.640$, implying a singlino mass parameter much smaller than the Higgsino masses. Correspondingly, we find that the lightest neutralino, $\chi_1$, is mostly singlino-like with a mass of $m_{\chi_1} = 97.6\,$GeV. The second- and third-lightest neutralino, $\chi_2$ and $\chi_3$, are mostly Higgsino-like and have masses of $m_{\chi_2} = 248\,$GeV and $m_{\chi_3} = 323\,$GeV, respectively. The singlino-Higgsino mixing is smaller than for BP1, but $\chi_3$ still has a singlino fraction of $\sim 0.3$. Note that due to the mass spectra, the singlet-like CP-odd state $a$ is allowed to decay into pairs of $\chi_1$'s, while such decays are kinematically forbidden for the CP-even state $h$.

Due to their mostly-singlet nature, the direct production cross sections of $a$ and $h$ are much smaller than those of SM Higgs bosons of the same mass, rendering them difficult to detect with conventional search strategies. The CP-even state $h$ dominantly decays into pairs of $b$-quarks with a branching ratio of ${\rm BR}(h \to b\bar{b}) = 0.909$. The CP-odd state $a$ is kinematically allowed to decay into pairs of neutralinos $\chi_1$. Similar to the case of BP1, the (somewhat smaller but still sizable) Higgsino components of $\chi_1$ and its large singlino component renders the corresponding branching ratio large, ${\rm BR}(a \to \chi_1 \chi_1) = 0.994$, since such decays proceed through both of the NMSSM's large couplings $\lambda$ and $\kappa$.

The heavier (doublet-like) CP-even state $H$ has large branching ratios into pairs of top quarks, neutralinos, and into a $Z$ boson and an $a$. The latter decay mode is particularly relevant for Higgs cascade searches, the corresponding branching ratio is ${\rm BR}(H \to Z a) = 0.249$. Together with the large branching ratio of $a$ into pairs of neutralinos $\chi_1$, we find a large projected signal strength in the corresponding mono-$Z$ final state, $\mu_{\rm Proj.}^{3000\,{\rm fb}^{-1}}(gg \to H \to Z a \to \ell^+ \ell^- \chi_1 \chi_1) = 2.14$. Further, although the branching ratio for ($H \to h h_{125})$‚ decays is rather small, the large branching ratio of $h$ into pairs of $b$-quarks renders BP2 in reach of Higgs+visible searches in the $b\bar{b}b\bar{b}$ final state with projected signal strength of $\mu_{\rm Proj.}^{3000\,{\rm fb}^{-1}}(gg \to H \to h_{125} h \to b\bar{b}b\bar{b}) = 1.64$.

The doublet-like CP-odd state $A$ mostly decays into pairs of top quarks or a SM-like Higgs and the light pseudo-scalar state ($A \to h_{125} a$). The branching ratio of the latter decay is large, ${\rm BR}(A \to h_{125} a) = 0.212$, and together with the large branching ratio of ($a \to \chi_1 \chi_1$) decays puts this point within the projected reach of mono-Higgs searches with a signal strength of $\mu_{\rm Proj.}^{3000\,{\rm fb}^{-1}}(gg \to A \to h_{125} a \to \gamma\gamma \chi_1 \chi_1) = 1.77$. The branching ratio for ($A \to Z h$) decays is small, ${\rm BR}(A \to Z h) = 5.56 \times 10^{-2}$, such that despite a rather large branching ratio of $h$ into pairs of $\tau$-leptons, ${\rm BR}(h \to \tau^+ \tau^-) = 8.78 \times 10^{-2}$ this benchmark point would only be probed in $Z$+visible Higgs cascade searches via ($gg \to A \to Z h$) if the sensitivity of the search is improved by approximately one order of magnitude beyond our projections.

In summary, as was the case for BP1 and BP2, benchmark point BP3 is challenging to probe via conventional search strategies. This is because the heavier doublet-like states $H$ and $A$ have no large branching ratios into pairs of SM particles except into pairs of top quarks, and because the production cross sections of the lighter states $a$ and $h$ are suppressed due to their singlet-like nature. The most promising probe of BP3 are $Z$+visible Higgs cascade searches with a projected signal strength of $\mu_{\rm Proj.}^{3000\,{\rm fb}^{-1}}(gg \to H \to Z a \to \ell^+ \ell^- \chi_1 \chi_1) = 2.14$. Although with somewhat smaller signal strengths, this benchmark point would also readily be observable in Higgs+visible searches via ($H \to h_{125} h$) decays, $\mu_{\rm Proj.}^{3000\,{\rm fb}^{-1}}(gg \to H \to h_{125} h \to b\bar{b}b\bar{b}) = 1.64$, and in mono-Higgs searches via ($A \to h_{125} a$) decays, $\mu_{\rm Proj.}^{3000\,{\rm fb}^{-1}}(gg \to A \to h_{125} a \to \gamma\gamma \chi_1 \chi_1) = 1.77$. 

\subsubsection*{BP4 - $Z$+visible}

This benchmark points features somewhat lighter doublet-like Higgs states $A$ and $H$ than BP1$-$BP3. Furthermore, the mixing angle of the CP-odd Higgs bosons is sizable, $P_A^{\rm S} = 0.550$. This leads to a larger split between the masses of the doublet-like states, $m_A = 533\,$GeV and $m_H = 460\,$GeV. The singlet-like states $a$ and $h$ have masses of $m_a = 157\,$GeV and $m_h = 298\,$GeV, respectively. The Higgsino mass parameter takes a comparatively small absolute value, $\mu = 141\,$GeV, while $\kappa$ has a large absolute value, $\kappa = -0.734$. Thus, $2|\kappa|/\lambda = 2.20$, corresponding to a singlino mass parameter much larger than the Higgsino mass parameter. Accordingly, we find that the two lightest neutralinos, $\chi_1$ and $\chi_2$, are mostly Higgsino-like with masses of $m_{\chi_1} = 96.6\,$GeV and $m_{\chi_2} = 142\,$GeV, respectively. The third-lightest neutralino, $\chi_3$, is mostly singlet-like with a mass of $m_{\chi_3} = 376\,$GeV. These masses imply that the singlet-like CP-even state $h$ can decay into pairs of $\chi_1$'s, while such decays are kinematically forbidden for the singlet-like CP-odd state $a$.

The direct production cross sections of the singlet-like states $a$ and $h$ are again suppressed by their singlet-like nature. Note that due to its sizable $A^{\rm NSM}$ component, the direct production cross section of $a$ is only suppressed by a relatively small factor of $\sigma(gg \to a)/\sigma_{ggh}(m_a) = 0.08$ with respect to the gluon fusion production cross section of a SM Higgs boson of the same mass, $\sigma_{ggh}(m_a)$. However, this state is still very challenging to detect at the LHC because its couplings to pairs of $W$ and $Z$ bosons vanish at tree-level for a CP-odd state. Thus, its branching ratios into pairs of $W$ and $Z$ bosons as well as into pairs of photons are much reduced compared to a SM Higgs boson of the same mass. The most promising conventional search channels for $a$ are via its dominant decay modes ($a \to b\bar{b}$) and ($a \to \tau^+ \tau^-$), but the current limits in the corresponding search channels are relatively weak. The largest signal strength in conventional searches is in the ($gg \to a \to \tau^+\tau^-$) search mode, $\mu_{\rm Curr.\,Lim.}^{<37\,{\rm fb}^{-1}}(gg \to a \to \tau^+\tau^-) = 0.117$. Recall that this signal strength is calculated with respect to the current limit. We expect the limit in this channel to improve in future runs of the LHC such that BP4 may be in reach of conventional search channels via the ($gg \to a \to \tau^+ \tau^-$) channel. The CP-even singlet-like state $h$ on the other hand still has only a  small doublet fraction and thus a small production cross section at the LHC. Moreover, its dominant decay mode is into pairs of neutralinos with a corresponding branching ratio of ${\rm BR}(h \to \chi_1 \chi_1) = 0.680$, rendering it virtually impossible to discover with conventional search strategies.

The heavier (doublet-like) CP-even state $H$ mostly decays into pairs of top quarks and via the ($H \to Z a$) channel. The latter decay has a large branching ratio of ${\rm BR}(H \to Z a) = 0.569$. Together with the sizable branching ratio of $a$ into pairs of $\tau$-leptons, BP4 is rendered within reach of $Z$+visible Higgs cascade searches via the ($gg \to A \to Z h$) channel. The projected signal strength is $\mu_{\rm Proj.}^{3000\,{\rm fb}^{-1}}(gg \to H \to Z a \to \ell^+ \ell^- \tau^+ \tau^-) = 3.95$.

The doublet-like CP-odd state $A$ has large branching ratios into pairs of top quarks and into a SM-like Higgs boson and a light CP-odd state $a$, ${\rm BR}(A \to h_{125} a) = 0.111$. In combination with the large branching ratio of $a$ into pairs of $b$-quarks, ${\rm BR}(a \to b\bar{b}) = 0.885$, this makes BP4 accessible for Higgs+visible Higgs cascade searches with signal strengths of $\mu_{\rm Proj.}^{3000\,{\rm fb}^{-1}}(gg \to A \to h_{125} a \to b\bar{b}b\bar{b}) = 3.68$ and $\mu_{\rm Proj.}^{3000\,{\rm fb}^{-1}}(gg \to A \to h_{125} a \to \gamma\gamma b\bar{b}) = 2.17$. This benchmark point is difficult to probe in mono-Higgs and mono-$Z$ channels despite the sizable branching fraction of $h$ into pairs of neutralinos. The most promising mode is via the ($gg \to A \to Z h$) channel; however, the corresponding branching ratio of $A$, ${\rm BR}(A \to Z h) = 0.06$ is not sufficiently large to put the point in the projected reach of the LHC. The projected signal strength in this channel is $\mu_{\rm Proj.}^{3000\,{\rm fb}^{-1}}(gg \to A \to Z h \to \ell^+ \ell^- \chi_1 \chi_1) = 0.189$. 

In summary, although passing all current limits arising through conventional search strategies, BP4 may be probed by conventional searches if their sensitivity can be improved significantly in future runs of the LHC. The ($gg \to a \to \tau^+\tau^-$) mode is the most promising decay channel with a current signal strength of $\mu_{\rm Curr.\,Lim.}^{<37\,{\rm fb}^{-1}}(gg \to a \to \tau^+\tau^-) = 0.117$. Furthermore, this benchmark point can be probed by Higgs cascade searches. We find the largest projected signal strength in $Z$+visible searches, $\mu_{\rm Proj.}^{3000\,{\rm fb}^{-1}}(gg \to H \to Z a \to \ell^+ \ell^- \tau^+ \tau^-) = 3.95$. This point is also within reach of Higgs+visible search channels via ($A \to h_{125} a$) decays, with signal strengths of $\mu_{\rm Proj.}^{3000\,{\rm fb}^{-1}}(gg \to A \to h_{125} a \to b\bar{b}b\bar{b}) = 3.68$ and $\mu_{\rm Proj.}^{3000\,{\rm fb}^{-1}}(gg \to A \to h_{125} a \to \gamma\gamma b\bar{b}) = 2.17$.

\newpage
%*********************************************************
\section{Trilinear couplings in the Higgs basis}\label{app:tricoup}
%*********************************************************

\begin{table}[h!]
 	\begin{center}
 	\setlength{\extrarowheight}{4.5pt}
 		\begin{tabular}{rl}
 			\hline\hline
 			{\normalsize $\left(\Phi^i \Phi^j \Phi^k\right)$} : & {\normalsize $\sqrt{2} g_{\Phi^i \Phi^j \Phi^k}$}\\
 			\hline
 			$\left(H^{\rm SM} H^{\rm SM} H^{\rm SM}\right)$ : & $3 \left(m_Z^2 c_{2\beta}^2 + \lambda^2 v^2 s_{2\beta}^2 \right)/v$ \\
 			$\left(H^{\rm SM} H^{\rm SM} H^{\rm NSM}\right)$ : & $ - 3 \left(m_Z^2 - \lambda^2 v^2 \right) s_{2\beta}c_{2\beta}/v$ \\
 			$\left(H^{\rm SM} H^{\rm SM} H^{\rm S}\right)$ : & $2 \lambda \mu \left( 1 - \frac{M_A^2}{4\mu^2} s_{2\beta}^2 - \frac{\kappa}{2 \lambda} s_{2\beta} \right)$\\
 			$\left(H^{\rm SM} H^{\rm NSM} H^{\rm NSM}\right)$ : & $ m_Z^2/v \left( 2 s_{2\beta}^2 - c_{2\beta}^2 \right) - \lambda^2 v \left( s_{2\beta}^2 - 2 c_{2\beta}^2 \right) $ \\
 			$\left(H^{\rm SM} H^{\rm NSM} H^{\rm S}\right)$ : & $- \lambda \mu c_{2\beta} \left( \frac{M_A^2}{2\mu^2} s_{2\beta} + \frac{\kappa}{\lambda} \right)$ \\
 			$\left(H^{\rm SM} H^{\rm S} H^{\rm S}\right)$ : & $2 \lambda v \left( \lambda - \kappa s_{2\beta} \right) $ \\
 			$\left(H^{\rm NSM} H^{\rm NSM} H^{\rm NSM}\right)$ : & $3 s_{2\beta} c_{2\beta} \left( m_Z^2/v - \lambda^2 v \right) $ \\
 			$\left(H^{\rm NSM} H^{\rm NSM} H^{\rm S}\right)$ : & $ \lambda \frac{M_A^2}{2\mu} s_{2\beta}^2 + \mu\left( 2\lambda + \kappa s_{2\beta} \right) $ \\
 			$\left(H^{\rm NSM} H^{\rm S} H^{\rm S}\right)$ : & $ -2 v \lambda \kappa c_{2\beta} $ \\
 			$\left(H^S H^S H^S\right)$ : & $ 2\kappa \left( A_\kappa + 6 \kappa \mu / \lambda \right) $ \\
 			\hline
 			$\left(H^{\rm SM} A^{\rm NSM} A^{\rm NSM}\right)$ : & $- \left[ \left(m_Z^2 - \lambda^2 v^2 \right) c_{2 \beta}^2 - \lambda^2 v^2\right]/v$ \\
 			$\left(H^{\rm SM} A^{\rm NSM} A^{\rm S}\right)$ : & $\lambda \left( \frac{M_A^2}{2\mu} s_{2\beta} - \frac{3 \kappa \mu}{\lambda} \right)$ \\
 			$\left(H^{\rm SM} A^{\rm S} A^{\rm S}\right)$ : & $ 2 \lambda v \left( \lambda + \kappa s_{2\beta} \right)$ \\
 			$\left(H^{\rm NSM} A^{\rm NSM} A^{\rm NSM}\right)$ : & $s_{2\beta} c_{2\beta} \left( m_Z^2 - \lambda^2 v^2 \right)/v$ \\
 			$\left(H^{\rm NSM} A^{\rm NSM} A^{\rm S}\right)$ : & $0$ \\
 			$\left(H^{\rm NSM} A^{\rm S} A^{\rm S}\right)$ : & $ 2v \lambda \kappa c_{2\beta}$ \\
 			$\left(H^{\rm S} A^{\rm NSM} A^{\rm NSM}\right)$ : & $ \lambda \frac{M_A^2}{2\mu} s_{2\beta}^2 + \mu\left( 2\lambda + \kappa s_{2\beta} \right) $ \\
 			$\left(H^{\rm S} A^{\rm NSM} A^{\rm S}\right)$ : & $ - 2 v \lambda \kappa $ \\
 			$\left(H^{\rm S} A^{\rm S} A^{\rm S}\right)$ : & $ - 2 \kappa \left( A_\kappa - 2\kappa\mu / \lambda \right)$ \\
 			\hline \hline
 		\end{tabular}
 		\caption{Trilinear couplings in the NMSSM Higgs sector in the extended Higgs basis.}
 		\label{tab:tricoup_NMSSM}
 	\end{center}
\end{table}

\newpage

%*********************************************************
\section{Additional Figures}\label{app:ratios}
%*********************************************************
\begin{figure}[h!]
	\includegraphics[width=.49\linewidth]{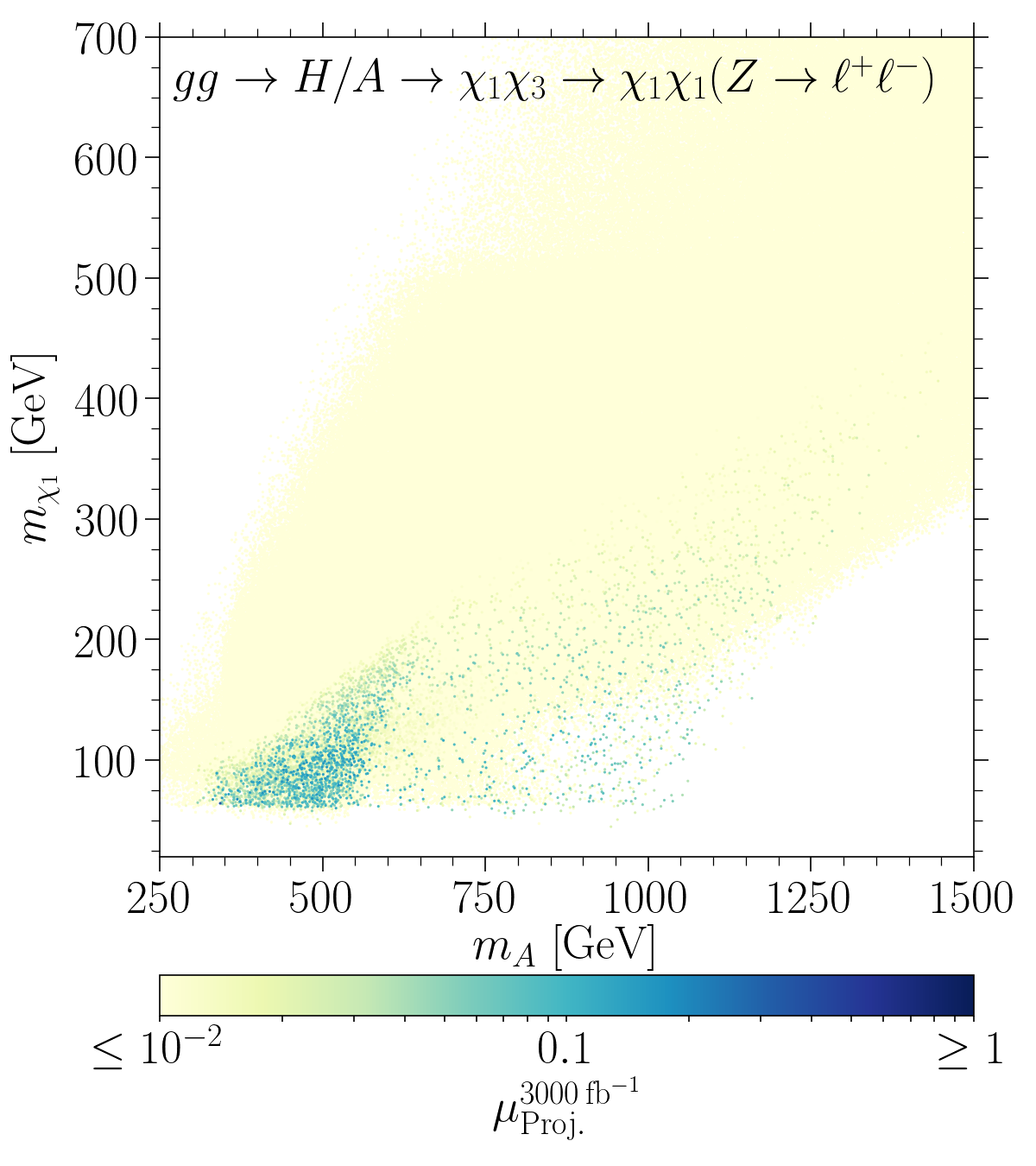}
	\includegraphics[width=.49\linewidth]{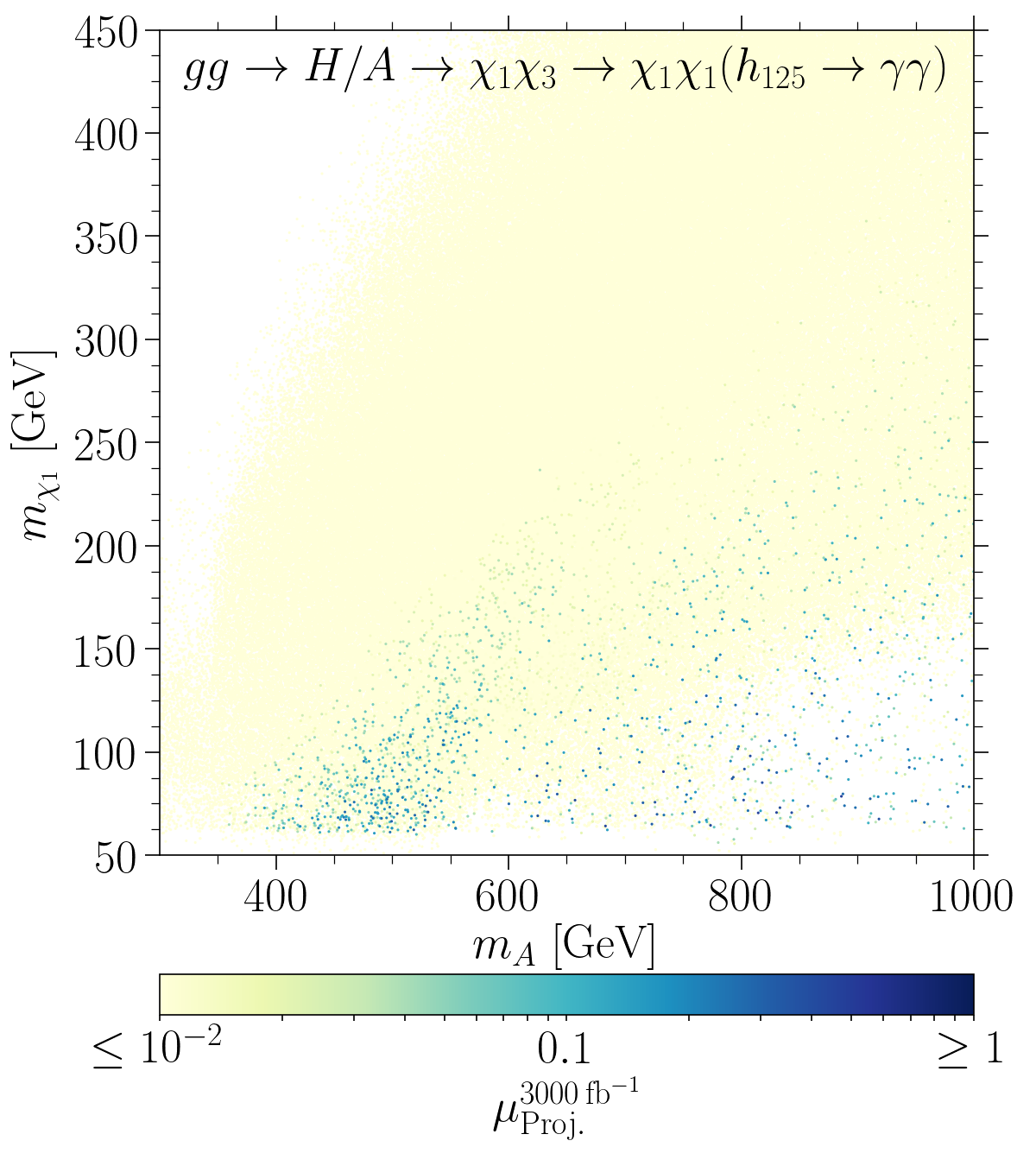}
	\caption{Same as Fig.~\ref{fig:variousreach1}, but for the mono-$Z$ (left) and mono-Higgs (right) final states arising through decays of the parent heavy Higgs into a pair of neutralinos where one of the neutralinos subsequently radiates off a $Z$ or a Higgs, cf. diagram (c) in Fig.~\ref{fig:Hdiagrams}.}
	\label{fig:variousreach2}
\end{figure}

\bibliographystyle{JHEP.bst}
\bibliography{nmssmbib}

\end{document}